# High-quality, deep formation and identification of oxygen-vacancy defects in silicon carbide

Ryo Shiotani[1,†], Fu-Chieh Chuang[1,†], Mitsuaki Kaneko[2,†], Tetsuri Nishikawa[1,3], Hiroshi Abe[4], Taiga Matsuoka[2], Shigehito Miki[5], Kazuki Okajima[1], Takeshi Ohshima[4,6], Yoshihiko Kanemitsu[1], Tsunenobu Kimoto[2], Naoya Morioka[1,3,*], and Norikazu Mizuochi[1,3]

1. Institute for Chemical Research, Kyoto University, Uji 611-0011, Japan
2. Department of Electrical, Electronic, and Digital Science and Engineering, Kyoto University, Kyoto 615-8510, Japan
3. Center for Spintronics Research Network, Institute for Chemical Research, Kyoto University, Uji 611-0011, Japan
4. National Institutes for Quantum Science and Technology (QST), Takasaki 370-1292, Japan
5. Advanced ICT Research Center, National Institute of Information and Communications Technology (NICT), Kobe 651-2492, Japan
6. Department of Materials Science, Tohoku University, Sendai 980-8579, Japan

[†]These authors contributed equally.

[*]e-mail: morioka.naoya.8j@kyoto-u.ac.jp

# Abstract

Optically addressable paramagnetic defects in silicon carbide (SiC) are attractive for scalable quantum technologies such as quantum sensing. PL5, PL6, and PL7 are promising defects for their high brightness and spin-readout contrast. Although generation by oxygen implantation has been reported very recently, nonuniform spin and optical properties and the limited generation depth of implanted defects hinder both high-sensitivity sensing and conclusive microscopic identification. Here we engineer high-quality defects by decoupling oxygen incorporation and defect formation, realizing highly homogeneous spin-resonance frequencies as well as narrow optical linewidths and long $T_2^*$ coherence time. Further, we demonstrate defect formation beyond the ion range utilizing oxygen diffusion, exceptional among impurities in SiC. We further provide hyperfine-interaction statistics of $^{17}$O and host nuclei across many PL6 defects, making the oxygen-vacancy origin conclusive. Together, the microscopic identification and the demonstrated formation route establish a foundation for developing the oxygen-vacancy spin defect into a versatile quantum platform.

Silicon carbide (SiC) has emerged as a scalable host for optically addressable spin defects in quantum technologies,[1–3] owing to its mature material growth, wafer-scale semiconductor processing, and the discovery of multiple defect species with favorable optical and spin properties.[4–9] Among these defects, PL5, PL6, and PL7 (PL5–PL7)[4,10] in 4H-SiC have attracted particular interest because they combine very bright near-infrared emission with high spin contrast at room temperature,[11] making them promising candidates for quantum sensing via optically detected magnetic resonance (ODMR).[12–14] PL6 is particularly notable because its high photon count rate (140–560 kcps)[11,15] and ODMR contrast (23–32%)[11] potentially bring SiC spin defects into a performance regime previously associated mainly with diamond nitrogen-vacancy centers.[16]

PL5–PL7 were first observed in as-grown high-purity semi-insulating (HPSI) 4H-SiC, but their microscopic origins have long remained under debate.[17–19] Implantation of various ion species has generated them, including carbon[11], nitrogen,[20,21] and helium,[22] albeit with ensemble yields significantly lower than those of divacancies. By contrast, electron irradiation followed by annealing, which efficiently generates divacancies, has not been reported to yield PL5–PL7,[23] suggesting an involvement of extrinsic impurities or complex defect configurations arising from implantation damage. Recently, oxygen implantation was shown to efficiently generate four distinct optically addressable species, PL5, PL6,[15,24] PL7′,[24] and PL8′.[24]

PL7′ can be assigned to previously known PL7 in HPSI material based on the zero-field splitting (ZFS) parameters.[25] PL8′ may correspond to the defect previously misassigned as "room-temperature ODMR-active PL1" found in ion-implanted and HPSI samples,[10,11,26,27] although PL1 divacancy is reported ODMR-inactive at room temperature.[10] Because PL8′ is unrelated to previously reported PL8,[28] we hereafter refer to it as PL1*.

For high-sensitivity ensemble quantum sensing, spin defects must be generated throughout a large sensing volume,[29] including regions deep below the surface. This remains difficult for PL5–PL7: in HPSI wafers, they have been reported mainly near the surface,[17,26] whereas ion implantation confines defect formation to the projected ion range. In addition, implantation-induced PL5 and PL6 exhibited large variations in their optical and spin properties,[15,30] which can degrade ensemble sensing performance.[29]

These variations also complicate the microscopic assignment of PL5–PL7 and PL1*. Oxygen-vacancy complexes ($O_cV_{Si}$) have been proposed as candidates by comparing experimentally observed hyperfine couplings to nearby host nuclear spins, zero-field splitting (ZFS) parameters, and zero-phonon lines (ZPLs) with first-principles calculations.[18,19] However, other complex configurations, including divacancy-carbon antisite pairs, have been predicted to show similar properties.[18] The reported variability in implantation-induced defects leaves the possibility that nominal PL5–PL7 and PL1* signals arise from a heterogeneous population of several defect complexes, including $O_cV_{Si}$ and modified divacancy-type configurations, rather than from a single $O_cV_{Si}$ structure.

Further, although six-line hyperfine splitting associated with implanted $^{17}O$ nuclei ($I = 5/2$) has been observed in individual PL5, PL6, and PL1* defects,[15,24] the measured couplings are smaller than DFT predictions for $O_cV_{Si}$.[18,31] The hyperfine-coupling measurements on single defects alone cannot determine whether oxygen is part of the defect or occupies a distant site, as its strength depends on the defect's spin-density distribution rather than simply distance from the vacancy.[32,33] Statistical analysis is therefore required to distinguish between these possibilities.

Here, we introduce a defect-engineering strategy that separates oxygen doping, lattice-damage recovery, and vacancy generation, enabling deep and high-quality formation of oxygen-containing spin defects in 4H-SiC (Fig. 1a). Using oxygen-doped 4H-SiC followed by electron irradiation and annealing, we generate PL5, PL6, PL7, and PL1$^*$ from single defects to ensembles. The resulting defects show markedly uniform spin resonance frequencies and reproduce the resonance frequencies reported for PL5–PL7 in HPSI 4H-SiC.[25,26] In addition, the created defects exhibit narrower photoluminescence (PL) linewidths than those of implantation-induced defects, and a relatively long $T_2^*$ time was observed. We also demonstrate PL5–PL7 and PL1$^*$ defect formation in the crystal beyond the projected ion implantation range via oxygen diffusion followed by electron irradiation and annealing, which breaks the limitation of oxygen incorporation through ion implantation. This result also proves the generation of these defects from separately introduced oxygen and vacancies. We further establish oxygen as a structural constituent of the PL6 defect through a statistical analysis of high-resolution hyperfine spectra from multiple PL6 defects created with $^{17}$O. We also extract the previously unreported perpendicular hyperfine coupling assuming axial symmetry, suggesting a nearly isotropic interaction consistent with DFT predictions for $O_CV_{Si}$.[18,31] These findings identify PL6's microscopic origin as an oxygen vacancy and provide a new foundation for materials engineering for SiC-based quantum sensors and devices.

## Results and Discussion

Oxygen is a deep double donor in SiC,[34,35] but it is an uncommon impurity, and controlled oxygen incorporation during epitaxial growth remains challenging.[34] Although oxygen ion implantation provides a practical route to introduce oxygen, it also generates a high density of lattice defects and substantially degrades crystallinity, which do not recover sufficiently at the post-annealing temperatures used to form PL5–PL7 and PL1$^*$. We therefore introduce a material-engineering strategy termed "decoupled engineering" that separates oxygen introduction and spin-defect formation (Fig. 1a). In this approach, oxygen-doped 4H-SiC is first prepared by oxygen implantation, followed by high-temperature annealing (HTA) to restore crystallinity and reduce residual damage. Vacancies are then introduced by electron irradiation, followed by post-irradiation annealing, to form spin defects.

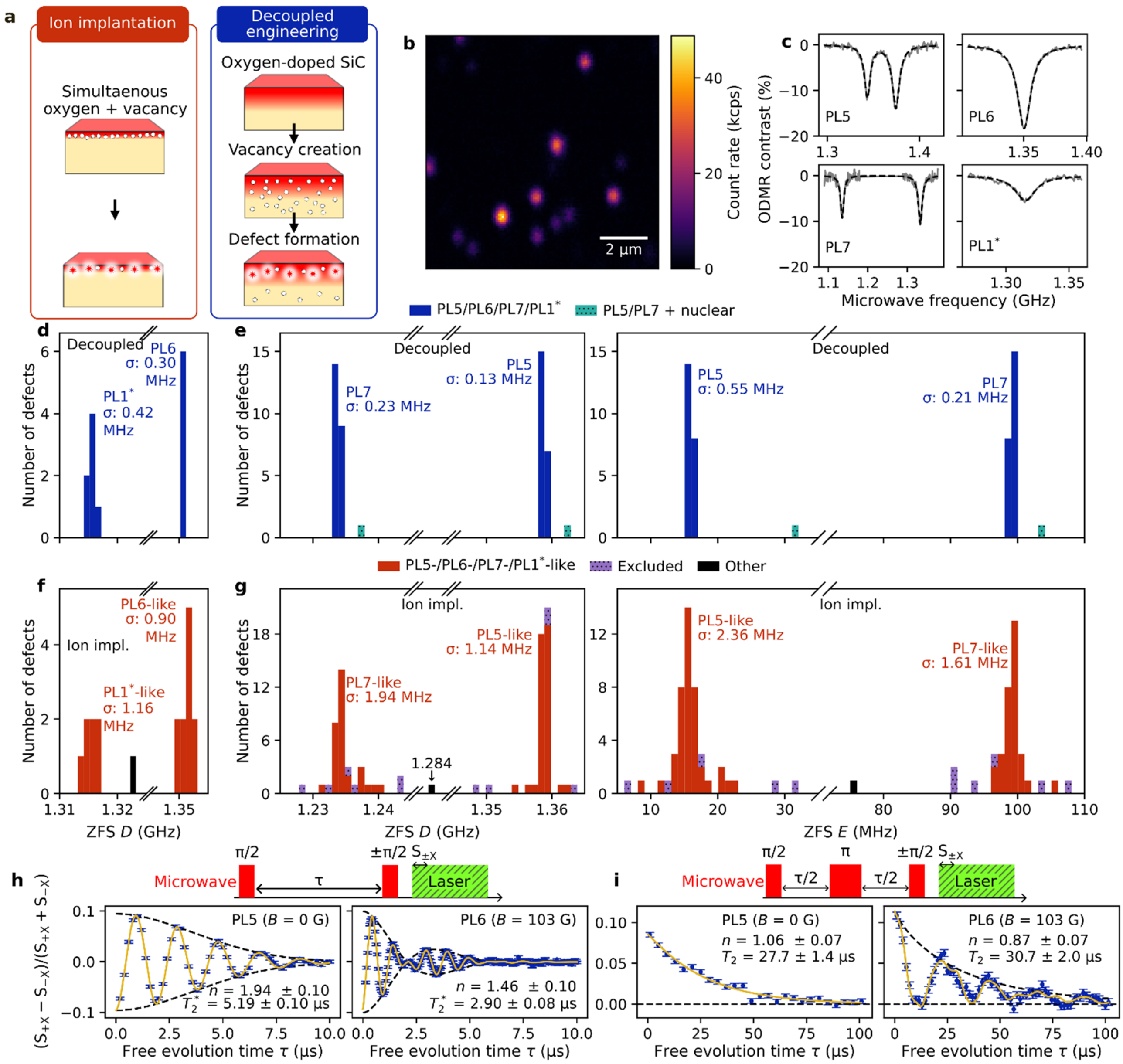

**Fig. 1: Single defect formation of PL5, PL6, PL7, and PL1* by decoupled defect engineering**

**a**, Defect formation by ion implantation (left) and decoupled engineering (right): decoupled oxygen doping and vacancy creation. **b**, Confocal map of defects created by decoupled engineering (Sample A1). **c**, Representative zero-field continuous-wave ODMR spectra of PL5–PL7 and PL1* in Sample A1. **d**, Histogram of the zero-field splitting (ZFS) parameter $D$ for PL6 and PL1* in Sample A1. **e**, $D$ (left) and $E$ (right) for PL5 and PL7 in Sample A1. Two defects in hatched teal are considered PL5 and PL7 with a strongly coupled nuclear spin (see Supplementary Note 4). σ values are the sample standard deviations, excluding the two strongly nuclear-spin-coupled defects. **f**, **g**, Histograms of ZFS parameters for defects created by the ion implantation method (Sample B) exhibiting small and large excitation-polarization dependence, respectively. Hatched purple indicates data excluded by a 95% Mahalanobis-distance criterion,[36] with black denoting those outside the main distributions and attributed to other defect species. $\sigma$ values are sample standard deviations for PL5-, PL6-, PL7-, and PL1*-like defects. The bin width of **d**–**g** is 1 MHz. **h**, Ramsey signals for $T_2^*$ of PL5 (left, 0.5-MHz detuning) and PL6 (right, 1.0-MHz detuning). Solid lines are fits with a stretched-exponential decay, $\exp[-(\tau/T_2^*)^n]$, and one (PL5) or two (PL6) oscillatory components. **i**, Hahn echo for $T_2$ of PL5 (left) and PL6 (right). Solid yellow lines are fits with $\exp[-(\tau/T_2)^n]$, including electron-spin-echo envelope modulation[37,38] for PL6 (see Supplementary Note 6). In **h** and **i**, the broken line indicates the decay envelope, and blue points show experimental data calculated using $S_{\pm x}$, the photon counts measured at the start of the laser pulse following the final $\pm\pi/2$ pulses. Error bars indicate photon-shot noise.

**Table 1: Zero-field splitting parameters of PL5, PL6, PL7, and PL1* at room temperature.**

[a] Mean values of Fig. 1d and 1e (Sample A1), excluding defects with strong hyperfine coupling. [b] Mean values of Fig. 1f and 1g (Sample B), excluding the data corresponding to other defects and the outliers. $E$ values for PL6 and PL1* were assumed to be zero but not experimentally determined because they could not be distinguished from the hyperfine splitting by measurements at zero field. [c] Values measured by the photoelectrical detection of magnetic resonance technique in high-purity semi-insulating (HPSI) 4H-SiC taken from Ref. [25]. [d] Value reported as PL1 created by carbon ion implantation.[11] [e] Value reported as PL1 in HPSI 4H-SiC.[26]

| | Decoupled engineering | | | Ion implantation | | Previous reports | |
|---|---|---|---|---|---|---|---|
| | ODMR (MHz)[a] | $D$ (MHz)[a] | $E$ (MHz)[a] | $D$ (MHz)[b] | $E$ (MHz)[b] | $D$ (MHz) | $E$ (MHz) |
| PL5 | 1342.96, 1374.58 | 1358.77 | 15.81 | 1358.9 | 16.0 | 1358.1[c] | 16.3[c] |
| PL6 | 1350.44 | 1350.44 | 0 | 1351.2 | n.d. | 1350.6[c] | 0.0 |
| PL7 | 1134.66, 1332.97 | 1233.81 | 99.16 | 1234.8 | 99.2 | 1233.6[c] | 99.0[c] |
| PL1* | 1315.52 | 1315.52 | 0 | 1315.3 | n.d. | 1317[d,e] | 0.0 |

We first apply this strategy to form individual defects. We prepared O-doped SiC using box-profile oxygen implantation followed by HTA at 1750 °C. At this stage, no fluorescent defects were observed in the implanted region; sparse surface defects were removed by sacrificial oxidation. Vacancies were then introduced by low-fluence electron irradiation followed by post-irradiation annealing at 900 °C (Sample A1, see also Methods and Supplementary Note 1). This process yielded isolated PL spots in confocal microscopy (Fig. 1b), indicating the formation of individual optically active defects. Zero-field ODMR measurements revealed four recurring species, assigned to PL5, PL6, PL7, and PL1*. Among 103 measured isolated spots, 60 showed ODMR; 58 were well assigned to these four species, comprising 22 PL5, 6 PL6, 23 PL7, and 7 PL1* defects. Representative ODMR spectra are shown in Fig. 1c, and the single-defect nature of the corresponding defects was confirmed by second-order autocorrelation (Supplementary Note 2). The distribution of apparent ZFS parameters $D$ and $E$ of 60 spots, extracted from the ODMR spectra, is summarized in Figs. 1d and 1e, respectively. Except for two spots, the distribution is very narrow for each defect type, and Table 1 summarizes their mean values. ZFSs of PL5–PL7 agree very well with those in HPSI,[25] and $D$ of PL1* agrees with the resonance frequency of "ion-implantation-induced PL1" (Table 1).[11] The remaining two ODMR-active defects, exhibiting PL5- and PL7-like spectra with slightly shifted resonance frequencies, can be assigned to PL5 and PL7 coupled to a nearest-neighbor $^{13}C$ nuclear spin (Supplementary Note 3).

By contrast, defects generated by oxygen implantation followed only by annealing up to 900 °C (Sample B), without HTA, could not be assigned to PL5–PL7 or PL1* unambiguously because of the large variation in their resonance frequencies. The distributions of $D$ and $E$ in Sample B are shown in Figs. 1f and 1g, respectively. To distinguish candidates for (PL5, PL7) and (PL6, PL1*), we plot defects with relatively large and small excitation-polarization dependence of their fluorescence separately.[27,30] The ZFS values of defects in Sample B (detailed analysis procedures are provided in Supplementary Note 4) lie near those of PL5–PL7 and PL1* in Sample A1, but they exhibit substantially broader distributions. The standard deviations are sub-MHz in Sample A1 and MHz in Sample B, indicating that HTA and the associated recovery of implantation-induced lattice damage are crucial for generating homogeneous defects. PL saturation curves for both samples confirmed that comparable optical properties can be achieved by either fabrication process (see Supplementary Note 5). Isolated PL5–PL7 and PL1* defects were generated at the surface just after HTA without sacrificial oxidation (Sample A2), whose results are presented in Supplementary Note 4.

Single defects generated by the decoupled engineering strategy also exhibited improved inhomogeneous dephasing times ($T_2^*$). Whereas previously reported room-temperature $T_2^*$ of single PL5 and PL6 are approximately 2 μs,[24] a PL5 and PL6 generated using our method exhibited $T_2^* = 5.19 \pm 0.10$ μs and $2.90 \pm 0.08$ μs, respectively (Fig. 1h). These longer $T_2^*$ suggest reduced low-frequency local noises, plausibly owing to the recovery of implantation-induced lattice damage and associated defect-related noise sources by HTA. The Hahn-echo coherence times ($T_2$) for PL5 and PL6 (Fig. 1i) were comparable to previously reported values.[24] Additional information on $T_2^*$ and $T_2$ measurements, including statistics for $T_2^*$ of PL6 are described in Supplementary Note 6.

Next, we examine ensemble formation using oxygen-doped 4H-SiC. After HTA, the sample was irradiated with a higher electron fluence and annealed at 800 °C to generate spin ensembles (Sample A3). Here, we omitted the sacrificial oxidation step because the ensemble defect dominates over the sparse surface defects. ODMR spectra confirmed the formation of PL5–PL7 and PL1*, with contrasts reaching approximately 5% (Fig. 2a). These contrasts are roughly an order of magnitude higher than those reported for PL5–PL7 ensembles in HPSI 4H-SiC,[10,26] indicating efficient ensemble formation in the oxygen-doped material.

To assess the role of HTA, we compared Sample A3 with a control sample that underwent oxygen implantation followed by the same electron irradiation and 800 °C annealing, but without HTA (Sample C1). ODMR spectra measured in a side-view geometry to probe the near-surface implanted region revealed higher ODMR contrasts and narrower linewidths in Sample A3 (Fig. 2b) than in Sample C1 (see also Supplementary Note 7). Consistent with the single-defect measurements, this improvement indicates that HTA reduces the inhomogeneity in the local defect environment by recovering implantation-induced lattice damage. The enhanced ODMR contrast is also attributed to the reduced formation of other near-infrared-emitting defects that contribute to the background PL.

PL spectra further support this interpretation. In the sample without HTA (Sample C1), the divacancy (PL1, PL2, and PL3) emissions were much stronger than PL5 and PL6,[4] whereas Sample A3, which underwent HTA, showed preferential formation of PL5–PL7 and PL1* (Fig. 2c). A sample prepared by oxygen implantation followed only by 800 °C annealing, without HTA or electron irradiation (Sample C2), showed a spectrum similar to Sample C1, suggesting little additional effect of electron irradiation without HTA. Although a previous oxygen implantation study reported strong suppression of PL1 and PL2,[24] further optimization of oxygen concentration, electron fluence, and annealing conditions may improve selectivity also in the decoupled engineering process. Importantly, even without such optimization, markedly improved optical and spin homogeneity in PL5–PL7 and PL1* ensembles have already been achieved.

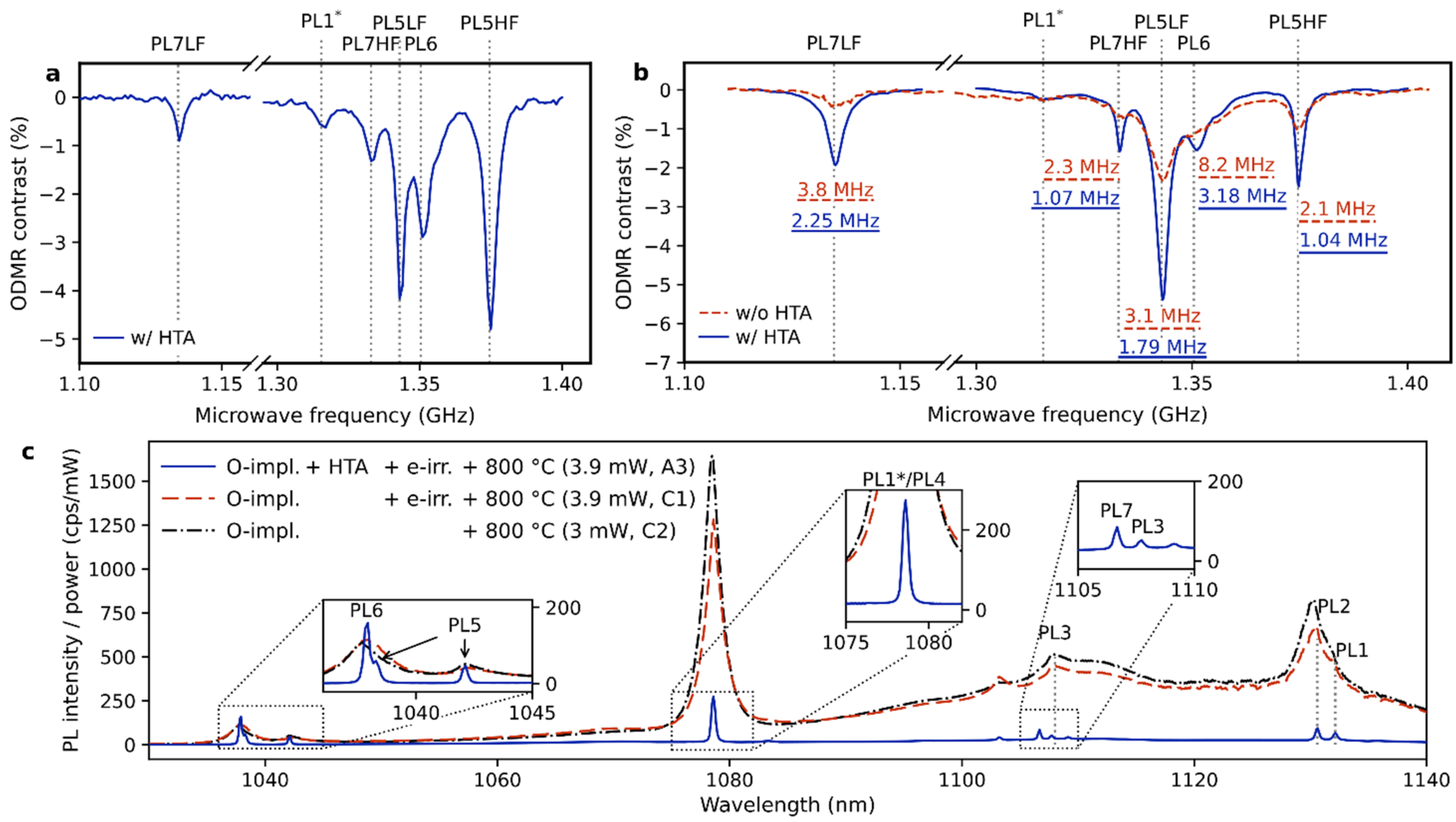

**Fig. 2: Ensemble generation of PL5, PL6, PL7, and PL1* by decoupled defect engineering.**

**a**, Continuous-wave ODMR (cw-ODMR) spectrum of defect ensembles fabricated by decoupled defect engineering (Sample A3), which underwent high-temperature annealing (HTA) after oxygen ion implantation. **b**, cw-ODMR spectra measured at the surface of the samples with and without HTA (Samples A3 and C1, respectively) in a side-view geometry. Values indicate the FWHM linewidths of the PL5–PL7 resonances; a reliable FWHM of PL1* could not be estimated. In **a** and **b**, the vertical dotted lines show the mean resonance frequencies of single defects fabricated by decoupled engineering (Sample A1; Table 1). **c**, Low-temperature photoluminescence spectra of defect ensembles prepared by different engineering methods: Samples A3 (solid blue), C1 (broken red), and C2 (dash-dotted black). Sample C2 was prepared only by oxygen implantation followed by defect-forming annealing, without HTA or electron-irradiation steps. Spectra were measured at a base-plate temperature of 4.3 K under 905-nm excitation and normalized to the excitation power (3.9, 3.9, and 3.0 mW for A3, C1, and C2, respectively). Insets show enlarged views around the zero-phonon lines of PL6 and PL5 (left), PL1* and PL4 (center), and PL7 and PL3 (right).[4,10,26,24] FWHMs by Lorentzian fits are 0.21, 0.21, and 0.25 nm for PL5, PL6, and PL7 in Sample A3, respectively.

---

In addition, the decoupled-engineered sample also exhibited narrow ZPL linewidths. The full widths at half maximum (FWHM) of PL5–PL7 ZPLs were 0.21–0.25 nm, corresponding to approximately two detector pixels and thus close to the instrumental resolution limit. These narrow linewidths enabled clear separation of PL3 from PL7, which have closely spaced ZPLs, and also resolved the second PL5 peak overlapping the spectral region near PL6.[24] PL1* and PL4 are reported to have closely spaced ZPLs[4,24] and remain unresolved here, with a 0.36-nm FWHM, implying a separation below this value. The observed linewidths are comparable to those of divacancies generated by electron irradiation and 800 °C annealing in a nitrogen-doped SiC epilayer (Supplementary Note 8), suggesting that narrower linewidths may be attributed to lattice-damage recovery during HTA. These results demonstrate that PL5–PL7 and PL1* ensembles formed by the decoupled engineering exhibit superior magnetic-resonance and optical properties.

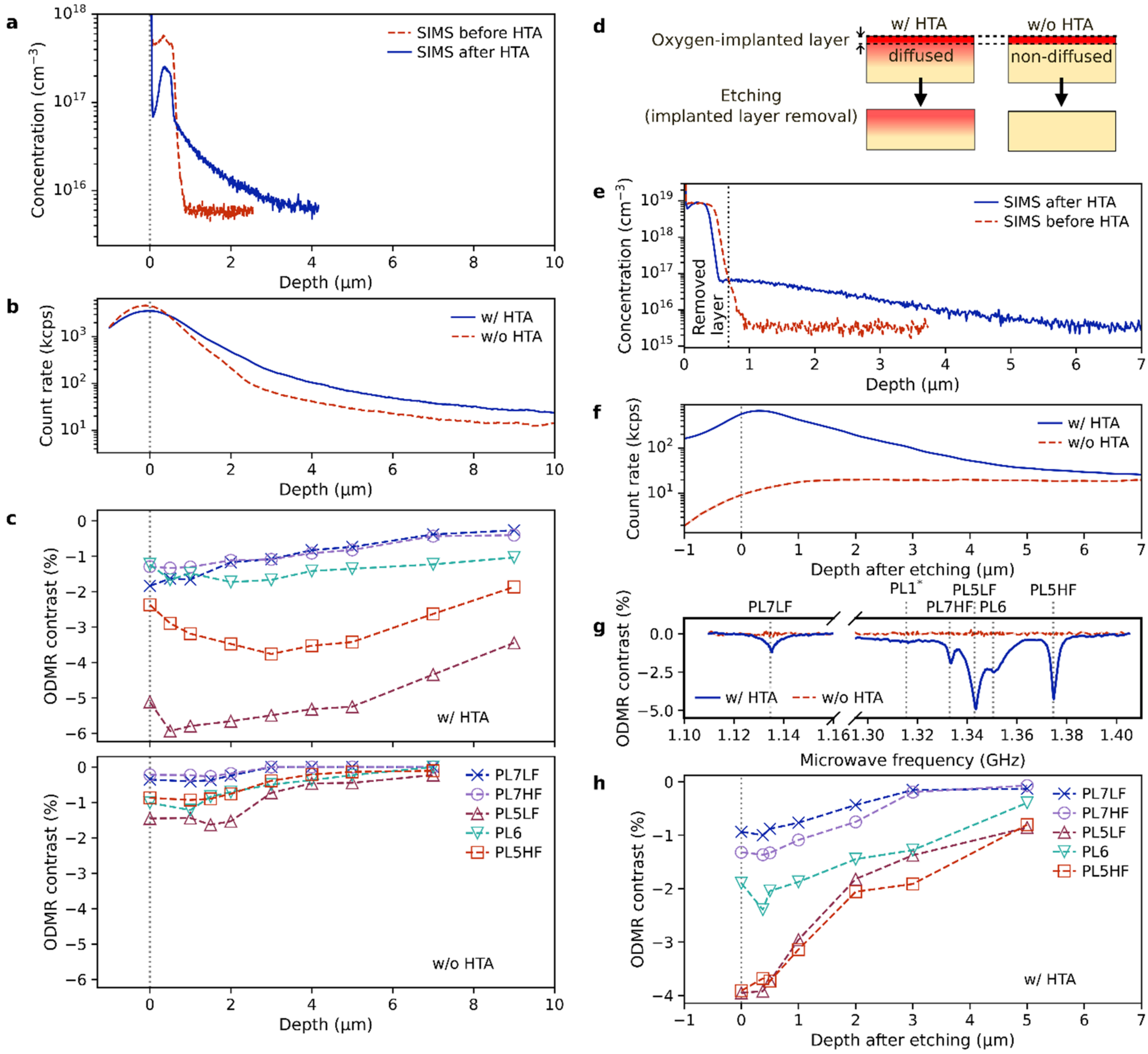

**Fig. 3: Defect formation beyond projected ion range by oxygen diffusion.**

**a**, Secondary ion mass spectrometry (SIMS) depth profiles of oxygen in oxygen-implanted samples before (broken red) and after (solid blue) high-temperature annealing (HTA) at 1750 °C. The detection limit is about $5\times10^{15}$ $cm^{-3}$. **b**, Photoluminescence (PL) intensity depth profiles of samples with (solid blue, Sample A3) and without (broken red, Sample C1) HTA between oxygen implantation and electron irradiation. **c**, Depth dependence of continuous-wave ODMR (cw-ODMR) contrasts of PL5–PL7 measured in Samples A3 (upper; w/ HTA) and C1 (lower; w/o HTA). ODMR contrasts were determined by fitting the cw-ODMR data with a sum of Lorentzian functions. The standard errors are smaller than the data points. The assignment of each resonance is indicated in Fig. 2a. **d**, Schematic of the implantation-layer-removal experiment. Samples D1 and D2 were prepared with and without HTA at 1900 °C, respectively. Both samples were subjected to etching to remove the implanted layers. **e**, SIMS depth profiles of oxygen measured before (broken red, corresponding to D2) and after (solid blue, corresponding to D1) HTA at 1900 °C. The dotted line indicates the etching depth, corresponding to the removal depth of the implanted layers. **f**, PL depth profiles measured in sample D1 (solid blue), which underwent HTA, and D2 (broken red), which did not undergo HTA. The detailed analysis procedure is described in Supplementary Note 9. **g**, Cw-ODMR spectra measured in Samples D1 (solid blue; w/ HTA) and D2 (broken red; w/o HTA). The vertical dotted lines show the mean resonance frequencies of single defects fabricated by decoupled engineering (Sample A1; Table 1). **h**, Depth dependence of cw-ODMR contrasts of PL5–PL7 measured in sample D1, analyzed as in **c**.

Having established the improved optical and spin properties of the ensembles, we next demonstrate that this approach enables PL5–PL7 and PL1* formation beyond the projected ion range. Most impurities introduced into SiC exhibit negligible thermal diffusion.[36] However, oxygen is an important exception and can diffuse appreciably in SiC at high temperature. Fig. 3a shows secondary-ion mass spectrometry (SIMS) data, showing oxygen diffusion over more than 2 μm after HTA at 1750 °C compared with the as-implanted profile. We examine whether oxygen diffusion during HTA could expand the PL5–PL7 and PL1* formation region beyond the implantation profile.

Fig. 3b compares the PL intensity depth profile of samples with and without HTA (Samples A3 and C1, respectively), measured from the cleaved side surface. The PL intensity of the HTA-treated sample decayed more gradually and remained stronger in deeper regions than in the sample without HTA (Fig. 3b). To confirm that the deep emission originates from PL5–PL7, we also measured the depth dependence of the ODMR contrast. As shown in Fig. 3c, the HTA-treated sample (Sample A3) showed measurable ODMR contrast down to depths of approximately 5 μm, followed by a gradual decrease at greater depths. In contrast, the control sample without HTA (Sample C1) showed a rapid decrease in ODMR contrast at 1 μm. These results demonstrate that oxygen diffusion during HTA enables the formation of PL5–PL7 spin ensembles well beyond the projected ion range.

To obtain more direct evidence of deep defect formation, we performed an implantation-layer-removal experiment (Fig. 3d). Samples D1 and D2 were prepared by box-profile oxygen implantation into the top 500 nm of 4H-SiC at $1 \times 10^{19}\ \mathrm{cm}^{-3}$. Only D1 was annealed at 1900 °C to diffuse oxygen, as seen in the SIMS profiles (Fig. 3e). We then removed approximately 0.68 μm from the surface of both samples by dry etching and thermal oxidation to remove the initially ion-implanted layer, followed by electron irradiation and annealing at 800 °C to form defects.

Cross-sectional PL measurements showed strong emission only for the HTA-treated sample D1 (Fig. 3f). ODMR measurements confirmed the formation of PL5–PL7 and PL1* only in Sample D1, with signals detected down to 5 μm (Figs. 3g and 3h). These results directly demonstrate that their formation is associated with oxygen, and that diffused oxygen can also serve as a source of defect formation. These findings establish that PL5–PL7 can be generated by electron irradiation and annealing in oxygen-doped SiC outside the ion-implantation range. This deep-formation capability is essential for creating large-volume spin ensembles for quantum sensing and may be enhanced by optimizing the oxygen doping profile, HTA, and irradiation conditions.

Finally, we aim to assign the atomic structure of PL5–PL7 and PL1* and deterministically identify oxygen as the constituent based on the statistical study of the hyperfine splitting using $^{17}$O ($I = 5/2$) implantation. We generated single defects with $^{17}$O using a normal ion implantation method (Sample E1) and decoupled engineering (Sample E2). For detailed sample fabrication, see Methods and Supplementary Note 1. Magnetic fields were applied along the defect axis to observe hyperfine splitting (Supplementary Note 10).

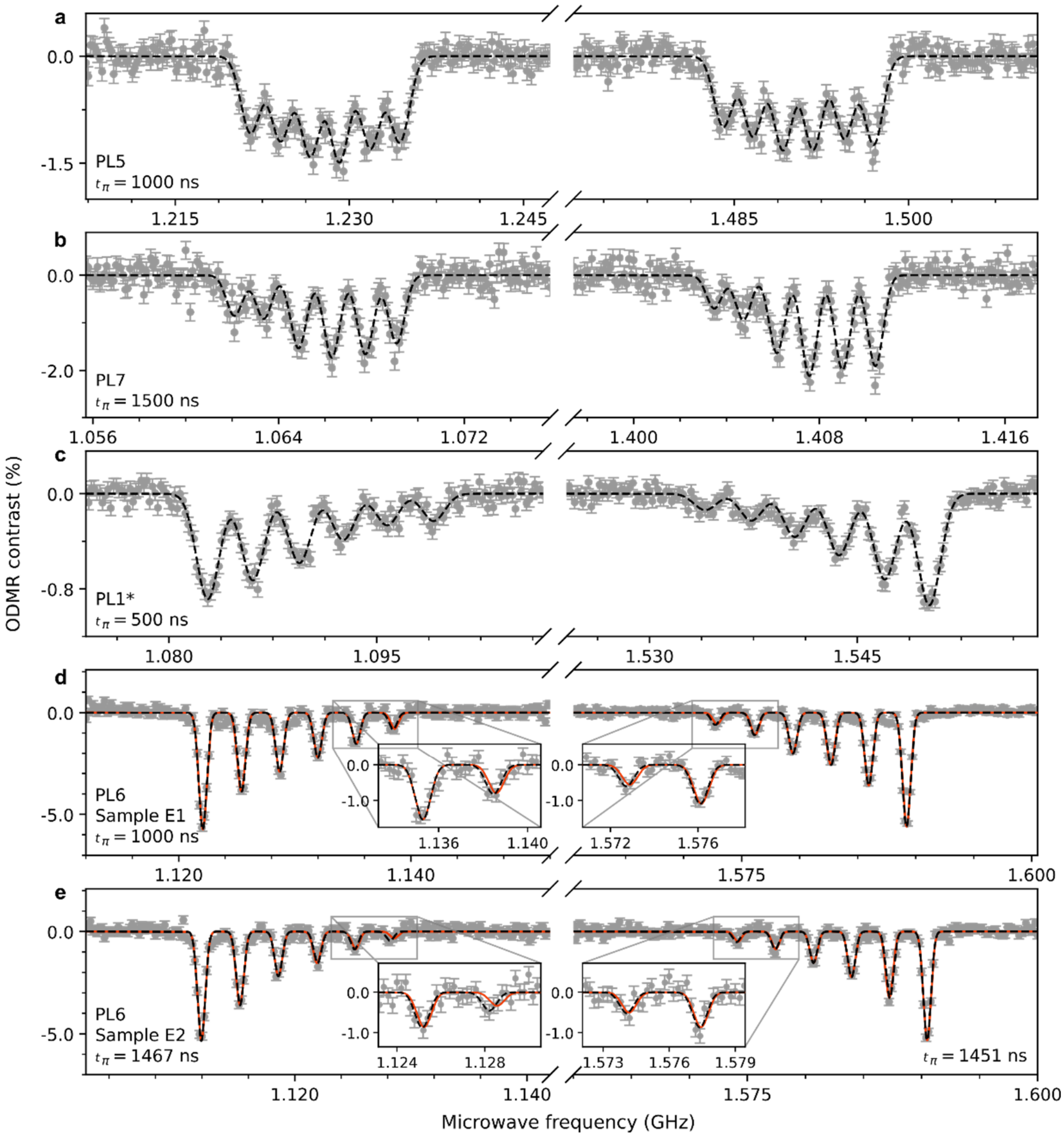

**Fig. 4: $^{17}$O hyperfine signatures of PL5, PL6, PL7, and PL1$^*$.**

**a-d,** Pulsed ODMR spectra of single defects formed by $^{17}$O implantation (Sample E1). **e,** PL6 formed by decoupled engineering. The $\pi$-pulse durations ($t_\pi$) are indicated. Hyperfine lines in each branch were fitted with Gaussian functions, assuming a common linewidth but unconstrained peak positions for six peaks. Black dashed curves show the summed fits. Red curves in **d** and **e** show the summed fits constrained to six equally spaced lines. Insets show enlarged views of data and fits. The error bars represent photon shot noise. The applied magnetic field along the principal *z*-axis for PL5's and PL7's ODMR measurements is 46.5 G and 49.5 G, respectively, estimated from the resonance frequencies assuming negligible perpendicular fields. PL6 and PL1$^*$ were measured under magnetic fields of 80.4 G (**c, d**) and 82.4 G (**e**), aligned with the principal *z*-axis of the defects (see Supplementary Note 10 for magnetic-field alignment and characterization). Zero-field ODMR and autocorrelation measurements confirm the assignments in **a-d** to single PL5, PL7, PL1$^*$, and PL6 defects (Supplementary Note 11). The low-contrast peaks favor fits with unconstrained peak positions (black dashed curves) over equally spaced fits (red curves), indicating unequal hyperfine splitting that can be explained by the perpendicular hyperfine coupling $A_\perp$ of PL6 (Supplementary Note 13).

---

Figs. 4a–4c shows ODMR spectra of single PL5, PL7, and PL1$^*$ defects in Sample E1, resolving the sextet likely arising from the $^{17}$O nuclear spin. Notably, the previously unresolved splitting of PL7 is resolved here. We observed

$m_I$-dependent ODMR contrast of PL6 and PL1*. Our simulation in Supplementary Note 14 suggests that the origin is dynamic nuclear-spin polarization (DNP).[10,37,38] Using the $A_{\parallel}S_zI_z$ term of the hyperfine interaction, we obtain $A_{\parallel} = 2.607 \pm 0.010$, $1.734 \pm 0.007$, and $3.249 \pm 0.011$ MHz, for PL5, PL7, and PL1*, respectively (Supplementary Note 12 for details). For PL6, our high-resolution spectra first reveal unequal hyperfine splittings (Fig. 4d and 4e) yielding $A_{\parallel} = 3.288 \pm 0.004$ and $3.295 \pm 0.005$ MHz and the previously unreported $|A_{\perp}| = 3.19 \pm 0.26$ and $3.46 \pm 0.27$ MHz for Samples E1 and E2, respectively, assuming axial symmetry (Supplementary Note 13).[39,40] The observed $A_{\parallel}$ for PL5, PL6, and PL1* agrees with previous reports[15,24] but are slightly smaller than the first-principles predictions for $O_cV_{Si}$ (5.1–5.2 MHz for axial defects).[18,31]

Here, we establish the reproducibility of this hyperfine fingerprint over a substantially larger number of single PL6 defects. We measured 8 and 19 defects in Samples E1 and E2, respectively. Owing to additional coupling to surrounding $^{13}C$ and $^{29}Si$ ($I = 1/2$), most defects exhibited more than six peaks. One example is shown in Fig. 5a, where we can find one $I = 5/2$ and three $I = 1/2$ nuclear spins. Similarly, we successfully identified $I = 5/2$ splitting in 24 out of 27 investigated defects in Samples E1 and E2 (Supplementary Note 15; two showed linewidths too broad to resolve, and one may originate from $I = 0$ oxygen isotope that possibly existed before implantation).

The resolved $I = 5/2$ and $I = 1/2$ splittings are summarized in Fig. 5b and 5c. Some defects were measured with a slightly misaligned magnetic field to mitigate the weak ODMR contrast originating from DNP. The identified $I = 5/2$ hyperfine splitting averaged 3.308 MHz with a standard deviation of 23 kHz. This highly uniform hyperfine fingerprint provides strong experimental evidence that oxygen occupies a well-defined site intrinsic to the defect complex and remains consistent across engineering strategies. The hyperfine coupling to surrounding $^{13}C$ and $^{29}Si$ is also consistent with the previous experimental results from $^{16}O$-implanted PL6[24] and DFT results of $O_CV_{Si}$,[18] supporting the $O_CV_{Si}$ origin.

For PL7, the newly identified smallest $^{17}O$ hyperfine parameter, together with recent discussion on ZPL (PL7′, PL3a)[24,25] and large $E$[25], identify PL7 as $O_CV_{Si}$ (*hk*),[18,31] in addition to the recently proposed assignment of PL6, PL1*, and PL5 as *hh*, *kk*, and *kh* configurations, respectively.[18] Although the measured hyperfine couplings are smaller in magnitude than DFT predictions, the agreement in the $|A_{\perp}|/A_{\parallel}$ ratio further supports the $O_CV_{Si}$ origin.

In summary, we have demonstrated the generation of single and ensemble PL5, PL6, PL7, and PL1* defects in oxygen-doped SiC through decoupled engineering, achieving highly uniform spin-resonance properties and narrow optical linewidths. Statistical analysis of the $^{17}O$ hyperfine interaction provides strong evidence of oxygen occupying a well-defined site. Together with the requirement for both oxygen and vacancy-generating irradiation revealed from the decoupled engineering, our hyperfine results indicate $O_CV_{Si}$ as the origin. Decoupling oxygen incorporation from vacancy generation provides flexible routes to engineer defects for different applications. Oxygen diffusion or doping during growth could enable thick defect-containing SiC layers beyond the implantation range for high-sensitivity quantum sensing, while spatially selective vacancy creation by focused ion beam[41] or laser writing[42] could enable localized, high-quality defect generation for localized sensing[43,44] and spin–photon interfaces[45,46]. Together with its

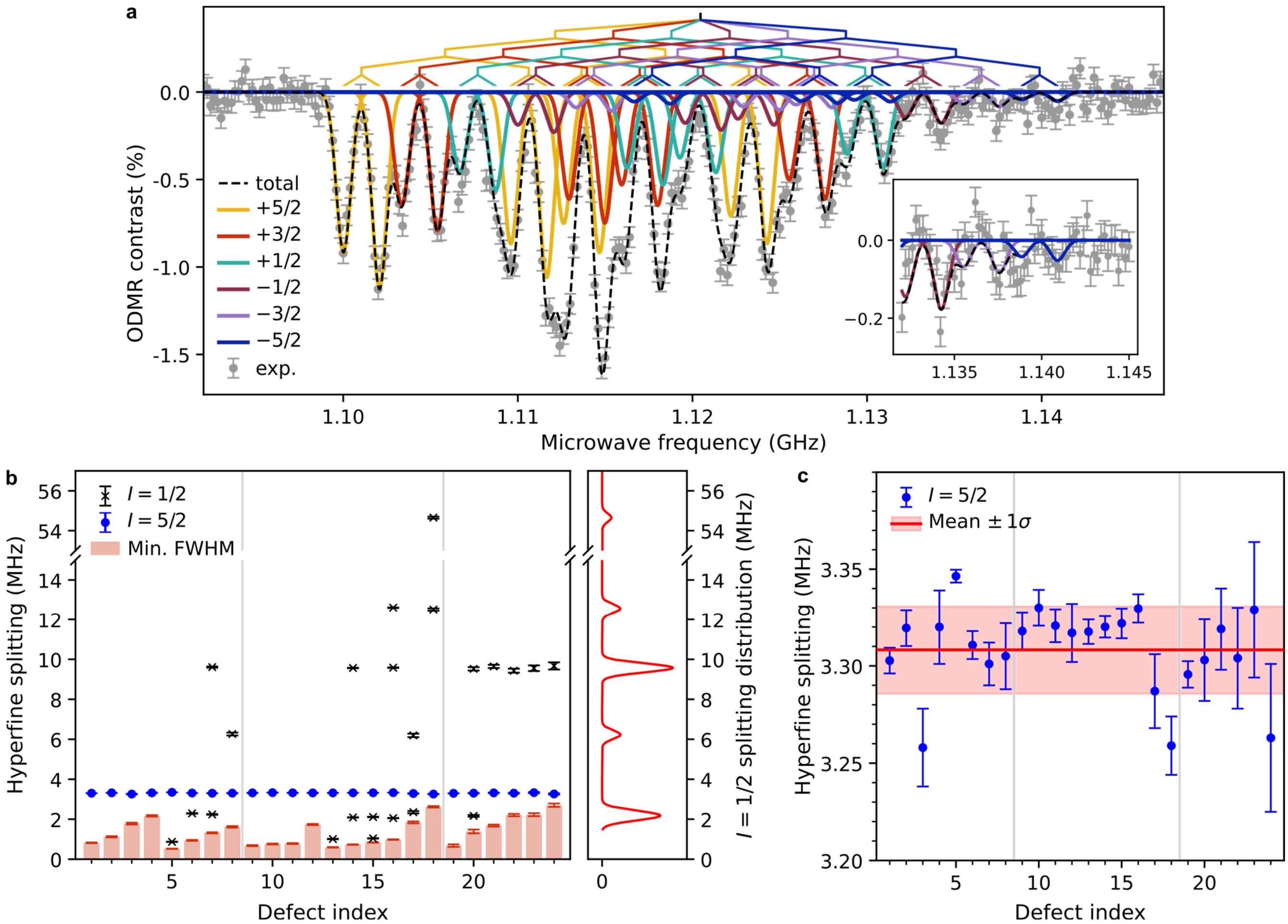

**Fig. 5: Hyperfine-splitting statistics of PL6 coupled to $^{17}$O and host nuclear spins.**

**a,** Pulsed ODMR spectra for the $m_S = 0 \leftrightarrow -1$ branch of single PL6 generated in $^{17}$O-doped 4H-SiC by decoupled engineering (Sample E2). The inset shows a longer-integration measurement of the low-contrast region arising from the $m_I$-dependent $^{17}$O contrast. Dots and error bars show the experimental data and photon shot noise, respectively. Solid lines are Gaussian fits assuming equally spaced $I = 5/2$ hyperfine components and a common linewidth. The spectrum is well reproduced by a model comprising one $I = 5/2$ nucleus and three $I = 1/2$ nuclei, with hyperfine splittings of 3.330 ± 0.007 MHz and 12.605 ± 0.014, 9.582 ± 0.014, and 2.055 ± 0.008 MHz, respectively. The hyperfine-splitting diagram above follows this order from top to bottom. For the fit, the inset data were used in place of the main-spectrum data in the low-contrast region (Supplementary Note 15). **b,** Summary of the extracted hyperfine splittings from 24 PL6 defects in $^{17}$O samples E1 and E2. Blue circules and black crosses with error bars indicate the fitted hyperfine splittings and their standard errors for the $I = 5/2$ and $I = 1/2$ nuclei, respectively. The fitting assumed equally spaced hyperfine components for $I = 5/2$. Defect indices 1–8 and 9–18 correspond to defects in Samples E1 and E2, respectively, measured under an aligned magnetic field. Defect indices 19–24 correspond to defects in Sample E2 measured under a misaligned magnetic field. Magnetic-field conditions are provided in Supplementary Note 10. The fitted minimum Gaussian linewidth and its standard error for each defect are shown by a bar and an error bar, respectively, as their resolution limits. The distribution of the extracted $I = 1/2$ hyperfine splitting is shown on the right after Gaussian smoothing with an FWHM of 0.5 MHz. **c,** Enlarged view of $I = 5/2$ hyperfine splittings from **b**.

---

compatibility with wafer-scale fabrication and the favorable optical and spin properties, these capabilities establish a promising defect-engineering route toward scalable quantum technologies.

# References

1. Castelletto, S. *et al.* Silicon Carbide Photonics Bridging Quantum Technology. *ACS Photonics* **9**, 1434–1457 (2022).
2. Roberts, H., Abudayyeh, H., Li, X. & Li, X. Quantum Sensing with Spin Defects Beyond Diamond. *ACS Nano* **19**, 22528–22575 (2025).
3. Son, N. T. *et al.* Developing silicon carbide for quantum spintronics. *Appl. Phys. Lett.* **116**, 190501 (2020).
4. Koehl, W. F., Buckley, B. B., Heremans, F. J., Calusine, G. & Awschalom, D. D. Room temperature coherent control of defect spin qubits in silicon carbide. *Nature* **479**, 84–87 (2011).
5. Widmann, M. *et al.* Coherent control of single spins in silicon carbide at room temperature. *Nat. Mater.* **14**, 164–168 (2015).
6. Wang, J.-F. *et al.* Coherent Control of Nitrogen-Vacancy Center Spins in Silicon Carbide at Room Temperature. *Phys. Rev. Lett.* **124**, 223601 (2020).
7. Wolfowicz, G. *et al.* Vanadium spin qubits as telecom quantum emitters in silicon carbide. *Sci. Adv.* **6**, eaaz1192 (2020).
8. Shafizadeh, D. *et al.* Spin-active chlorine-related centers in 4H-SiC with telecom-band emissions. *Appl. Phys. Lett.* **129**, 104001 (2026).
9. Anisimov, A. N., Mavridou, K., Mathews, A. V., Helm, M. & Astakhov, G. V. Telecom-band quantum memory with chlorine defects in silicon carbide. Preprint at https://doi.org/10.48550/arXiv.2605.03717 (2026).
10. Falk, A. L. *et al.* Polytype control of spin qubits in silicon carbide. *Nat. Commun.* **4**, 1819 (2013).
11. Li, Q. *et al.* Room-temperature coherent manipulation of single-spin qubits in silicon carbide with a high readout contrast. *Natl. Sci. Rev.* **9**, nwab122 (2022).
12. Zhou, Y. *et al.* Self-Protected Thermometry with Infrared Photons and Defect Spins in Silicon Carbide. *Phys. Rev. Appl.* **8**, 044015 (2017).
13. Quan, W.-K., Liu, L., Luo, Q.-Y., Liu, X.-D. & Wang, J.-F. Fiber-coupled silicon carbide divacancy magnetometer and thermometer. *Opt. Express* **31**, 15592–15598 (2023).
14. Li, P. *et al.* Non-invasive bioinert room-temperature quantum sensor from silicon carbide qubits. *Nat. Mater.* **24**, 1913–1919 (2025).
15. Chen, Y. *et al.* Oxygen-vacancy origin of robust surface-proximity spin centers in silicon carbide. *Innov. Phys.* **1**, 100013 (2026).
16. Jelezko, F. & Wrachtrup, J. Single defect centres in diamond: A review. *Phys. Status Solidi A* **203**, 3207–3225 (2006).
17. Ivády, V. *et al.* Stabilization of point-defect spin qubits by quantum wells. *Nat. Commun.* **10**, 5607 (2019).
18. Zhao, X., Liu, M., Chen, Y., Zhang, Q. & Duan, C.-K. Towards identifying the PL6 center in SiC: From first-principles screening to hyperfine validation of competing defect candidates. *Phys. Rev. Mater.* **9**, 116204 (2025).
19. Bai, R., Liu, D., Chen, S. & Wu, Y.-N. Origin of the unidentified color centers in 4H-SiC from first principles. *Phys. Rev. B* **111**, 014106 (2025).
20. Lin, W.-X. *et al.* Temperature dependence of divacancy spin coherence in implanted silicon carbide. *Phys. Rev.*

*B* **104**, 125305 (2021).

21. Wörnle, R. *et al.* RF-free driving of nuclear spins with color centers in silicon carbide. Preprint at https://doi.org/10.48550/arXiv.2601.21499 (2026).
22. He, Z.-X. *et al.* Robust single modified divacancy color centers in 4H-SiC under resonant excitation. *Nat. Commun.* **15**, 10146 (2024).
23. Son, N. T., Shafizadeh, D., Ohshima, T. & Ivanov, I. G. Modified divacancies in 4H-SiC. *J. Appl. Phys.* **132**, 025703 (2022).
24. Hu, Q.-C. *et al.* High-Yield Engineering and Identification of Oxygen-Related Modified Divacancies in 4H-SiC. *Adv. Mater.* **38**, e73419 (2026).
25. Morioka, N., Nishikawa, T., Abe, H., Ohshima, T. & Mizuochi, N. Photoelectrical detection and characterization of divacancy and PL5–PL7 spins in silicon carbide. *Phys. Rev. B* **113**, 104426 (2026).
26. Shafizadeh, D., Son, N. T., Abrikosov, I. A. & Ivanov, I. G. Evolution of the optically detected magnetic resonance spectra of divacancies in 4 H -SiC from liquid-helium to room temperature. *Phys. Rev. B* **111**, 165201 (2025).
27. Zhou, J.-Y. *et al.* Experimental Determination of the Dipole Orientation of Single Color Centers in Silicon Carbide. *ACS Photonics* **8**, 2384–2391 (2021).
28. Yan, F.-F. *et al.* Room-temperature coherent control of implanted defect spins in silicon carbide. *Npj Quantum Inf.* **6**, 38 (2020).
29. Barry, J. F. *et al.* Sensitivity optimization for NV-diamond magnetometry. *Rev. Mod. Phys.* **92**, 015004 (2020).
30. Eserin, S. *et al.* Efficient formation and identification of single emitters in 4H-SiC following maskless heavy ion implantation. Preprint at https://doi.org/10.48550/arXiv.2608.27281 (2026).
31. Kobayashi, T., Shimura, T. & Watanabe, H. Oxygen-vacancy defect in 4H-SiC as a near-infrared emitter: An ab initio study. *J. Appl. Phys.* **134**, 145701 (2023).
32. Gali, A., Fyta, M. & Kaxiras, E. Ab initio supercell calculations on nitrogen-vacancy center in diamond: Electronic structure and hyperfine tensors. *Phys. Rev. B* **77**, 155206 (2008).
33. Ivády, V. *et al.* Identification of Si-vacancy related room-temperature qubits in 4H silicon carbide. *Phys. Rev. B* **96**, 161114 (2017).
34. Dalibor, T. *et al.* Oxygen in silicon carbide: shallow donors and deep acceptors. *Mater. Sci. Eng. B* **61–62**, 454–459 (1999).
35. Gali, A. *et al.* Isolated oxygen defects in 3C- and 4H-SiC: A theoretical study. *Phys. Rev. B* **66**, 125208 (2002).
36. Roccaforte, F., Fiorenza, P., Vivona, M., Greco, G. & Giannazzo, F. Selective Doping in Silicon Carbide Power Devices. *Materials* **14**, 3923 (2021).
37. Jacques, V. *et al.* Dynamic Polarization of Single Nuclear Spins by Optical Pumping of Nitrogen-Vacancy Color Centers in Diamond at Room Temperature. *Phys. Rev. Lett.* **102**, 057403 (2009).
38. Ivády, V. *et al.* Theoretical model of dynamic spin polarization of nuclei coupled to paramagnetic point defects in diamond and silicon carbide. *Phys. Rev. B* **92**, 115206 (2015).
39. Kothandaraman, G., Pratt, D. W. & Tinti, D. S. Optically detected magnetic resonance spectra of the lowest triplet state of 1-iodonaphthalene in durene at zero and high magnetic fields. *J. Chem. Phys.* **63**, 3337–3348

(1975).
40. Yamauchi, S. & W. Pratt, D. Oxygen-17 hyperfine structure in the zero-field odmr spectrum of triplet state benzophenone. *Chem. Phys. Lett.* **60**, 145–149 (1978).
41. Wang, J. *et al.* Scalable Fabrication of Single Silicon Vacancy Defect Arrays in Silicon Carbide Using Focused Ion Beam. *ACS Photonics* **4**, 1054–1059 (2017).
42. Chen, Y.-C. *et al.* Laser Writing of Scalable Single Color Centers in Silicon Carbide. *Nano Lett.* **19**, 2377–2383 (2019).
43. Yamazaki, Y. *et al.* Quantum electrometry in a silicon carbide power device. Preprint at https://doi.org/10.48550/arXiv.2603.13836 (2026).
44. Sasaki, K. *et al.* Magnetic field imaging by hBN quantum sensor nanoarray. *Appl. Phys. Lett.* **122**, 244003 (2023).
45. Knaut, C. M. *et al.* Entanglement of nanophotonic quantum memory nodes in a telecom network. *Nature* **629**, 573–578 (2024).
46. He, Z.-X. *et al.* Coherent Spin-Photon Interface of single PL6 Color Centers in Silicon Carbide. Preprint at https://doi.org/10.48550/arXiv.2602.06421 (2026).

# Methods

## Sample preparation.

All samples doped or implanted with oxygen were prepared from several nitrogen-doped *n*-type 4H-SiC epitaxial wafers with net donor concentrations ranging from $4 \times 10^{13}$ to $4 \times 10^{14}\ \mathrm{cm}^{-3}$. Detailed information on the epitaxial wafers is provided in Supplementary Note 1.

All high-temperature annealing (HTA) treatments in this study were performed in Ar for 20 min with a carbon cap. All defect-formation annealing treatments following irradiation or implantation were performed in Ar for 30 min without a cap. All electron irradiations were performed at an energy of 2 MeV. All sacrificial oxidation treatments were performed in dry $O_2$ at 1300 °C for 10 hr, consuming approximately 180 nm of SiC, followed by HF etching to remove the grown $SiO_2$.

Samples A1, A2, A3, C1 and C2 were implanted with $^{16}O$ ions to form a 600-nm-deep box profile with a concentration of $5 \times 10^{17}\ \mathrm{cm}^{-3}$. A1, A2, and A3 underwent HTA at 1750 °C, whereas C1 and C2 did not undergo HTA. For A1 only, the surface layer was removed by sacrificial oxidation after HTA. Then, A1 was irradiated with electrons to a fluence of $2 \times 10^{13}\ \mathrm{cm}^{-2}$ and subsequently annealed at 900 °C. A2 was examined immediately after HTA. C2 was annealed at 800 °C after $^{16}O$ implantation to form defects. A3 and C1 were irradiated with electrons to $1 \times 10^{17}\ \mathrm{cm}^{-2}$ followed by annealing at 800 °C.

Sample B was implanted with $^{16}O$ ions at 60 keV to a fluence of $1 \times 10^{12}\ \mathrm{cm}^{-2}$, through 40-nm-diameter apertures patterned in an electron beam resist, followed by successive annealing treatments at 300, 500, 600, and 900 °C, in this order.

Samples D1 and D2 were implanted with $^{16}O$ ions to form a 500-nm-deep box profile with a concentration of $1 \times 10^{19}$ cm$^{-3}$. D1 underwent HTA at 1900 °C, whereas D2 did not. The ion-implanted layers of D1 and D2 were then removed by reactive ion etching (approximately 500 nm), and sacrificial oxidation. Both samples were subsequently subjected to electron irradiation to a fluence of $1 \times 10^{17}$ cm$^{-2}$, followed by annealing at 800 °C.

Sample E1 was implanted with $^{17}O$ ions at 60 keV to a fluence of $1 \times 10^{12}$ cm$^{-2}$, through 40-nm-diameter resist apertures, followed by annealing at 800 °C. Sample E2 was implanted with $^{16}O$ ions to form a 420-nm-deep box profile with a concentration of $5 \times 10^{17}$ cm$^{-3}$, followed by HTA at 1750 °C, sacrificial oxidation, electron irradiation ($1 \times 10^{14}$ cm$^{-2}$), and annealing at 800 °C. For $^{17}O$ isotope implantation, $^{17}O$-enriched $O_2$ gas containing 70.4 at. % $^{17}O$ was used as the source gas, and the generated ions were mass-selected prior to implantation.

The fabricated SiC sample was mounted on a custom-made printed circuit board (PCB), and an aluminum wire with a diameter of 25 μm was placed on the sample surface to deliver the microwave.

## Measurement setups.

Room-temperature measurements were performed using two different home-built confocal laser-scanning microscope (CLSM) setups, CLSM1 and CLSM2. Data for the $^{17}O$ samples (E1 and E2) and for the $^{16}O$ sample A1 shown in Figs. 1b–1e were acquired using CLSM2. All other data for the $^{16}O$ samples were acquired using CLSM1. Both CLSMs share similar optical configurations. PL5–PL7 and PL1$^*$ were measured using 940 nm excitation lasers (Sacher Lasertechnik, Serval Plus TEC-420-0915-1500 for CLSM1; Thorlabs M9-940-0200 for CLSM2). The excitation laser was modulated using an acousto-optic modulator, and its polarization was controlled by a half-wave plate. The laser was focused onto the sample using an oil-immersion objective lens (Nikon, CFI Plan Apo Lambda D 100X Oil, NA = 1.45). The photoluminescence (PL) was collected through the same objective, separated from the excitation light by a 1000-nm long-pass dichroic mirror (Thorlabs, DMLP1000), and further filtered using 937-nm (Semrock, FF01-937/LP-25) and 950-nm (Thorlabs, FELH0950) long-pass filters for CLSM1 and CLSM2, respectively, followed by 1000-nm and 1050-nm (Thorlabs, FELH1000 and FELH1050) long-pass filters common to both setups. The PL was then focused using an air objective lens (Shibuya Optical, M-iPLAN-APO NIR10, NA = 0.28) and coupled to a single-mode fiber (HI1060), with the fiber core serving as the confocal pinhole. The polarization in the fiber was adjusted using a manual paddle fiber polarization controller. The PL was detected using home-built superconducting nanowire single-photon detectors. For the autocorrelation measurements, we inserted a fiber-beam splitter (Thorlabs, TW1064R5A1A) before the polarization controller to construct a Hanbury Brown–Twiss interferometer. Photon detection events were processed using a Time Tagger 20 (Swabian Instruments). The microwave was pulsed using a microwave switch. Control pulses for laser modulation and microwave switching were generated by timing-control electronics (Tektronix, AWG5014C for CLSM1; Swabian Instruments, Pulse Streamer 8/2 for CLSM2).

Low-temperature photoluminescence spectroscopy was performed using a laser microspectroscopy system equipped with a liquid-helium flow cryostat with a base-plate temperature of approximately 4.3 K. Excitation was provided by a 905-nm laser through an NIR objective lens (Olympus, LCPLN20XIR, NA = 0.45). The fluorescence was separated from the excitation light using 980-nm long-pass edge filters, passed through a 100-μm pinhole, and spectrally dispersed by an imaging Czerny-Turner spectrometer (SOL Instruments, MS5004i, focal length = 520 mm) equipped with a 400 grooves/mm grating blazed at 1.2 μm. The spectra were recorded using an NIR camera (Andor, iDUS DU491A-1.7). The instrumental spectral resolution was approximately 0.22 nm.

## Acknowledgements

We thank H. Tsuchida and K. Murata (Central Research Institute of Electric Power Industry) for providing the SiC epitaxial samples used in part of this work. This work was supported by JST PRESTO (Grant No. JPMJPR245C), JSPS KAKENHI (Grant No. JP25K01262), research grant from the Asahi Glass Foundation, research grant from the Mazda Foundation, MEXT Q-LEAP (Grant No. JPMXS0118067395), the Spintronics Research Network of Japan, and JST ASPIRE (Grant No. JPMJAP24C1). This work was partly supported by JSPS KAKENHI (Grant No. 23K19120). A part of this work was conducted in Nanotechnology Hub in Kyoto University, supported by "Advanced Research Infrastructure for Materials and Nanotechnology in Japan (ARIM)" of the Ministry of Education, Culture, Sports, Science and Technology (MEXT) (Proposal No. JPMXP1226KT1249). K.O. acknowledges support from JST SPRING (Grant No. JPMJSP2110). The SNSPD devices were fabricated using the facilities of the Advanced ICT Device R&D Promotion Center, NICT.

## Author contributions

NMo conceived the project and, together with NMi, supervised the work. MK fabricated lattice-recovered oxygen-doped SiC and discovered oxygen diffusion, which led NMo to develop the concept of deep defect formation. RS carried out the principal work on defect engineering, while FCC carried out the principal work on the hyperfine interaction.

RS and TN created single defects using electron-beam lithography and ion implantation and characterized the resulting defects from $^{16}O$ implantation. NMo, RS, FCC, TN, and NMi coordinated the $^{17}O$ implantation. MK fabricated oxygen-doped SiC samples, including high-dose $^{16}O$ implantation, high-temperature annealing, surface etching, sacrificial oxidation, and doping-profile characterization, with support from TK and TM. RS and TN characterized surface defects after high-temperature annealing. HA and TO performed electron irradiation. RS, NMo, and TN performed defect-formation annealing. RS and NMo performed room-temperature characterization of ensemble defects. NMo, MK, and TN performed low-temperature photoluminescence measurements. NMo, FCC, and RS performed ODMR measurements on single defects created by decoupled engineering from $^{16}O$. RS and NMo measured the spin coherence times. FCC and NMo characterized the hyperfine interaction with $^{17}O$. KO supported the measurements and annealing processes. SM designed and fabricated the SNSPDs, which SM, RS, TN, and NMo characterized with support from YK. TN, FCC, and NMo constructed confocal microscope setups. TN performed DNP simulations with guidance from NMo. RS, FCC, TN, NMo, and NMi wrote the manuscript. All authors discussed the results and reviewed the manuscript.

Supplementary Information

# High-quality, deep formation and identification of oxygen-vacancy defects in silicon carbide

(Dated: October 23, 2026)

Ryo Shiotani[1,†], Fu-Chieh Chuang[1,†], Mitsuaki Kaneko[2,†], Tetsuri Nishikawa[1,3], Hiroshi Abe[4], Taiga Matsuoka[2], Shigehito Miki[5], Kazuki Okajima[1], Takeshi Ohshima[4,6], Yoshihiko Kanemitsu[1], Tsunenobu Kimoto[2], Naoya Morioka[1,3,*], and Norikazu Mizuochi[1,3]

1. Institute for Chemical Research, Kyoto University, Uji 611-0011, Japan
2. Department of Electrical, Electronic, and Digital Science and Engineering, Kyoto University, Kyoto 615-8510, Japan
3. Center for Spintronics Research Network, Institute for Chemical Research, Kyoto University, Uji 611-0011, Japan
4. National Institutes for Quantum Science and Technology (QST), Takasaki 370-1292, Japan
5. Advanced ICT Research Center, National Institute of Information and Communications Technology (NICT), Kobe 651-2492, Japan
6. Department of Materials Science, Tohoku University, Sendai 980-8579, Japan

† These authors contributed equally.

* e-mail: morioka.naoya.8j@kyoto-u.ac.jp

## Supplementary Note 1: Detailed information on sample preparation

Key points in the sample fabrication are described in the Methods. The summary is given in Supplementary Table 1. Here we provide additional information on the epitaxial layer, electron-beam lithography, and carbon cap.

### 1.1 Epitaxial layers

We used five nitrogen-doped *n*-type epitaxial 4H-SiC layers (Epi1–5) in this study, whose thickness, net donor concentration, and correspondence with the Sample ID are summarized in Supplementary Table 2. For the oxygen-related study, we used Epi1–4. Epi5 was used as a control study of optical linewidth of divacancies (see Supplementary Note 8).

### 1.2 Sample preparation for cross-section measurements

Samples A1 and A2 were cleaved to expose the cross-section after high-temperature annealing or oxygen ion implantation, respectively. Electron irradiation and defect-formation annealing were performed after cleavage. Samples D1 and D2 were cleaved after the whole process, i.e., after the defect-formation annealing at 800 °C.

### 1.3 Electron-beam lithography for masked implantation

A polymethyl methacrylate (PMMA) electron-beam lithography resist was spin-coated to a thickness of approximately 400 nm, followed by electron-beam exposure to make circular apertures with a diameter of 40 nm. After oxygen ion implantation, the resist was removed by organic solvents and piranha cleaning.

## 1.4 Carbon cap

The carbon cap (C-cap) was formed by spin-coating a photoresist, followed by baking at 750 °C in Ar. The C-cap was removed by oxidation in dry $O_2$ at 900 °C for 30 min.

**Supplementary Table 1 | Summary of sample fabrication steps.**

| Sample | O implantation conditions | HTA (°C) | Surface removal | Electron irradiation fluence ($cm^{-2}$) | Defect-formation annealing temperature (°C) |
|---|---|---|---|---|---|
| A1 | Box,<br>600 nm, $5 \times 10^{17}$ $cm^{-3}$ | 1750 | Ox. | $2 \times 10^{13}$ | 900 |
| A2 | Box,<br>600 nm, $5 \times 10^{17}$ $cm^{-3}$ | 1750 | — | — | — |
| A3 | Box,<br>600 nm, $5 \times 10^{17}$ $cm^{-3}$ | 1750 | — | $1 \times 10^{17}$ | 800 |
| B | Mask,<br>60 keV, $1 \times 10^{12}$ $cm^{-3}$ | — | — | — | 300, 500, 600, and 900<br>successively treated in this order |
| C1 | Box,<br>600 nm, $5 \times 10^{17}$ $cm^{-3}$ | — | — | $1 \times 10^{17}$ | 800 |
| C2 | Box,<br>600 nm, $5 \times 10^{17}$ $cm^{-3}$ | — | — | — | 800 |
| D1 | Box,<br>500 nm, $1 \times 10^{19}$ $cm^{-3}$ | 1900 | RIE + Ox. | $1 \times 10^{17}$ | 800 |
| D2 | Box,<br>500 nm, $1 \times 10^{19}$ $cm^{-3}$ | — | RIE + Ox. | $1 \times 10^{17}$ | 800 |
| E1 | $^{17}$O mask,<br>60 keV, $1 \times 10^{12}$ $cm^{-3}$ | — | — | — | 800 |
| E2 | $^{17}$O Box,<br>420 nm, $5 \times 10^{17}$ $cm^{-3}$ | 1750 | Ox. | $1 \times 10^{14}$ | 800 |

**Supplementary Table 2 | Information about epitaxial wafers.**

| Epi ID | Thickness (μm) | Net donor concentration ($cm^{-3}$) | Sample ID |
|---|---|---|---|
| Epi1 | 11* | $2 \times 10^{14}$ $cm^{-3}$ | A1, A2, A3, C1, C2 |
| Epi2 | 52 | $4 \times 10^{14}$ $cm^{-3}$ | B |
| Epi3 | 150 | $2 \times 10^{14}$ $cm^{-3}$ | D1, D2 |
| Epi4 | 100* | $4 \times 10^{13}$ $cm^{-3}$ | E1, E2 |
| Epi5 | 50 | $7 \times 10^{14}$ $cm^{-3}$ | F (see Supplementary Note 8) |

* including a buffer layer

# Supplementary Note 2: Second-order autocorrelation function $g^{(2)}(\tau)$ of defects in Fig. 1c.

Here we show the $g^{(2)}(\tau)$ data for the PL5, PL6, PL7, and PL1* defects shown in Fig. 1c. The obtained $g^{(2)}(\tau)$ curves were fitted using a three-level model

$$g^{(2)}(\tau) = 1 - a\exp\left(-\frac{|\tau|}{\tau_1}\right) + b\exp\left(-\frac{|\tau|}{\tau_2}\right), \quad \text{(S1)}$$

where $\tau_1$ and $\tau_2$ denote time constants of the antibunching and bunching components, respectively, and $a$ and $b$ are their amplitudes. The results are shown in Supplementary Fig. 1. The emission from each defect exhibited $g^{(2)}(0) < 0.5$, confirming that these defects are all single-photon emitters.

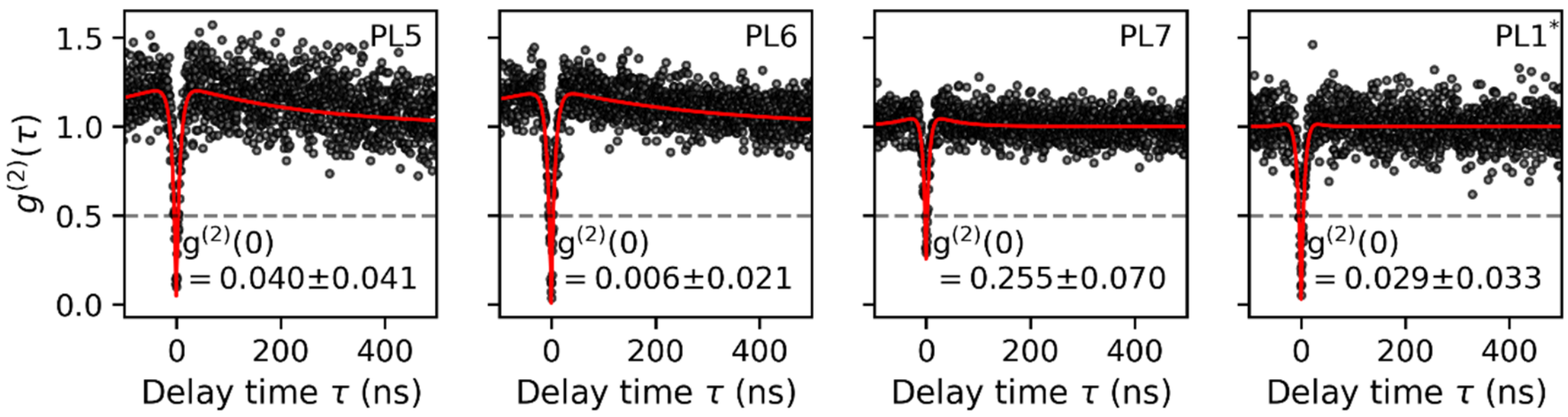

**Supplementary Fig. 1 | $g^{(2)}(\tau)$ of PL5–PL7 and PL1* defects in Fig. 1c.**

## Supplementary Note 3: Resonance frequency shift attributed to strong interaction with host nuclear spins in ODMR spectra of PL5 and PL7

We found two defects whose apparent zero-field splitting (ZFS) parameters ($D$ and $E$) fell outside the corresponding distributions observed for PL5 and PL7 (Fig. 1f). The corresponding ODMR spectra are shown in Supplementary Fig. 2. We examined whether these apparent $D$ and $E$ values could be accounted for by a strong hyperfine interaction with a nearby $I = 1/2$ nuclear spin. Assuming the $O_CV_{Si}$ microscopic structure for PL5 and PL7, first-principles calculations predict that the strongest hyperfine interaction of $O_CV_{Si}$ arises from a nearest-neighbor $^{13}C$ nucleus.[1] The orientation of the principal axis of this hyperfine tensor, however, has not been explicitly reported for $O_CV_{Si}$. We therefore assumed, by analogy with the tetrahedral nearest-neighbor geometry of the 4H-SiC divacancy $V_CV_{Si}$, that the hyperfine principal $z$ axis forms an angle of 70.5° with the ZFS principal axis.[2]

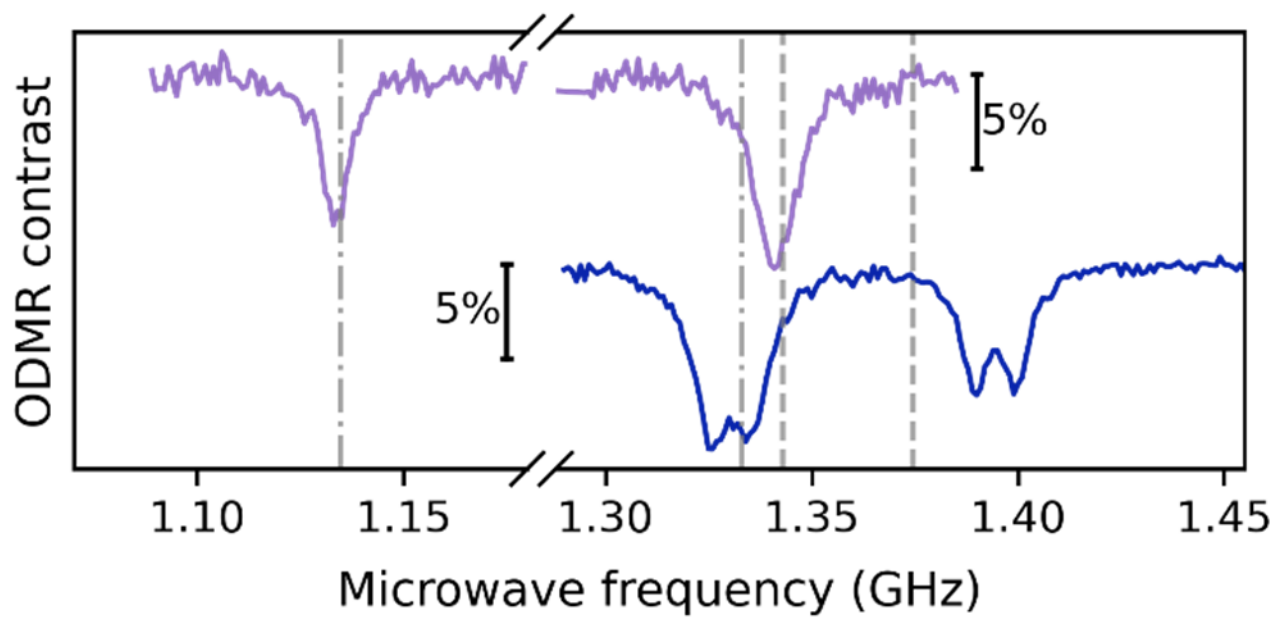

**Supplementary Fig. 2 | Cw-ODMR spectra of defects in Sample A1 showing anomalous apparent ZFS.** Broken and dash-dotted vertical lines indicate the mean resonance frequencies of normal PL5 and PL7 in the same sample, respectively.

The nearest-neighbor $^{13}C$ sites are not all equivalent with respect to the *c*-axis. For the basal $O_CV_{Si}$ model considered in first-principles calculations, the nearest-neighbor $^{13}C$ site approximately aligned with the *c*-axis exhibits an essentially axially symmetric hyperfine tensor. In contrast, the off-axis sites show a small anisotropy between the two transverse principal values. We neglected the transverse anisotropy and approximated the tensor as axially symmetric,

$$A_{x'x'} = A_{y'y'} = A_{\perp}, \qquad A_{z'z'} = A_{\parallel},$$

where $(x', y', z')$ denote the principal axes of the hyperfine tensor. At zero magnetic field, the spin Hamiltonian in the ZFS principal-axis frame $(x, y, z)$ can be written as

$$H = DS_z^2 + E\left(S_x^2 - S_y^2\right) + \mathbf{S}^{\mathrm{T}}\mathbf{A}\mathbf{I},$$

where $\mathbf{A}$ denotes the hyperfine tensor expressed in the ZFS principal-axis frame. We define a unit vector $\mathbf{n}$ along the hyperfine principal $z'$ axis, expressed in the ZFS frame, using the polar angle $\theta$ and azimuthal angle $\phi$, $\mathbf{n} = (\sin\theta\cos\phi, \sin\theta\sin\phi, \cos\theta)^{\mathrm{T}}$. Under the axial approximation, the hyperfine tensor in its principal-axes frame can be written as

$$\mathbf{A}' = A_{\perp}\mathbf{1} + (A_{\parallel} - A_{\perp})\mathbf{e}_{z'}\mathbf{e}_{z'}^{\mathrm{T}},$$

where $\mathbf{1}$ is the identity operator and $\mathbf{e}_{z'}$ is the unit vector along the $z'$ axis. The corresponding hyperfine tensor in the ZFS principal-axis frame is given by

$$\mathbf{A} = A_{\perp}\mathbf{1} + (A_{\parallel} - A_{\perp})\mathbf{n}\mathbf{n}^{\mathrm{T}}.$$

For each defect species, $D$ and $E$ were fixed to the mean values obtained from the corresponding normally split centers in Sample A1 (Table 1). Because the crystallographic site of the coupled $^{13}C$ nucleus cannot be identified from the present experiment, we considered the two inequivalent nearest-neighbor orientations in the idealized tetrahedral geometry, represented by $(\theta, \phi) = 0°$ and $\pm 120°$. For each orientation, the Hamiltonian was numerically diagonalized to obtain $A_{\perp}$ and $A_{\parallel}$ that reproduce the apparent $D$ and $E$ of the shifted defects.

The results are summarized in Supplementary Table 3. The sign of $A_{\parallel}$ is not uniquely determined by the present analysis, and thus both positive- and negative-$A_{\parallel}$ results are shown. First-principles calculations for $O_CV_{Si}$, however, predict positive values for $A_{\parallel}$ and $A_{\perp}$.[1] We therefore regard the positive solutions as the physically relevant ones. The zero-field ODMR spectrum is independent of the sign of $A_{\perp}$, and thus only $|A_{\perp}|$ is reported.

The positive-$A_{\parallel}$ solutions are consistent with the first-principles calculations for $O_CV_{Si}$(*kh*) and (*hk*) configurations.[1] Thus, within the assumptions described above, the anomalous apparent ZFS values of PL5 and PL7 can be consistently explained by a strong hyperfine interaction with a nearest-neighbor $^{13}C$ nuclear spin.

**Supplementary Table** 3 | **Hyperfine parameters reproducing the anomalous apparent ZFS in Sample A1.** PL5 showed an additional splitting due to nuclear spin coupling, which was averaged to obtain apparent *D* and *E*.

| | Apparent ZFS (MHz) | | | $(\lvert A_{\perp}\rvert, A_{\parallel})$ (MHz) | |
|---|---|---|---|---|---|
| | *D* | *E* | Assumed $(\theta, \phi)$ | $A_{\parallel} > 0$ | $A_{\parallel} < 0$ |
| PL5 | 1362.6 | 32.0 | (70.5°, 0°) | (43.3, 115) | (37.8, −119) |
| | | | (70.5°, ±120°) | (46.5, 112) | (31.1, −117) |
| PL7 | 1237.4 | 103.9 | (70.5°, 0°) | (53.0, 100) | (43.9, −109) |
| | | | (70.5°, ±120°) | (62.2, 88.8) | (56.5, −96.2) |

## Supplementary Note 4: Analysis procedure for ZFS parameters and measurements of isolated PL5–PL7 and PL1* defects generated at the surface just after HTA without sacrificial oxidation

ZFS parameters were determined from continuous-wave ODMR (cw-ODMR) spectra measured at zero applied magnetic field. The electron spin Hamiltonian of $S = 1$ at zero magnetic field is given by

$$H = DS_z^2 + E(S_x^2 - S_y^2).$$

We take $E > 0$ as a convention. The zero-field spin resonances appear at $f_\pm = D \pm E$, where $f_+$ and $f_-$ denote the higher and lower frequencies, respectively. For spectra in which the two electron-spin transitions are resolved, the ZFS parameters are therefore given by

$$D = \frac{f_+ + f_-}{2}, \qquad E = \frac{f_+ - f_-}{2}. \tag{S2}$$

Sample A1 was fabricated using decoupled defect engineering. For PL5 and PL7, two resonance frequencies $f_+$ and $f_-$ extracted from Lorentzian fits were used to calculate $D$ and $E$ by Eq. (S2). For PL6 and PL1*, defects exhibiting a single resolvable resonance frequency were considered to have a negligibly small $E$, and the resonance frequency was taken as $D$. Some defects assigned to PL6 and PL1* showed multiple resonance frequencies, which are likely associated with hyperfine splitting arising from nearby nuclear spins. Assuming that these splittings occurred symmetrically around $D$, the mean of resonance frequencies was taken as $D$. The obtained $D$ and $E$ parameters were plotted in Figs. 1d and 1e. Because $g^{(2)}$ measurements were not performed for every defect, we cannot exclude the possibility that some individual spots contained more than one emitter. As discussed in the main text and Supplementary Note 3, all spectra were assigned as PL5, PL6, PL7, or PL1*.

We also measured ODMR of defects that were observed at the surface after HTA before sacrificial oxidation (Sample A2). Most defects could be classified as PL5, PL6, PL7, or PL1* based on the resonance frequencies, except for one defect discussed later. As in Sample A1, the $g^{(2)}$ measurements were not performed for every defect.

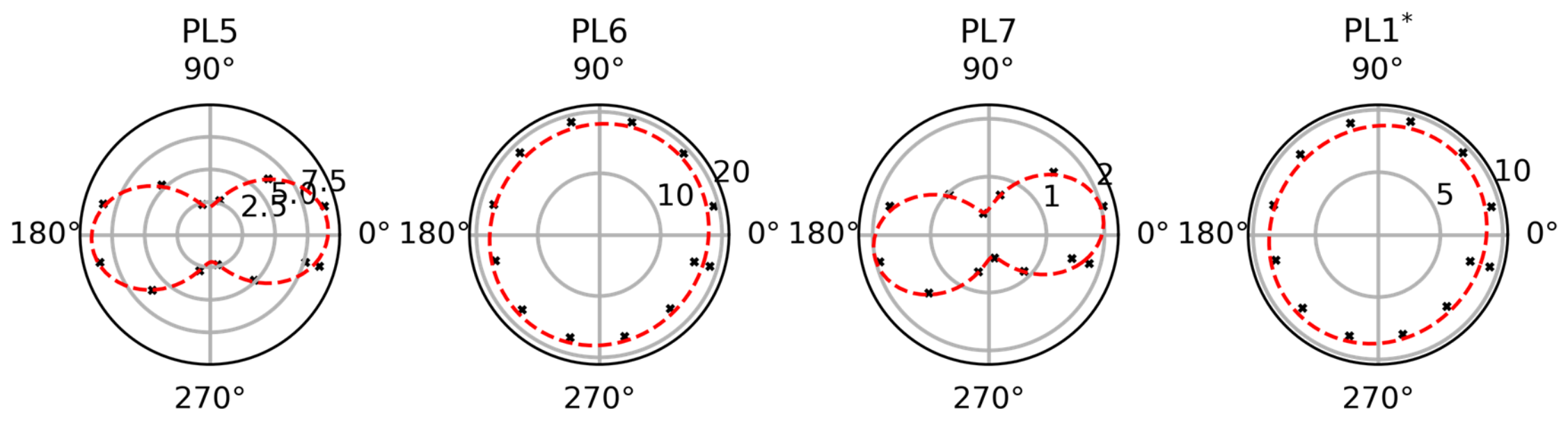

**Supplementary Fig. 3 | Excitation-polarization dependence of fluorescence intensity for PL5, PL6, PL7, and PL1* in Sample A1.** The fluorescence intensity is given in units of kcps.

In contrast, the large variation in resonance frequencies in Sample B made straightforward assignment to PL5, PL6, PL7, or PL1* difficult. We therefore first classified the defects in Sample B into PL5/PL7 and PL6/PL1* candidate groups based on their excitation-polarization dependence, independently of their ODMR spectra. A study of defects created by ion implantation reported that PL5 exhibited strong polarization dependence, whereas PL6 and ion-implantation-induced PL1 exhibited negligible polarization dependence when observed from the *c*-face.[3] Our measurements on Sample A1, fabricated by decoupled engineering, confirmed that PL5 and PL7 exhibited polarization-dependent fluorescence, whereas PL6 and PL1* showed negligible polarization dependence (Supplementary Fig. 33).

To quantify the polarization dependence, the degree of polarization (DoP) was determined by fitting the polarization-dependent fluorescence intensity using

$$I(\theta) = A\cos^2(\theta - \phi) + B, \qquad \mathrm{DoP} = \frac{\mathrm{A}}{A + B}$$

where $\theta$ is the excitation-polarization angle, $A$ is the amplitude of the polarization-dependent fluorescence component, $\phi$ is the polarization angle at which this component reaches its maximum, and $B$ is the polarization-independent offset. The classification threshold is DoP = 0.30, which is the minimum value for PL5 and PL7 in Sample A1 and consistent with the value reported by Eserin *et al.*[4]; i.e., defects with DoP < 0.30 and DoP ≥ 0.30 are PL6/PL1* and PL5/PL7 candidates, respectively.

Within these candidate groups, the defects were further classified as PL5, PL6, PL7, or PL1* based on their ODMR spectral shapes. As in Samples A1 and A2, the $g^{(2)}$ measurements were not performed for every spot. In addition, for some defects in Samples A2 and B, the frequency sweep did not cover the 1.1–1.2 GHz range, where the lower-frequency resonance of PL7 occurs. Therefore, for these defects, the presence of an additional resonance in the 1.1–1.2 GHz range cannot be excluded.

Several PL6/PL1* candidates in Samples A2 and B showed two resonance frequencies with splittings that could not be explained based on previously reported hyperfine interactions[5] or that show pronounced differences in contrast or linewidth between the two resonances, as shown in Supplementary Figs. 4a and 4b. Application of a magnetic field would be required to determine whether these resonances originate from zero-field splitting, hyperfine interactions, or other mechanisms. Because such measurements were not performed, the origin of these resonances remains uncertain. For assessing the robustness of the analysis, these defects were provisionally classified as group F. We evaluated the distribution of $D$ of PL6/PL1* candidates with and without taking group F into account. As summarized in Supplementary Fig. 5, Sample B showed a broader $D$ distribution for PL6/PL1* than Samples A1 and A2 regardless of whether group F was considered, and the inclusion of F does not significantly affect the evaluation of the distribution. Therefore, all data from Samples A2 and B, including group F, were used in the main text and the following analyses.

In Supplementary Fig. 5 and the following analysis, we used $D$ and $E$ values for defects evaluated as below, depending on the number of observed resonance frequencies.

We extracted the resonance frequencies by fitting with Lorentzian curves. For PL6/PL1* candidates exhibiting a single resonance frequency, the resonance frequency was taken as $D$. For those showing multiple resonances, excluding group F defects, the additional peaks were tentatively attributed to hyperfine splitting, and $D$ was determined as the mean value of the resonance frequencies. For group F defects, which exhibited two resonance frequencies of uncertain origin, their mean was used as an operational estimate of $D$. This approximation is justified because electron-spin splitting would occur symmetrically around $D$, whereas any shifts in the mean due to hyperfine coupling to nearby nuclear spins are expected to be negligible. For PL5/PL7 candidates exhibiting two resonance frequencies, the ZFS parameters were calculated using Eq. (S2). For defects in which each peak further split into a doublet (an example is shown in Supplementary Fig. 2), likely due to coupling to two nuclear spins, we averaged each doublet and used the resulting two frequencies to calculate $D$ and $E$ by Eq. (S2).

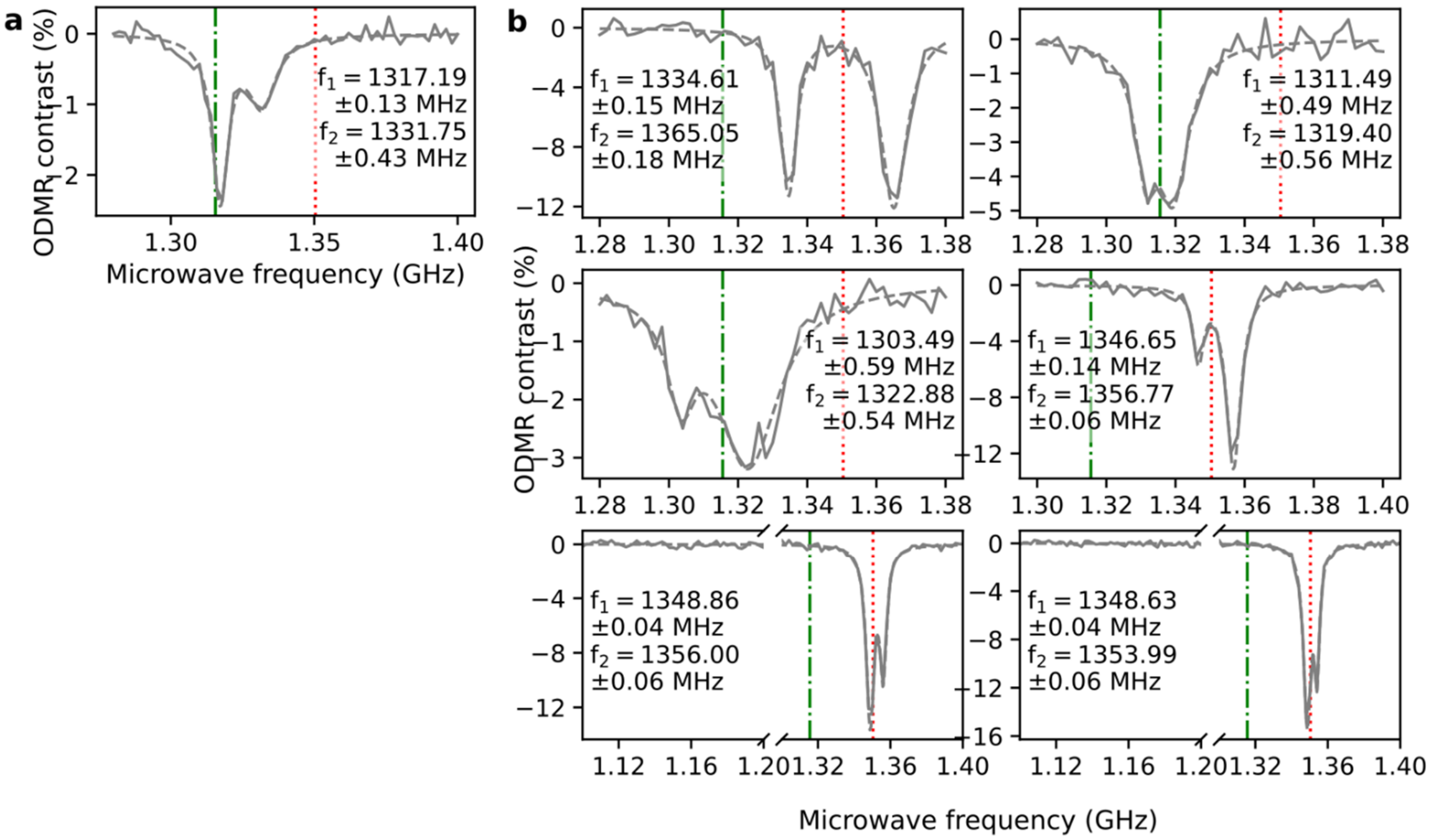

**Supplementary Fig. 4 | Zero-field cw-ODMR spectra of defects categorized as group F in a, Sample A2, and b, Sample B.** Dash-dotted green and broken red lines indicate the resonance frequencies of PL1* and PL6, respectively, calculated from the mean ZFS parameters obtained in Sample A1 (Table 1). $f_1$ and $f_2$ denote the resonance frequencies obtained from Lorentzian fits.

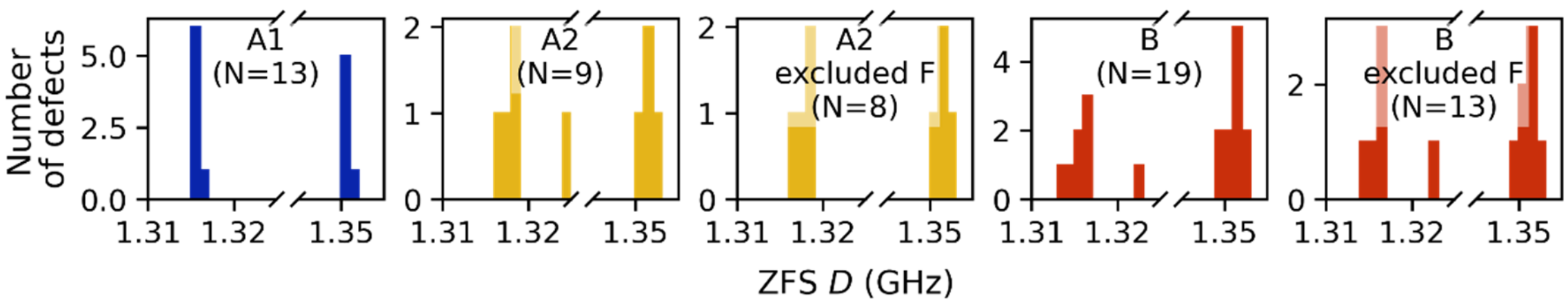

**Supplementary Fig. 5 | Distributions of ZFS parameter $D$ of PL6/PL1* for samples A1, A2, and B.** For Samples A2 and B, the $D$ distributions of PL6/PL1* candidates excluding group F defects are additionally shown.

Supplementary Fig. 6 shows the scatter plot of $D$ and $E$ for defects in Samples A1, A2, and B, revealing substantially narrower distributions in A1 and A2 than in B. For a reliable statistical comparison, we excluded outliers that could originate from different defects, as well as defects whose apparent ZFSs were shifted by strong hyperfine interactions (see Supplementary Note 3). First, one defect in Sample B that clearly lay outside the main distribution was excluded (displayed as "Others"). The remaining outliers, including defects with strong hyperfine interactions, were identified using the squared Mahalanobis distance.[6] Assuming a multivariate normal distribution, the squared Mahalanobis distance approximately follows a chi-squared distribution with one or two degrees of freedom, for PL6/PL1* and PL5/PL7 candidates, respectively. The corresponding chi-squared distributions were used to define a 95% confidence region.

Sample standard deviations (STDs) were calculated separately using 95% probability criteria and are summarized in Supplementary Table 4. For Sample A1, only two PL5 and PL7 defects with strong hyperfine coupling, discussed in Supplementary Note 3, were excluded. All STDs followed the order A1 < A2 < B. These results imply that the HTA step improves uniformity of the spin properties, likely through the recovery of implantation-induced lattice damage.

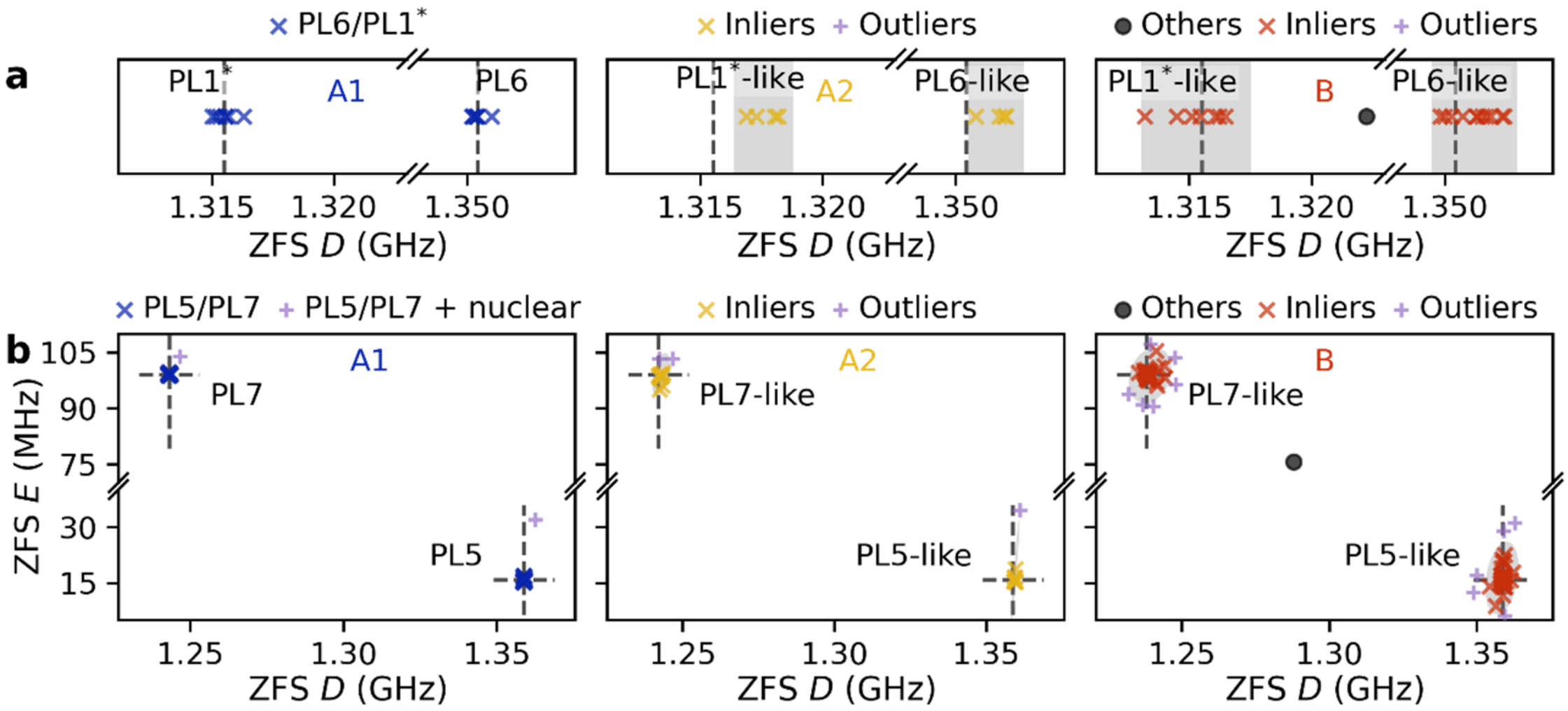

**Supplementary Fig. 6 | Scatter plots of zero-field splitting (ZFS) parameters. a**, $D$ for PL6/PL1* in Sample A1 (left) and PL6/PL1* candidates in Sample A2 (center) and B (right). **b**, $E$ vs. $D$ for PL5/PL7 in Sample A1 (left) and PL5/PL7 candidates in Sample A2 (center) and B (right). Gray shaded regions indicate the 95% probability intervals in a and 95% probability regions in b. Broken lines indicate the mean ZFS parameters obtained in Sample A1.

**Supplementary Table 4 | Sample standard deviations (MHz) of ZFS parameters.** Sample standard deviations of $\boldsymbol{D}$ and $\boldsymbol{E}$ for Sample A2 and B were calculated separately for the comparison groups defined using 95% probability criteria. The corresponding values for Sample A1 are shown for comparison.

| | | A1 (O impl. + HTA + e-irrad.) | A2 (O impl. + HTA; surface defects) | B (O impl.) |
|---|---|---|---|---|
| PL5 | $\sigma_D$ | 0.13 | 0.18 | 1.14 |
| | $\sigma_E$ | 0.55 | 1.11 | 2.36 |
| PL6 | $\sigma_D$ | 0.30 | 0.58 | 0.90 |
| PL7 | $\sigma_D$ | 0.23 | 0.44 | 1.94 |
| | $\sigma_E$ | 0.21 | 1.42 | 1.61 |
| PL1* | $\sigma_D$ | 0.42 | 0.62 | 1.16 |

## Supplementary Note 5: Comparison of excitation-power dependence of defects created using ion implantation and decoupled defect-engineering methods

The excitation-power dependence of fluorescence intensity was measured for single PL5, PL6, PL7, and PL1* created by decoupled engineering (Sample A1) and ion-implantation method (Sample B). For PL5 and PL7, the excitation polarization angle was set to the angle that maximized the fluorescence intensity.

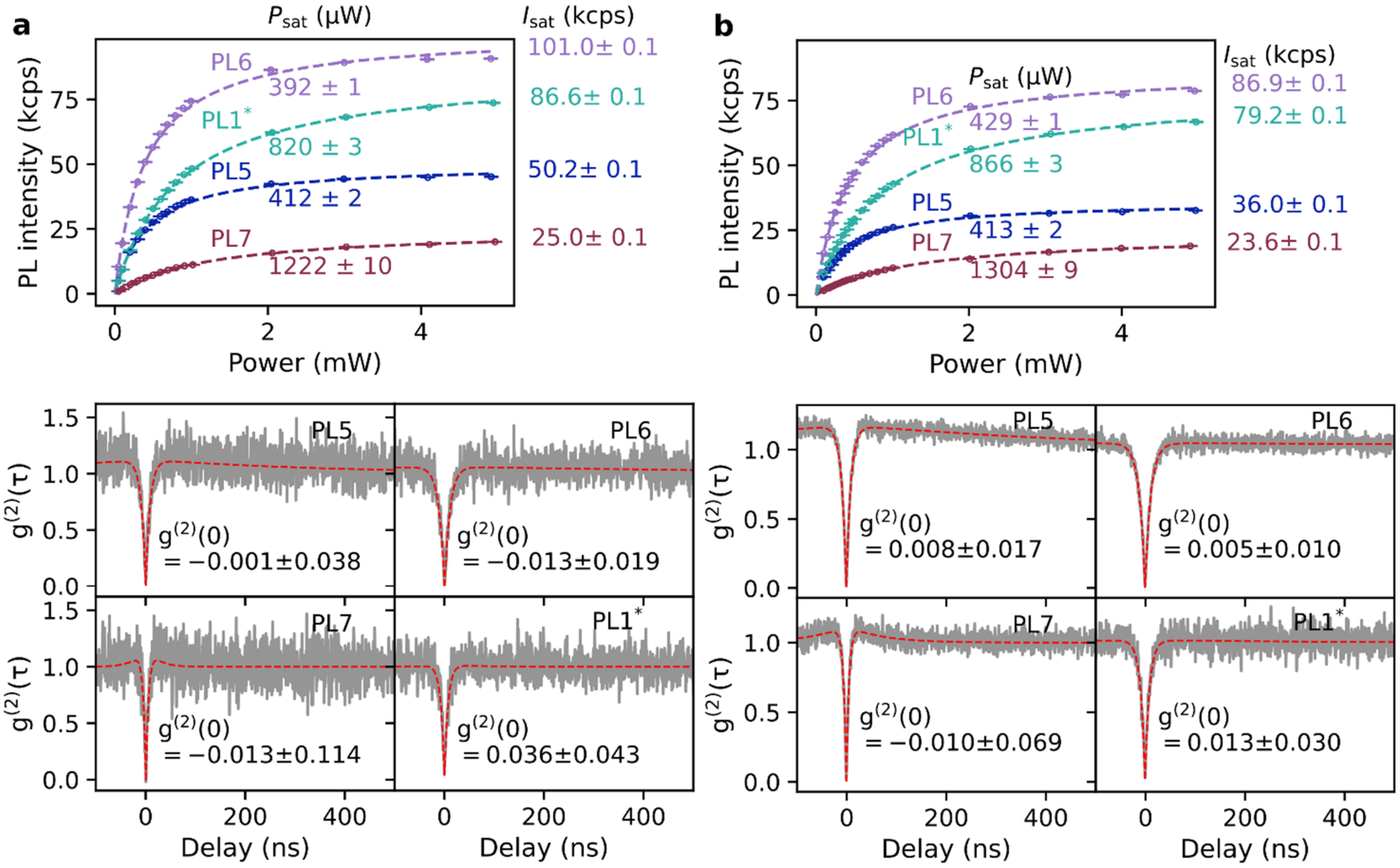

**Supplementary Fig. 7 | Excitation-power dependence of PL intensity of single defects in a, Sample A1 (decoupled engineering) and b, Sample B (ion implantation).** The $g^{(2)}(\tau)$ data of corresponding defects are shown in the bottom plot for each sample.

The power-dependent fluorescence intensities $I(P)$ were obtained as the peak intensity from the fitting of the confocal scan with a 2D Gaussian assuming constant background. The $I(P)$ was then fitted using the function

$$I(P) = \frac{I_{\text{sat}}P}{P + P_{\text{sat}}} \quad \text{(S3)},$$

to find the saturation count rate $I_{\text{sat}}$ and the half-saturation power $P_{\text{sat}}$.

The raw autocorrelation data $g_{\text{raw}}^{(2)}(\tau)$ were measured displayed after background correction[7] using

$$g^{(2)}(\tau) = \frac{g_{\text{raw}}^{(2)}(\tau) - (1 - \rho^2)}{\rho^2}, \quad \text{(S4)}$$

where $\rho = S/(S + B)$ with $S$ and $B$ denoting the count rates from the defect and the background, respectively.

The resulting power-dependent fluorescence and $g^{(2)}(\tau)$ for PL5, PL6, PL7, and PL1* created using the decoupled-engineering (Sample A1) and ion-implantation (Sample B) methods are summarized in Supplementary Figs. 7a and 7b, respectively. The background-corrected $g^{(2)}(0)$ values were close to zero for all defects, suggesting the background-subtracted fluorescence truly represents single-defect nature. The saturation count rates for PL5, PL6, PL7, and PL1* created by decoupled engineering were comparable to those obtained for defects created using the ion-implantation method.

## Supplementary Note 6: Details of $T_2$ measurements and statistics of $T_2^*$ for PL6

While the $T_2^*$ and $T_2$ measurements of PL5 were performed at zero magnetic field, those of PL6 were performed under an applied magnetic field aligned along the *c*-axis using a residual single silicon vacancy (V2) by maximizing the splitting of its two ODMR peaks.[8] In the V2 measurements, a 730-nm laser with an 805 nm long-pass dichroic mirror (DMLP805) was used. The magnetic field strength was estimated to be 103 G from the ODMR splitting of PL6. For both PL5 and PL6, we used the lower-frequency resonance for coherence time measurements.

We analyzed the Hahn-echo oscillation for PL6, whose frequencies varied with the applied magnetic field, assuming electron-spin-echo envelope modulation (ESEEM) caused by the precession of nearby nuclear spins.[9] $T_2$ decay curve for PL6 was fitted with a model that accounts for ESEEM arising from two host $^{13}$C and $^{29}$Si ($I = 1/2$) nuclear spins.[10] The lower nuclear precession frequencies obtained from the fit were $88.3 \pm 1.1$ and $88.0 \pm 1.3$ kHz. These values are consistent with the $^{29}$Si nuclear Larmor frequency expected at the applied magnetic field of 103 G, calculated from the $^{29}$Si nuclear gyromagnetic ratio[11] as 0.8465 kHz/G × 103 G = 87.2 kHz.

Further statistical measurements of $T_2^*$ were performed for randomly chosen 11 PL6 defects and summarized in Supplementary Fig. 8. The mean and median $T_2^*$ values were 1.64 μs and 1.13 μs, respectively. Several defects showed $T_2^*$ values exceeding previously reported room-temperature $T_2^*$ of PL6 created by oxygen ion implantation.[5]

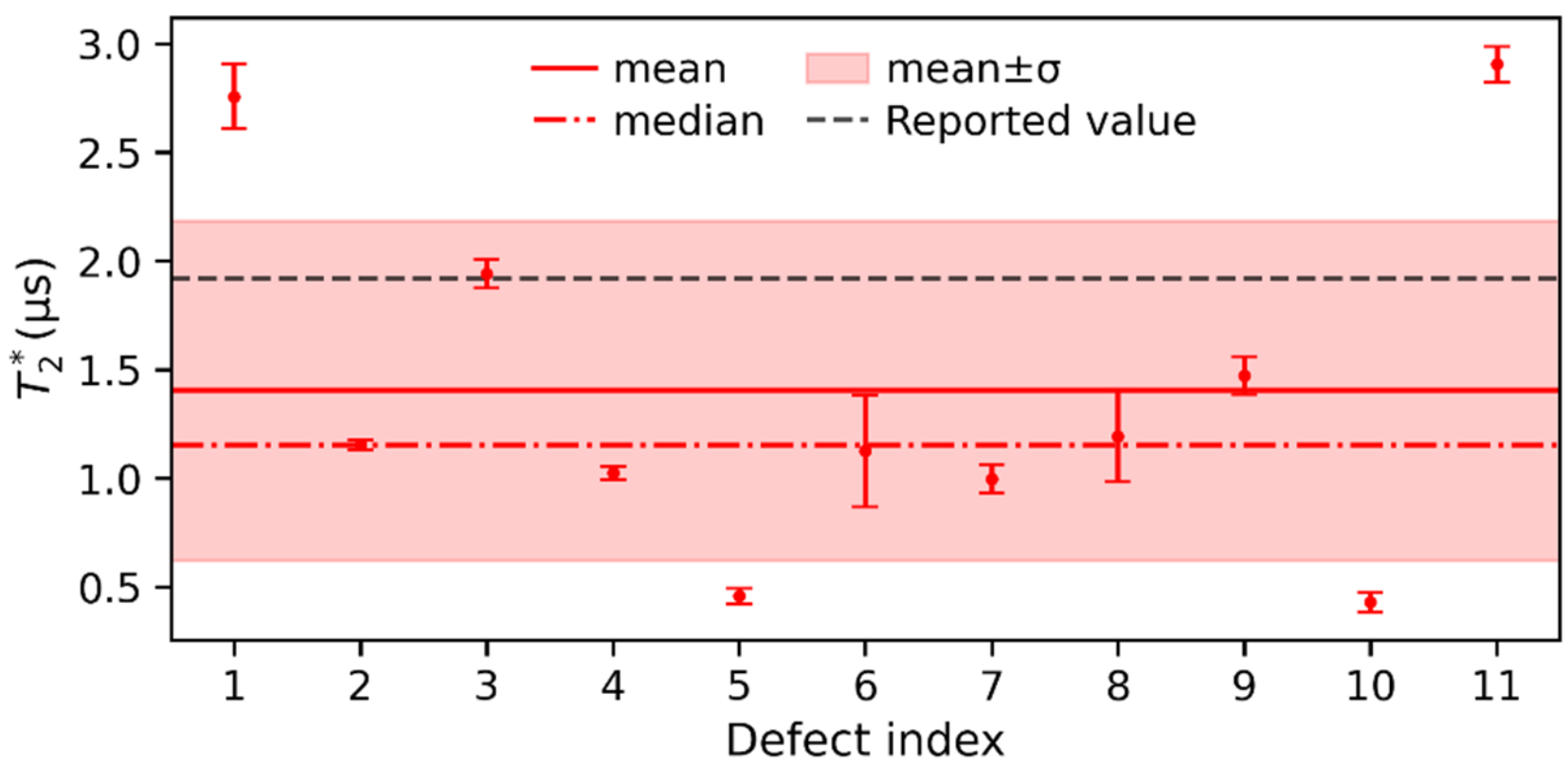

**Supplementary Fig. 8 | Statistics of $T_2^*$ of PL6 created by decoupled engineering.** The error bars represent standard errors. The black dashed line shows the previously reported $T_2^*$ for PL6 created by oxygen ion implantation.[5]

## Supplementary Note 7: Details of the side-view PL and ODMR measurements

The cleaved samples (see also Supplementary Note 1) were mounted onto a printed circuit board (PCB) with the cleaved surface facing upward to measure them in a side-view geometry (Inset of Supplementary Fig. 9). The in-plane rotation within the cleaved surface was corrected to determine the depth direction.

In Samples A3 and C1, positions located 30 μm and 27 μm away from the microwave wire were examined, respectively, for the ODMR in Fig. 2b and the depth-dependent PL and ODMR shown in Figs. 3b and 3c. To check the effect of distance from the wire, we show the additional ODMR spectrum for Sample C1 measured at the position located 30 μm from the wire, which was less bright than the 27 μm position, probably due to position-dependent cleaved-surface quality. The ODMR contrasts and linewidths of Sample C1 at 27 μm and 30 μm were comparable and were smaller and broader, respectively, than those in Sample A3 (Supplementary Fig. 9).

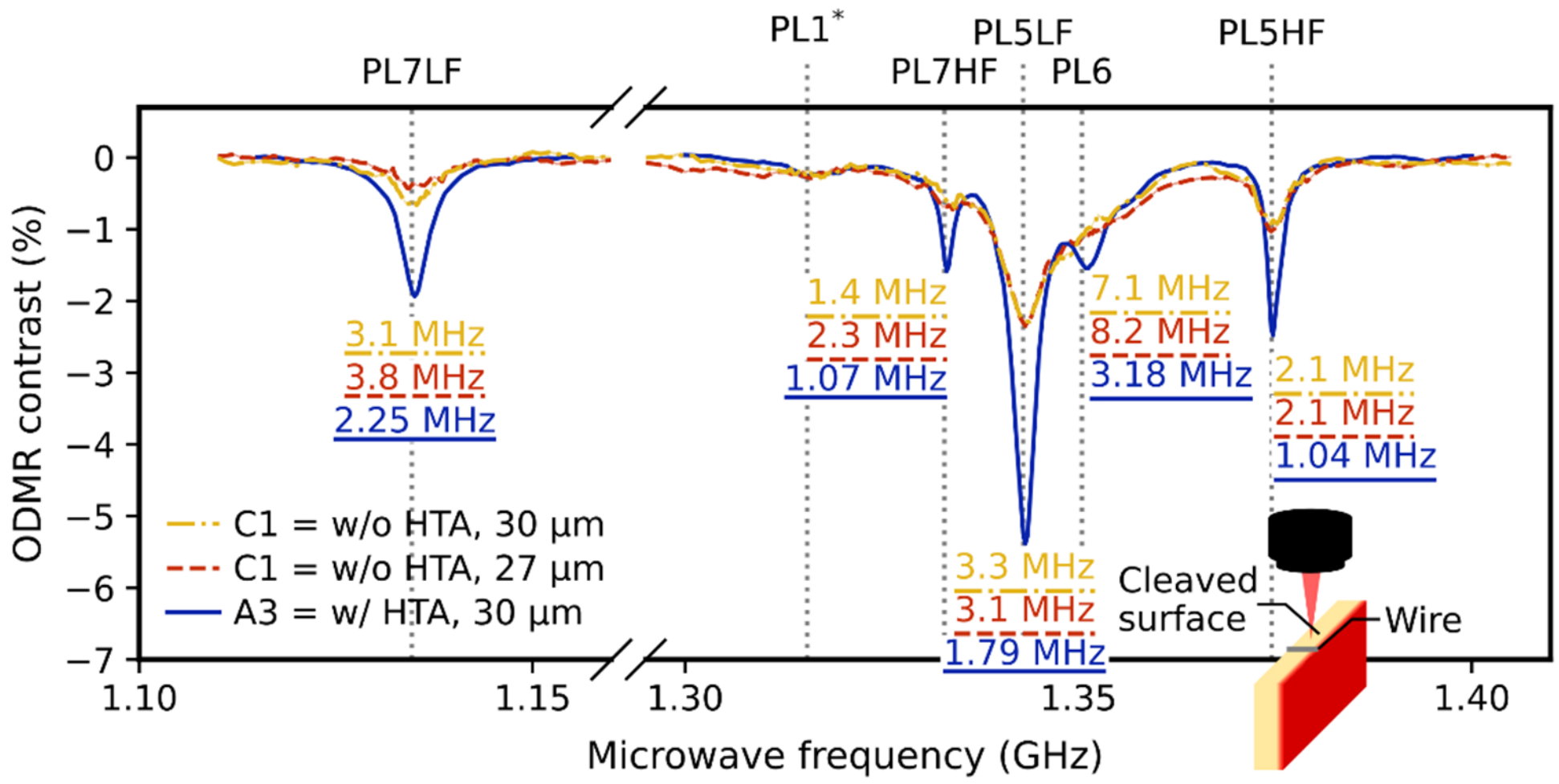

**Supplementary Fig. 9 | Comparison of side-view cw-ODMR spectra of Sample A3 and C1 at different distances from the wire antenna.** Data for A3 (30 µm) and C1 (27 µm) are the same as in Fig. 2b. Values indicate FWHMs obtained from the Lorentzian fits. The Inset shows a schematic of the side-view geometry.

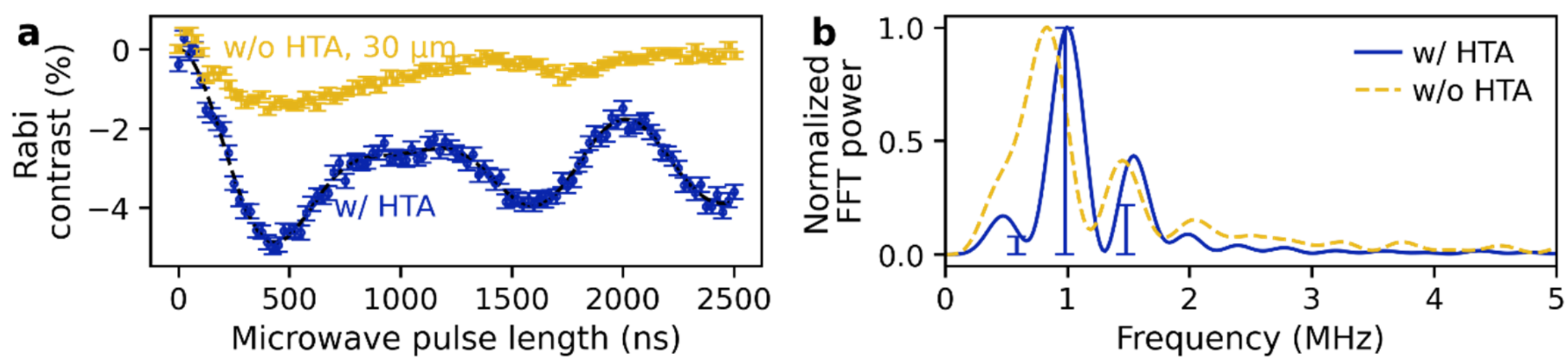

**Supplementary Fig. 10 | Rabi oscillations of Samples A3 and C1. a**, Rabi oscillations of PL5LF in Samples A3 (blue) and C1 (yellow) under the same microwave power as that used in the ODMR spectra shown in Supplementary Fig. 9. Broken lines represent a fit using a three-frequency Rabi oscillation model. **b**, Fast Fourier transform spectra calculated from the Rabi oscillations of Samples A3 (solid blue) and C1 (broken yellow) after linear detrending and zero padding without additional windowing. The power spectra were normalized to their respective maximum values, and only the frequency range around the dominant peak is shown. Vertical bars indicate the Rabi frequencies determined from the fit, with their heights representing the relative powers of the corresponding oscillation components.

Rabi oscillations of PL5LF of Samples A3 and C1 at 30 µm from the wire were compared in Supplementary Fig. 10, using the same microwave powers as those used in Supplementary Fig. 9. Three Rabi frequency components were observed from the fitting for Sample A3, as expected from the microwave antenna configuration along [0001].[12] However, the results of Sample C1 was poorly fitted, likely due to the broad PL6 resonance overlapping to PL5LF. We analyzed the Rabi results using fast Fourier transforms (FFTs) for both samples (Supplementary Fig. 10b). Three Rabi frequency components were resolved in the FFT spectrum of Sample A3, consistent with the frequencies and spectral powers obtained from the fitting. For Sample C1, two higher-frequency peaks appeared near the corresponding frequencies in Sample A3, together with a shoulder on the dominant peak that may correspond to the lowest-frequency component. The Rabi frequencies in Sample C1 do not exceed those in Sample A3, and thus the broader ODMR linewidths in Sample C1 cannot be explained by the power broadening.

## Supplementary Note 8: Optical linewidth comparison between decoupled engineered defects and divacancies created by electron irradiation.

We compare the optical linewidth of defects in Sample A3 with divacancies created by electron irradiation and annealing. A nitrogen-doped epitaxial sample (Epi5, see Supplementary Note 2) was electron-irradiated at 2 MeV to a fluence of $1 \times 10^{17}$ cm$^{-2}$, followed by 800 °C annealing for 30 min (Sample F).

Supplementary Fig. 11 shows photoluminescence (PL) spectra for Sample A3 and F. The PL spectrum of Sample F exhibits signals from the conventional divacancies PL1–PL4. FWHMs of ZPLs were extracted from Lorentzian fits in the wavelength domain. The FWHMs of PL1–PL3 in Sample F were 0.28–0.33 nm, whereas those of the PL5 doublet, PL6, and PL7 in Sample A3 were 0.21–0.34 nm, corresponding to only two to three detector pixels and are close to the instrumental resolution limit. In addition, the linewidths of PL1–PL3 in Sample A3 did not broaden compared to those in Sample F, whereas Samples C1 and C2 without HTA in Fig. 2c showed broad linewidths, indicating that their broader linewidths were associated with implantation-induced lattice damage.

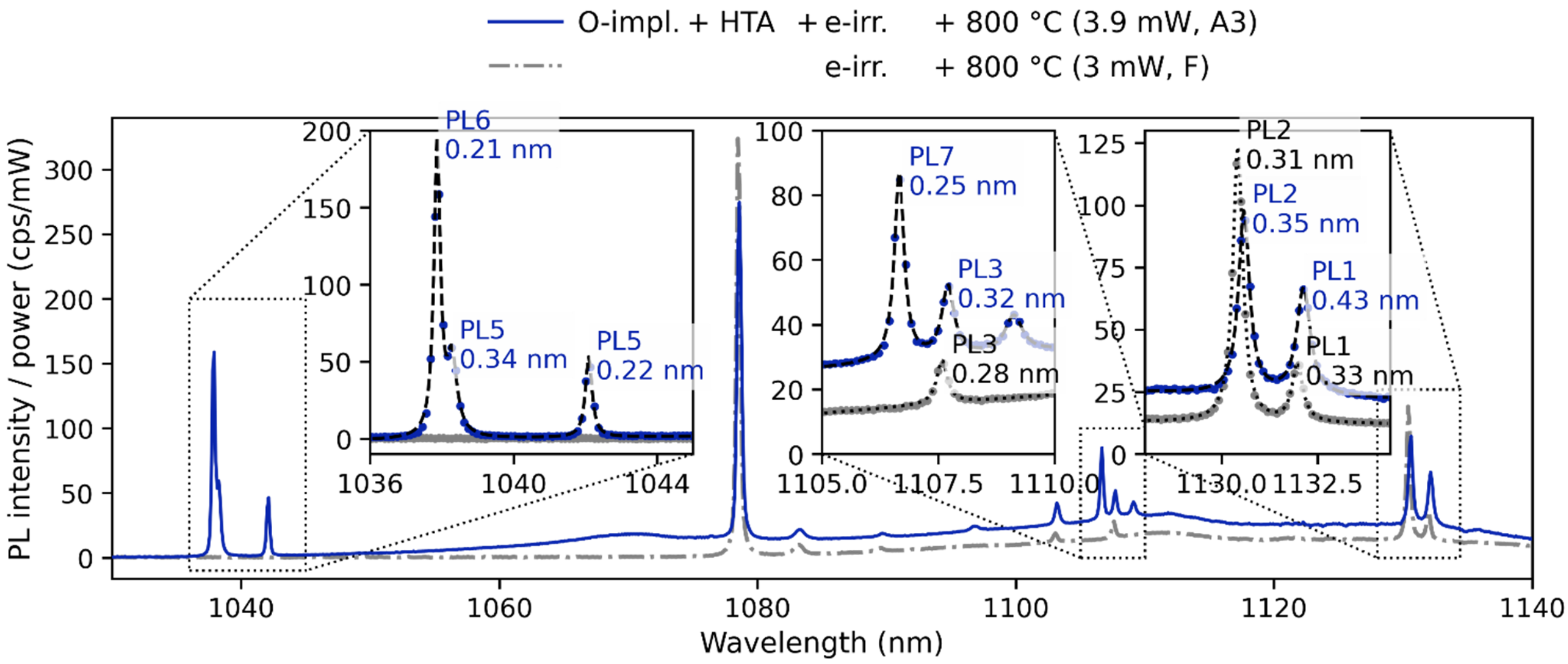

**Supplementary Fig. 11 | Low-temperature photoluminescence spectra of defect ensembles.** Solid blue and dash-dotted gray lines show the spectra of Samples A3 (also shown in Fig. 2c) and F, respectively. Broken (A3) and dotted (F) lines in the insets indicate Lorentzian fits in the wavelength domain, and the corresponding FWHMs are indicated for Samples A3 (blue) and F (black).

## Supplementary Note 9: Additional information on the implantation-layer-removal experiment.

Samples D1 and D2 were examined in a side-view geometry at positions 50 µm away from the microwave antenna. The rotation within the cleaved surface was corrected to define the depth direction. Supplementary Figs. 12a and 12b show side-view confocal *y*-*z* scan images of Samples D1 and D2, respectively. The depth origin (depth after etching = 0 in Figs. 3f and 3h) was determined from a line scan along the *y*-axis measured at the top-left corner of each sample in Supplementary Fig. 12. For Samples D1 and D2, the origin was defined as the *y* position of the maximum PL intensity and the midpoint of a sigmoid fit to the PL intensity, respectively.

The depth dependence of ODMR and fluorescence intensity in Figs. 3h and 3f, respectively, were measured at a piezo displacement of 3 µm from the cleaved surface, taking the sample tilt into account. The fluorescence intensity profile in Fig. 3f was obtained by averaging over a 4-µm-wide range in *x* at each depth after etching. The scan data before averaging are shown in Supplementary Fig. 13. The position of maximum PL intensity was slightly shifted from the defined depth origin, resulting in a small offset of the fluorescence from the depth origin in Fig. 3f.

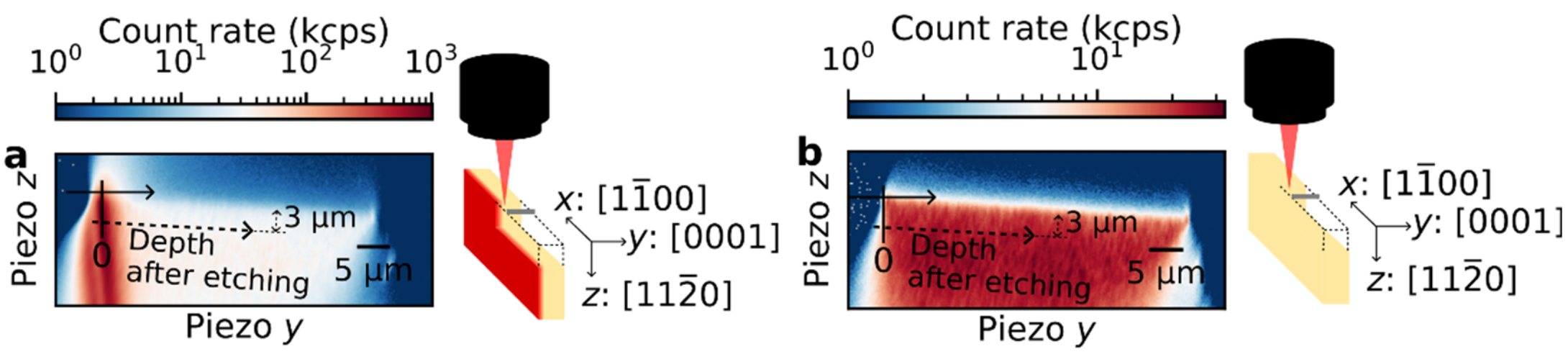

**Supplementary Fig. 12 | Cross-sectional photoluminescence images of a, Sample D1 and b, SampleD2 in the plane containing the laser-beam axis.** The $x$, $y$, and $z$ directions correspond to the three axes of the piezo stage. The solid and broken arrows indicate the line-scan direction used to define the depth origin and the defined depth direction, respectively.

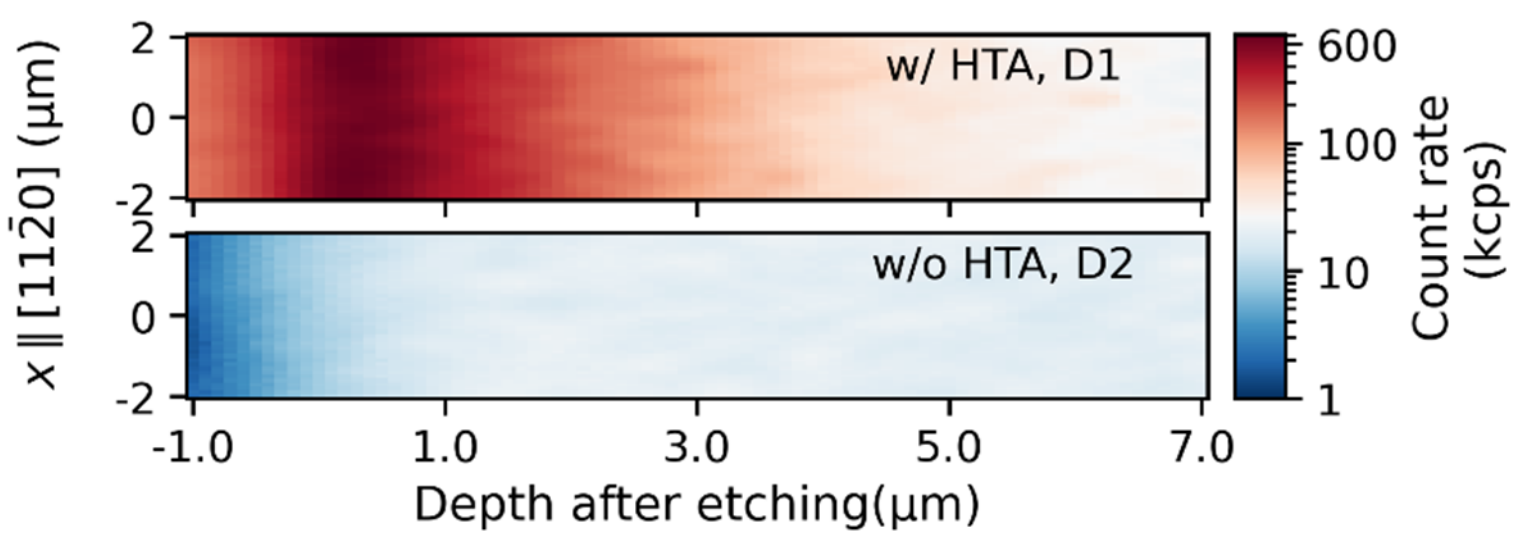

**Supplementary Fig. 13 | Cross-sectional photoluminescence maps of Samples D1 and D2**. Images were acquired on a plane parallel to the cleaved surface, at a piezo displacement of 3 µm into the sample from that surface.

## Supplementary Note 10: Magnetic field conditions and analysis methods for PL6 and PL1* in $^{17}$O isotope samples

### 10.1 Magnetic field alignment for ODMRs of PL6s and PL1* in $^{17}$O-implanted sample (Sample E1)

In the ODMR measurements of PL6s and PL1* in Sample E1, we used ODMR of silicon vacancies (V2, $S = 3/2$) to align the orientation of the static magnetic field along the crystal $c$-axis by maximizing the splitting between two resonances.[8] Achieved $139.99 \pm 0.34$ MHz splitting yields $4D$, where $D = 35$ MHz is the zero-field splitting of V2. Based on resonances of ensemble V2 in a marker region within 80 µm from the measured defect, $80.36 \pm 0.04$ G of magnetic field strength was characterized, assuming electron gyromagnetic ratio of V2, $\gamma_e = 2.803$ MHz/G.[13]

### 10.2 Magnetic field conditions for ODMRs of PL6 in $^{17}$O-doped sample processed with decoupled engineering (Sample E2)

In Sample E2, we aligned the magnetic field by maximizing the splitting using an isolated V2 spot near the region where PL6 defects were characterized. We obtained resonances at 161.002 ± 0.036 and 300.865 ± 0.030 MHz, yielding the splitting of 139.863 MHz. If we assume this value as $4D$ of this specific spot, we obtain $D = 34.966$ MHz and $B = 82.39$ G. Assuming $D = 35$ MHz, numerical diagonalization of the V2 Hamiltonian

$$H = DS_z^2 + \gamma_e B \cos\theta\, S_z + \gamma_e B \sin\theta\, S_x, \tag{S5}$$

yields the magnetic field strength $B = 82.39$ G and the misalignment polar angle $\theta = 1.4°$, which limit the accuracy of magnetic field alignment.

We also used an intentionally misaligned condition. The same fluorescence spot exhibited resonances at 163.72 ± 0.16 and 289.270 ± 0.051 MHz. Assuming $D = 35$ MHz, the extracted $B$ and $\theta$ are 80.1 G and 14.6°, respectively. Assuming $D = 34.966$ MHz, we obtained $B = 80.1$ G and $\theta = 14.5°$.

Supplementary Fig. 14 shows ODMR of two PL6s measured under aligned and misaligned conditions. The labeled defect indices correspond to those in Fig. 5b. Nearly uniform contrasts were obtained across six peaks under the misaligned field. Although two center peaks exhibited broadening, the uniform contrast is useful for observing higher-frequency hyperfine peaks. Hyperfine splittings extracted under two magnetic field conditions differ only by several tens of kHz, indicating that ODMRs under either aligned or misaligned magnetic field are appropriate for extracting hyperfine splittings. Therefore, some ODMR spectra for hyperfine-splitting statistics were measured under the misaligned magnetic field to boost the measurements.

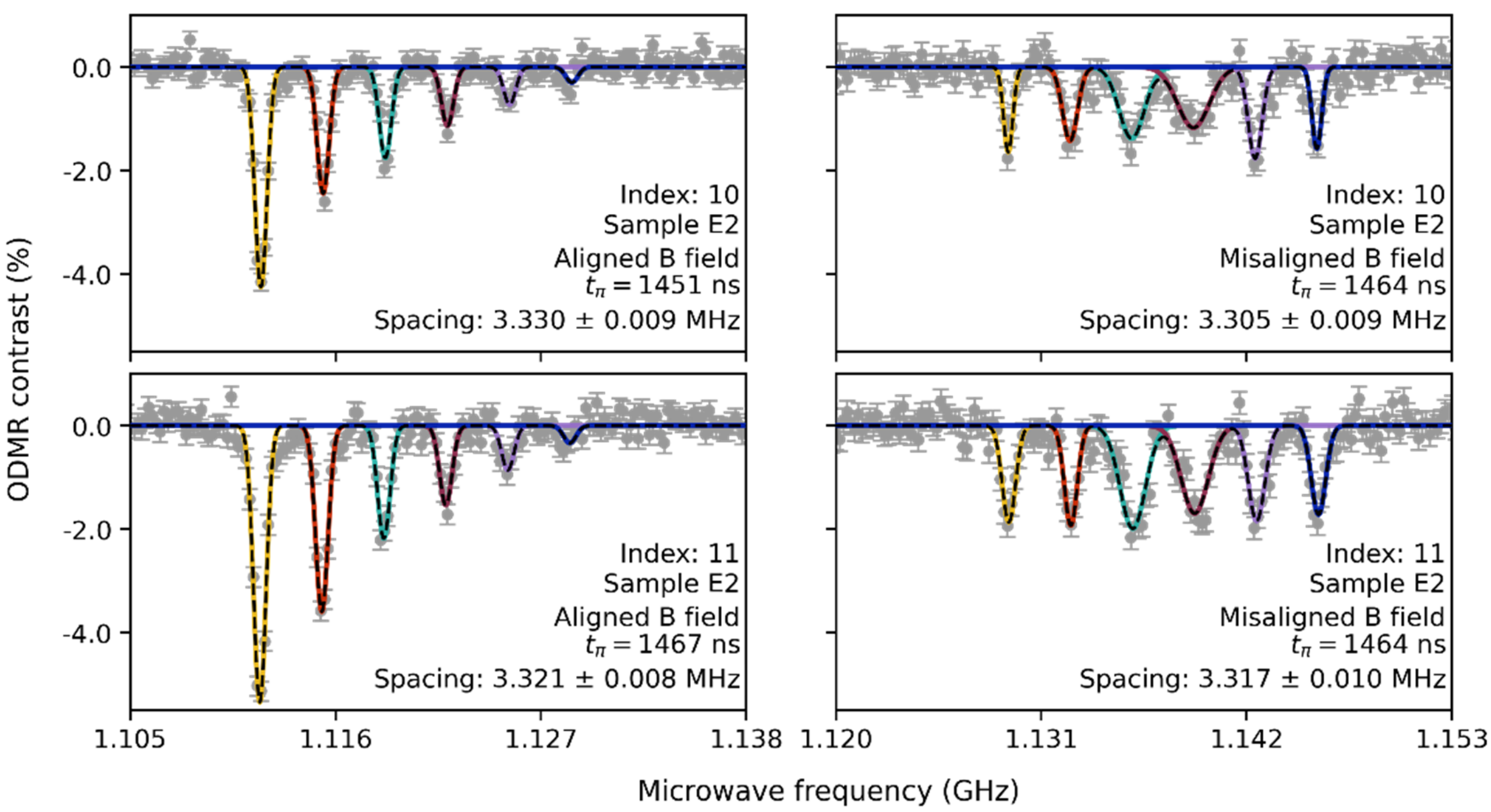

**Supplementary Fig. 14 | Effect of magnetic field misalignment on ODMR spectra of PL6s.** Left, aligned; right, misaligned. Two defects (indices 10 and 11) show similar spectra. Six equally spaced Gaussian functions were used for fitting, with a common linewidth for the aligned case and independent linewidths for the misaligned case.

## Supplementary Note 11 : Zero-field ODMR and $g^{(2)}(\tau)$ data for the defects shown in Fig. 4a–4d (Sample E1)

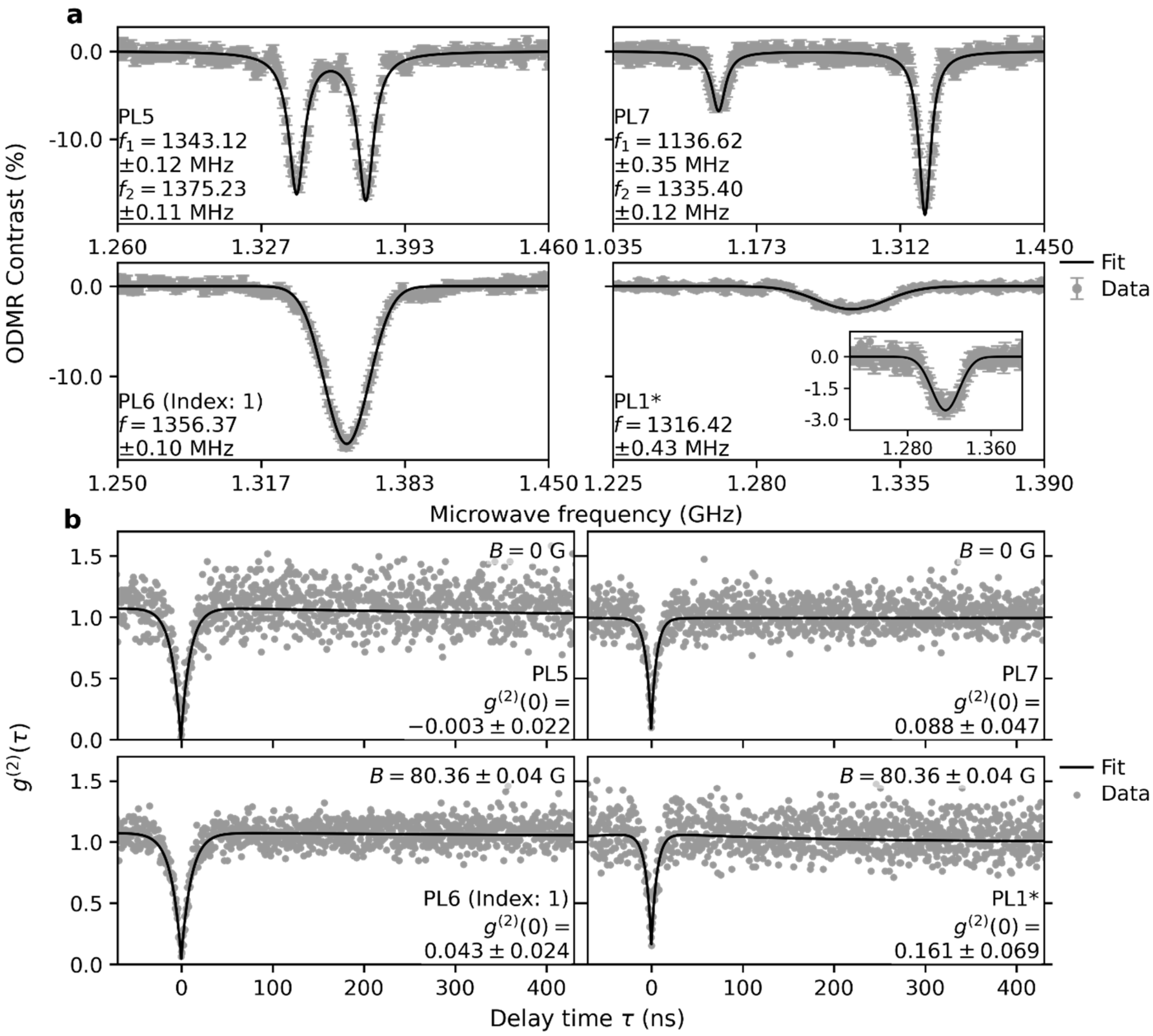

**Supplementary Fig. 15 | Zero-field pulsed ODMR and $g^{(2)}(\tau)$ data of the defects in Fig. 4 a-d created by $^{17}$O ion implantation (Sample E1).** The ODMR spectra were fitted using a Gaussian for PL6 and PL1* to consider broadening by the hyperfine interaction and using Lorentzians for PL5 and PL7 because the effect of hyperfine interaction is suppressed by a large $E$ value. $g^{(2)}(\tau)$ data were fitted using Eq. (S1).

**Supplementary Table 5 | ZFS of defects created by $^{17}$O ion implantation shown in Fig. 4a–4d**.

| | PL5 | PL6 (index 1) | PL7 | PL1* |
|---|---|---|---|---|
| $D$ (MHz) | 1359.18 ± 0.08 | 1356.37 ± 0.07 | 1236.01 ± 0.19 | 1316.42 ± 0.30 |
| $E$ (MHz) | 16.06 ± 0.08 | 0 | 99.39 ± 0.19 | 0 |

The zero-field ODMR spectra of PL5, PL6 (defect index 1 in Fig. 5b), PL7, and PL1* in the $^{17}$O-implanted sample (Sample E1) in Fig. 4 a-d are shown in Supplementary Fig. 15a. The extracted ZFS parameters using Eq. (S2) are summarized in Supplementary Table 5, exhibiting small differences from those of defects created by decoupled engineering in $^{16}$O-doped SiC (reported in Table 1) because of the implantation-induced variation. The raw $g^{(2)}(\tau)$ data for the corresponding defects shown in Supplementary Fig. 15b exhibit $g^{(2)}(0) < 0.5$, clearly indicating that they are single defects. $g^{(2)}(\tau)$ were recorded under zero magnetic field for PL5 and PL7 and under the aligned magnetic field characterized in Supplementary Note 10.1 for PL6 and PL1*.

## Supplementary Note 12: $A_{\parallel}$ of the $^{17}$O nucleus for PL5, PL7, and PL1*

From the sextet hyperfine lines of PL5, PL7, and PL1* in the $^{17}$O-implanted sample (Sample E1), we estimate the hyperfine parameters. Assuming the axial symmetry for the hyperfine tensor, the Hamiltonian of a $S = 1$ spin system interacting with a $I = 5/2$ nuclear spin can be written as

$$H = DS_z^2 + E\left(S_x^2 - S_y^2\right) + \gamma_{\mathrm{e}}\mathbf{B} \cdot \mathbf{S} + A_{\parallel}S_zI_z + A_{\perp}\left(S_xI_x + S_yI_y\right),$$

where $A_{\parallel} = A_{zz}$ and $A_{\perp}$ denote axial and transverse hyperfine parameters for $^{17}$O nuclear spin, respectively.

For PL5 and PL7, the transverse magnetic field strength is unknown because the magnetic fields were approximately set along the defect axis, assuming a tetrahedral structure. Also, the ODMR contrast is low for PL1*. Therefore, we are unable to reliably evaluate the effect of the transverse magnetic field or $A_{\perp}$. We thus ignore the flip-flop term and the transverse magnetic field component for PL5, PL7, and PL1*. The electron spin resonance appears at

$$\nu_{\pm,m_I} = D \pm \sqrt{E^2 + (A_{\parallel}m_{\mathrm{I}} + \omega_{\mathrm{e}})^2},$$

yielding

$$\left|A_{\parallel}m_I + \omega_{\mathrm{e}}\right| = \sqrt{[(\nu_+ - \nu_-)/2]^2 - E^2} \equiv Y(m_I). \tag{S6}$$

Therefore, from the slope of $Y = \sqrt{[(\nu_+ - \nu_-)/2]^2 - E^2}$ versus $m_I$, $A_{\parallel}$ is obtained and reported in the main text. The intercept equals to $\omega_{\mathrm{e}}$. The resonance frequencies were determined by Gaussian fitting to the data in Fig. 4, assuming a common linewidth for all hyperfine lines. We used $E$ values evaluated from zero-field ODMR summarized in Supplementary Table 5, assuming DNP is negligible at zero field.

From $\omega_{\mathrm{e}}$ evaluated from the intercept, the magnetic field along the ZFS *z* axis sensed by PL5 and PL7 are found to be $46.50 \pm 0.01$ G and $49.50 \pm 0.05$ G, respectively, assuming a gyromagnetic ratio of 2.803 MHz/G. Similarly, the estimated magnetic field is $80.49 \pm 0.01$ G for PL1*, which is consistent with the field condition evaluated by V2 in Supplementary Note 10.1.

## Supplementary Note 13: $A_{\parallel}$ and $|A_{\perp}|$ of $^{17}$O nucleus interacting with PL6

We found that spacings of six hyperfine peaks are unequal for PL6s in Figs.4d and 4e. These unequal splittings can be explained by the transverse hyperfine coupling $A_{\perp}$.[14,15]

The spin Hamiltonian is given by

$$H = DS_z^2 + \gamma_{\mathrm{e}}BS_z + PI_z^2 + A_{\parallel}S_zI_z + \frac{A_{\perp}}{2}(S_+I_- + S_-I_+), \tag{S7}$$

Here, we neglect the nuclear Zeeman interaction and assume an axially symmetric nuclear quadrupole interaction ($PI_z^2$) for simplicity. We treat the flip-flop term $A_\perp(S_+I_- + S_-I_+)/2$ in Eq. (S7) as a perturbation. The unperturbed eigenenergies and ODMR frequencies are then given by

$$E^{(0)}_{m_S,m_I} = Dm_S^2 + \omega_\mathrm{e} m_S + Qm_I^2 + A_\parallel m_S m_I,$$

$$\nu^{(0)}_{\pm,m_I} = E^{(0)}_{\pm1,m_I} - E^{(0)}_{0,m_I} = D \pm (\omega_\mathrm{e} + A_\parallel m_I), \tag{S8}$$

where $\omega_\mathrm{e} = \gamma_\mathrm{e} B$. Substituting the flip-flop term into the second-order perturbation formula

$$\Delta E^{(2)}_{m_S,m_I} = \frac{A_\perp^2}{4} \sum_{\substack{(m_S',m_I') \\ \neq (m_S,m_I)}} \frac{|\langle m_S', m_I' | S_+I_- + S_-I_+ | m_S, m_I\rangle|^2}{E^{(0)}_{m_S,m_I} - E^{(0)}_{m_S',m_I'}},$$

the corresponding second-order energy corrections are obtained as

$$\Delta E^{(2)}_{+1,m_I} = \frac{A_\perp^2}{8} \frac{35 - 4m_I(m_I+1)}{D + \omega_\mathrm{e} + A_\parallel m_I - P(2m_I+1)},$$
$$\Delta E^{(2)}_{0,m_I} = -\frac{A_\perp^2}{8} \left[ \frac{35 - 4m_I(m_I-1)}{D + \omega_\mathrm{e} + A_\parallel (m_I - 1) - P(2m_I - 1)} + \frac{35 - 4m_I(m_I+1)}{D - \omega_\mathrm{e} - A_\parallel (m_I + 1) + P(2m_I + 1)} \right],$$
$$\Delta E^{(2)}_{-1,m_I} = \frac{A_\perp^2}{8} \frac{35 - 4m_I(m_I-1)}{D - \omega_\mathrm{e} - A_\parallel m_I + P(2m_I-1)}.$$

Here, $D \pm \omega_\mathrm{e}$ are in the order of 1 GHz, whereas $A_\parallel(m_I \pm 1)$ and $P(2m_I \pm 1)$ are expected to be at most a few tens of MHz. We therefore neglect the $A_\parallel$ and $P$ terms in the denominators of the second-order corrections. The resulting corrections to the resonance frequencies are

$$\delta\nu^{(2)}_{\pm,m_I} = \Delta E^{(2)}_{\pm1,m_I} - \Delta E^{(2)}_{0,m_I} \simeq \frac{{A_\perp}^2}{4} \left[ \frac{35 - 4m_I(m_I \pm 1)}{2(D \mp \omega_\mathrm{e})} + \frac{35 - 4m_I^2}{D \pm \omega_\mathrm{e}} \right],$$

Therefore, the hyperfine-resolved ODMR resonance frequency for a given $m_I$ is

$$\begin{aligned} \nu_{\pm,m_I} &\simeq \nu^{(0)}_{\pm,m_I} + \delta\nu^{(2)}_{\pm,m_I} \\ &= D \pm (\omega_\mathrm{e} + A_\parallel m_I) + \frac{{A_\perp}^2}{4} \left[ \frac{35 - 4m_I(m_I \pm 1)}{2(D \mp \omega_\mathrm{e})} + \frac{35 - 4m_I^2}{D \pm \omega_\mathrm{e}} \right]. \end{aligned} \tag{S9}$$

We characterized two defects shown in Figs. 4d and 4e in Samples E1 and E2, respectively. The defect indices correspond to those on the horizontal axis in Fig. 5b. We measured pulsed ODMR for both of $|m_s = 0\rangle \leftrightarrow |m_s = -1\rangle$ and $|m_s = 0\rangle \leftrightarrow |m_s = +1\rangle$ transitions under the aligned magnetic field. We obtained twelve hyperfine-split resonance frequencies using Gaussian fittings without any constraints. We treated $D$, $\omega_\mathrm{e}$, $A_\parallel$, and $|A_\perp|$ as free parameters and fitted the data of twelve resonance frequencies vs. $m_I$ using Eq. (S9).

The obtained value of $A_\parallel$ and $|A_\perp|$ are reported in the main text. The analysis yielded $D = 1355.612 \pm 0.010$ and $1351.181 \pm 0.011$ MHz for Figs. 4d and 4e, respectively. The obtained $D$ from Fig. 4d agrees well with the zero-field ODMR data (Supplementary Table 5). The $D$ value from Fig. 4e (Sample E2) is slightly larger (0.74 MHz) than that obtained from the $^{16}$O case fabricated by decoupled engineering, which may be attributed to the isotope effect. This analysis yields estimated magnetic fields of 80.399 ± 0.003 and 82.429 ± 0.004 G for Samples E1 and E2, respectively, consistent with the results extracted from the V2 center discussed in Supplementary Note 10.

We visualize the quadratic frequency shift as a function of $m_I$ due to $A_\perp$ in Eq. (S9). We fitted the experimentally obtained resonance frequencies $\nu_{\pm,m_I}$ to the linear relation in Eq. (S8) to obtain $\nu_{\mathrm{linfit}}$, and plotted the residuals in Supplementary Fig. 16. We overlaid the theoretically calculated resonance frequencies by Eq. (S9) after subtracting the same $\nu_{\mathrm{linfit}}$ from the experimental data. To calculate the theoretical $\nu_{\pm,m_I}$, we used $D$, $\omega_{\mathrm{e}}$, $A_\parallel$, and $|A_\perp|$ obtained from the fitting analysis above. The theoretical curves reproduce the experimental trends well, supporting the interpretation of the deviations from equal spacing in terms of a nonzero $A_\perp$.

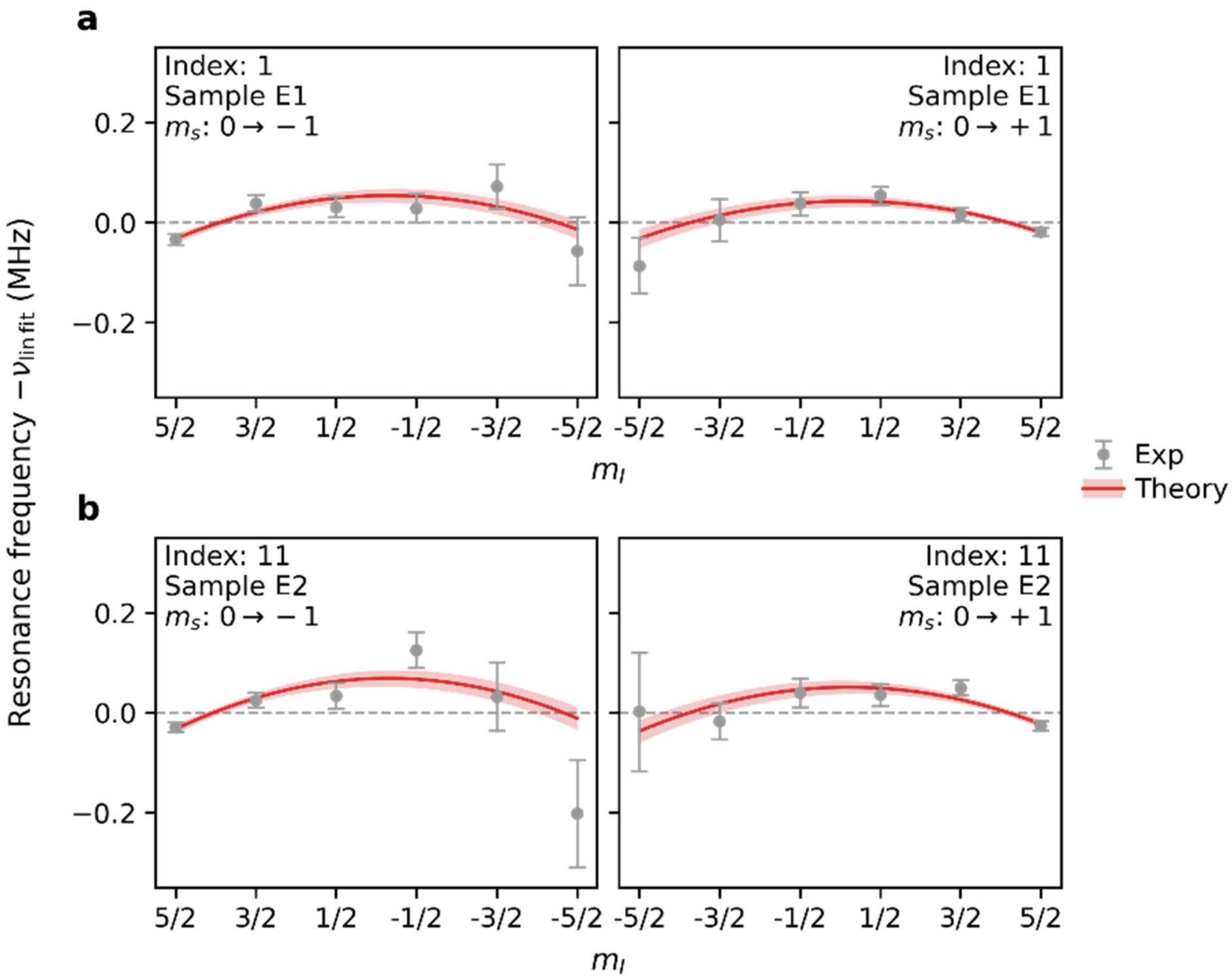

**Supplementary Fig. 16 | Deviation from linear dependence of hyperfine resonance frequencies on $m_I$. a**, The defect in Fig. 4d (Sample E1). **b**, The defect in Fig. 4e (Sample E2). Points represent the experimental resonance frequencies after subtracting $\nu_{\mathrm{linfit}}$. Error bars indicate the standard errors obtained from Gaussian fits. Red curves show the calculated resonance frequencies after subtracting the same $\nu_{\mathrm{linfit}}$, with shaded regions indicating their propagated $1\sigma$ uncertainties. The labeled defect indices 1 and 11 correspond to the horizontal axis in Fig. 5b.

## Supplementary Note 14: Numerical simulations of dynamic nuclear-spin polarization of $^{17}$O coupled with PL6.

We measured Rabi oscillations of hyperfine lines of PL6 in Sample E2 (defect index 10) under the aligned-field condition at the same microwave power (Supplementary Fig. 17). We found that Rabi frequencies are equal. Still, contrasts are $m_I$-dependent. These results indicate the identical electronic transition matrix but unequal nuclear spin populations, suggesting the occurrence of optically induced dynamic nuclear spin polarization (DNP).

The underlying mechanism is likely to involve electron-nuclear spin flip-flop transitions in the excited state,[16]but our magnetic field is far from the excited-state level anticrossing (ESLAC). To see if DNP is possible at a field far from ESLAC, we performed numerical simulations based on the Lindblad master equation implemented in QuTiP[17]. Each of the ground state (GS) and excited state (ES) manifolds comprises 18 product basis states ($|m_s, m_I\rangle$), arising from three electron-spin states and six nuclear-spin states. The metastable state (MS) manifold was described by the six nuclear-spin states only. Therefore, the total Hilbert space contains 42 basis states. The GS and ES Hamiltonians $H_0^{\mathrm{s}}$ (s = GS, ES) for an axially symmetric defect PL6 are given by:

$$H_0^{\mathrm{s}} = D^{\mathrm{s}} S_z^2 + \gamma_{\mathrm{e}} \mathbf{B} \cdot \mathbf{S} + A_{\parallel}^{\mathrm{s}} S_z I_z + \frac{A_{\perp}^{\mathrm{s}}}{2}(S_+ I_- + S_- I_+) - \gamma_{\mathrm{n}} \mathbf{B} \cdot \mathbf{I} + \frac{C_{\mathrm{q}}}{4I(2I-1)}\left(3I_z^2 - I(I+1)\right). \quad (10)$$

Here, $D^{\mathrm{s}}$ is the axial ZFS and $\gamma_{\mathrm{n}} = -0.577$ kHz/G is $^{17}$O nuclear-spin gyromagnetic ratio.[11] We assumed $\gamma_{\mathrm{e}} = 2.803$ MHz/G for the gyromagnetic ratio for electron spin. $A_{\perp}^{\mathrm{s}}$ and $A_{\parallel}^{s}$ are the transverse and axial hyperfine interaction constants, respectively. $C_{\mathrm{q}}$ is a quadrupole-interaction constant, assuming an axially symmetric interaction based on an axial oxygen-vacancy model. $\mathbf{B} = |B_0|(\sin\theta, 0, \cos\theta)$ is an external magnetic field, where $|B_0|$ is the magnetic field strength and $\theta$ is the angle between *c*-axis and the magnetic field. In the present calculations, we used $D^{\mathrm{GS}} = 1351.181$ MHz, $A_{\parallel}^{\mathrm{GS}} = 3.295$ MHz, and $A_{\perp}^{\mathrm{GS}} = 3.46$ MHz, obtained experimentally from PL6 shown in Figure 4e (Supplementary Note 13). We used $D^{\mathrm{ES}} = 940$ MHz as reported experimentally[18] and $C_{\mathrm{q}} = 7.04$ MHz from the first-principles study.[5] Here, we assumed the same constants of the quadrupole interaction for both GS and ES.

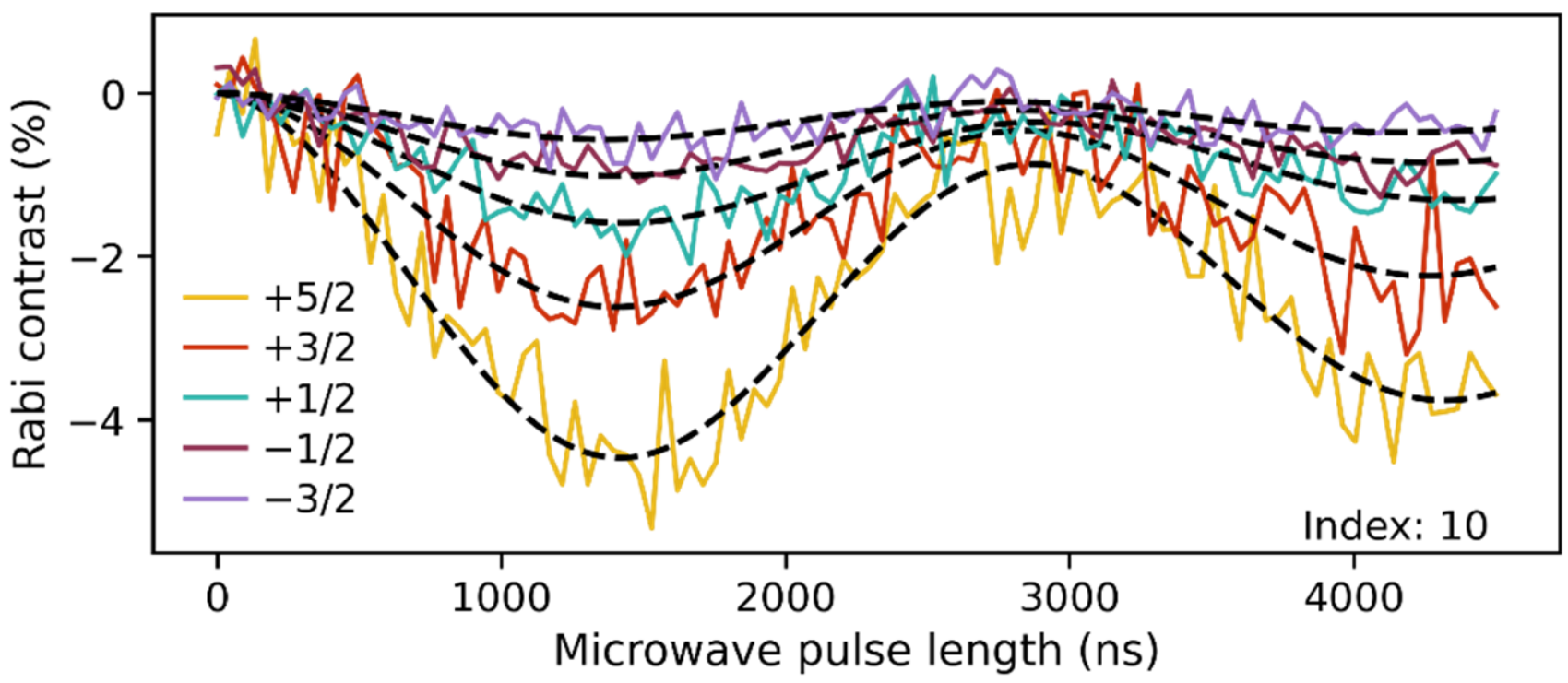

**Supplementary Fig. 17 | Rabi oscillations of individual hyperfine lines of PL6 coupled to 17O under the aligned magnetic field.** Measurements were performed on the $m_S = 0 \leftrightarrow -1$ transition for 5 hyperfine lines corresponding to $m_I = +5/2, \ldots, -3/2$ (defect index 10 in Fig. 5b; corresponding ODMR spectrum shown in Supplementary Fig. 14).

The spin dynamics involved in this simulation are as follows (Supplementary Fig. 18a): optical excitation from the GS to the ES, radiative decay from the ES to the GS, intersystem crossing (ISC) from the ES to the metastable state (MS), and ISC from the MS to the GS. We also considered the electron-spin $T_1$ relaxation ($T_{1,\mathrm{e}}$) within the GS, the electron-spin decoherence $T_2^*$ in the ES, and the $^{17}$O nuclear-spin $T_1$ relaxation ($T_{1,\mathrm{O}}$) through the whole system, whereas spin decoherence in the GS, such as $T_2$ and $T_2^*$, were not taken into consideration of our simulation. They were modeled as Lindblad dissipators. Although spin-dependent ES lifetimes have been reported,[19] individual ISC rates and the radiative decay rate have not been reported. Therefore, we assumed the MS lifetime to be 100 ns and the radiative decay rate to be 50 MHz and then set ISC rates to reproduce the Rabi contrast of 20% on PL6 measured in $^{16}$O-doped decoupled engineered SiC. We used $T_{1,\mathrm{e}} = 200$ µs for the GS[5] and $T_2^* = 5$ ns for the ES.[18,20] Although the $T_{1,\mathrm{O}}$ for $^{17}$O nuclear spin coupled with PL6 has remained unknown, the low-temperature $T_1$ of $^{29}$Si nuclear spin coupled to divacancy has been reported to be 4 s under optical excitation and around 2.5 hours in dark.[21] Also, we confirmed that our simulation results are not substantially affected when $T_{1,\mathrm{O}}$ is longer than 50 ms. Thus, in this simulation, we set $T_{1,\mathrm{O}}$ to be 1 s as a tentative value. The laser excitation rate was set to correspond to the power beyond saturation. The used rates are listed in Supplementary Table 6.

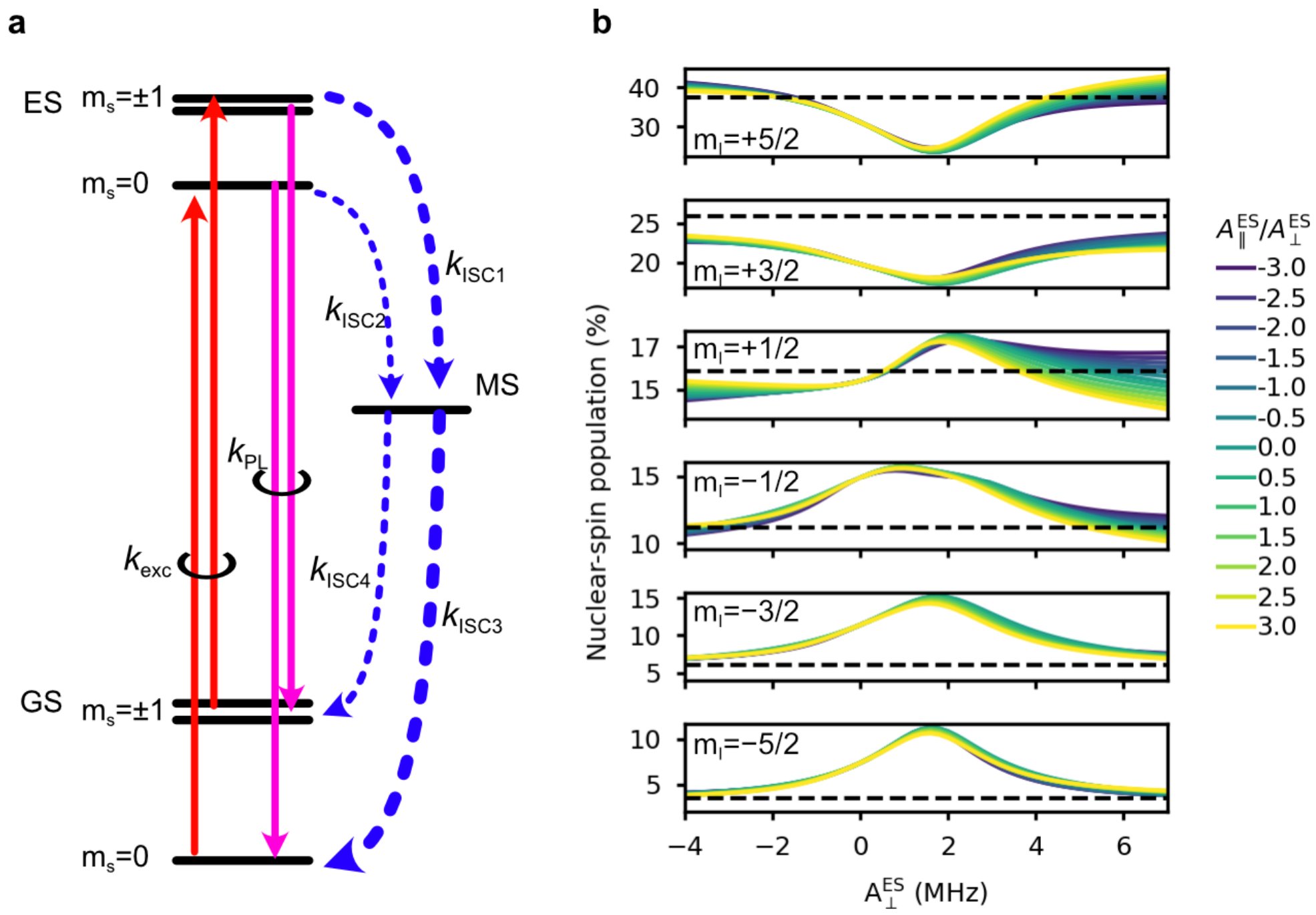

**Supplementary Fig. 18 | The evaluation of the DNP of $^{17}$O nuclear spin in PL6. a,** Simplified PL6's energy diagram and electron spin dynamics. Red and pink solid arrows show spin-conserving optical transitions of the laser excitation and radiative relaxation, respectively. Blue dashed arrows indicate the ISC relaxations. The corresponding transition rates are listed in Supplementary Table 6**. b,** Simulated DNP-induced nuclear-spin populations of $^{17}$O at $|B_0| = 82.429$ G and $\theta = 0°$ with various $A_{\perp}^{\mathrm{ES}}$ and $A_{\parallel}^{\mathrm{ES}}/A_{\perp}^{\mathrm{ES}}$. Dashed black lines show the experimentally obtained populations, estimated from the ODMR spectrum shown in Figure 4e.

**Supplementary Table 6 | Transition rates and spin-coherence times used in Lindblad master equations.**

| Transitions | Rate |
|---|---|
| $k_{\mathrm{exc}}$ (Laser excitation) | 500 MHz |
| $k_{\mathrm{PL}}$ (Radiative relaxation) | 50 MHz |
| $k_{\mathrm{ISC1}}$ (ISC from ES $m_s = \pm 1$ to MS) | 62.36 MHz |
| $k_{\mathrm{ISC2}}$ (ISC from ES $m_s = 0$ to MS) | 24.63 MHz |
| $k_{\mathrm{ISC3}}$ (ISC from MS to GS $m_s = 0$) | 7.6 MHz |
| $k_{\mathrm{ISC4}}$ (ISC from MS to GS $m_s = 0$) | 2.4 MHz |
| $T_{1,\mathrm{e}}$ for the GS | 200 µs |
| $T_2^*$ for the ES | 5 ns |
| $T_{1,\mathrm{O}}$ | 1 s |

We first analyze the electron-nuclear spin flip-flop transitions in the ES, which contribute to the DNP. Assuming $\theta = 0$, the two-dimensional subspace Hamiltonian for $|0, m_I\rangle$ and $|-1, m_I + 1\rangle$ is given by

$$H_{\mathrm{eff}} = \begin{pmatrix} -\Delta/2 & V \\ V & \Delta/2 \end{pmatrix}, \tag{11}$$

$$\Delta = D^{\mathrm{ES}} - (\gamma_{\mathrm{e}} + \gamma_{\mathrm{n}})B_0 - A_{\parallel}^{\mathrm{ES}}(m_I + 1) + \frac{3C_{\mathrm{q}}}{4I(2I-1)}(2m_I + 1), \tag{12}$$

$$V = \frac{V_{\perp}^{\mathrm{ES}}}{\sqrt{2}}\sqrt{I(I+1) - m_I(m_I + 1)} \tag{13}$$

The flip-flop probability $P$ from $|\mathrm{i}\rangle = |0, m_I\rangle$ to $|\mathrm{f}\rangle = |-1, m_I + 1\rangle$ is obtained by[20]

$$P = \int_0^T \rho(\tau)\langle \mathrm{f}|e^{-i2\pi H_{\mathrm{eff}}t}|\mathrm{i}\rangle \, dt, \tag{14}$$

where $\rho(t) = (1/\tau_{\mathrm{ES}}^*)\exp(-t/\tau_{\mathrm{ES}}^*)$ is the probability distribution function of the effective length of the ES evaluation time $\tau_{\mathrm{ES}}^*$.[20] By integrating from $t = 0$ to $\infty$ as an approximation, we obtain

$$P = \frac{2V^2}{\Delta^2 + 4V^2 + (2\pi\tau_{\mathrm{ES}}^*)^{-2}}.$$

By using $\tau_{\mathrm{ES}}^* = 5.1$ ns reported by Ivády et al.[20], we found that the probability for $I = 5/2$ nuclei is 5–9 times larger than those for $I = 1/2$ nuclei, such as $^{13}$C and $^{29}$Si, at $B_0 = 80$ G and the same hyperfine constants. Therefore, substantial DNP can be expected at magnetic fields far from the ESLAC.

Secondly, we calculated the DNP-induced nuclear-spin populations (Supplementary Fig. 18b). We found that larger $|A_{\perp}^{\mathrm{ES}}|$ induces the higher populations on $m_I = +5/2$ and $+3/2$ states and the smaller for the rest, and the larger $A_{\parallel}^{\mathrm{ES}}$ slightly increases the population on the $m_I = +5/2$ state. Also, compared with the populations estimated from the peak areas of ODMR spectrum shown in Supplementary Fig. 14, our calculations under the aligned magnetic field condition within the considered range of the ratio $|A_{\parallel}^{\mathrm{ES}}/A_{\perp}^{\mathrm{ES}}| \leq 3$ in Supplementary Fig. 18b suggested that $A_{\perp}^{\mathrm{ES}}$ should be larger than 4.5 MHz or less than −1 MHz. Here, we found that the simulated DNP remains finite even at $A_{\perp}^{\mathrm{ES}} = 0$ MHz and reaches a minimum at $A_{\perp}^{\mathrm{ES}} \cong 1.5$ MHz. This behavior cannot be explained by the ES-mediated DNP mechanism discussed above and may be attributed to additional DNP caused by the flip-flop transitions in the GS.

We next simulated the ODMR spectrum after numerous repetitions of the pulsed-ODMR sequence, where the unit sequence consists of a 350-ns laser pulse, a 500-ns dark interval, a 1-μs microwave $\pi$ pulse, and an 80-ns dark time, and compared it with the experimental results obtained on defect index 11 (Supplementary Fig. 14). Supplementary Fig. 19a shows the ODMR spectra under an aligned magnetic field, and our simulations reproduce the experimental $m_I$-dependent ODMR contrast very well, which also agree with that observed in the Rabi oscillations (Supplementary Fig. 17).

We further simulated the ODMR under a misaligned magnetic field as shown in Supplementary Fig. 19b, exhibiting approximately uniform contrasts and broadened linewidths in the two central peaks, which agree with the experimental spectrum. Also, we found that $A_{\perp}^{\mathrm{ES}}$ should have the positive sign to reproduce ODMR spectra under both aligned and misaligned magnetic field. We also calculated the DNP as a function of the magnetic-field angle ($\theta$) and found that by increasing $\theta$, the nuclear-spin population imbalance decreases and is slightly redistributed toward the $|m_I = \pm 1/2\rangle$ and $|m_I = \pm 3/2\rangle$ states (Supplementary Fig. 19c). In addition, by analyzing the GS eigenstates, we found that the magnetic-field misalignment induces state mixing within the GS due to $A_\perp^{\rm GS}$ and the quadrupole interaction, particularly between $|0,-1/2\rangle$ and $|0,+1/2\rangle$, enabling otherwise forbidden transitions that appear as weak additional peaks near the two central peaks (Supplementary Fig. 19d). Thus, $m_I$-dependent contrast disappears, and the two central peaks appear with apparently broadened linewidths.

Finally, we discuss the discrepancy between simulations and experiments. We found that our simulations cannot fully reproduce the experiments in the whole ODMR contrast and the contrast ratio across the six peaks, which may be attributed to differences in ISC rates and magnetic field angles and to decoherence. Also, since the DNP is very sensitive to the misalignment angle of the magnetic field, actual $A_\perp^{\rm ES}$ could be larger if we had a slight misalignment angle for the magnetic field in the experiments. To determine $A_\parallel^{\rm ES}$ and $A_\perp^{\rm ES}$ accurately, further investigation is required. In addition, we obtained almost the same linewidth in the simulations as in the experiments, although the applied $\pi$-pulse length was not consistent. This observation can be attributed to linewidth broadening in the experiments due to a short $T_2^*$ or unresolved small hyperfine splitting.

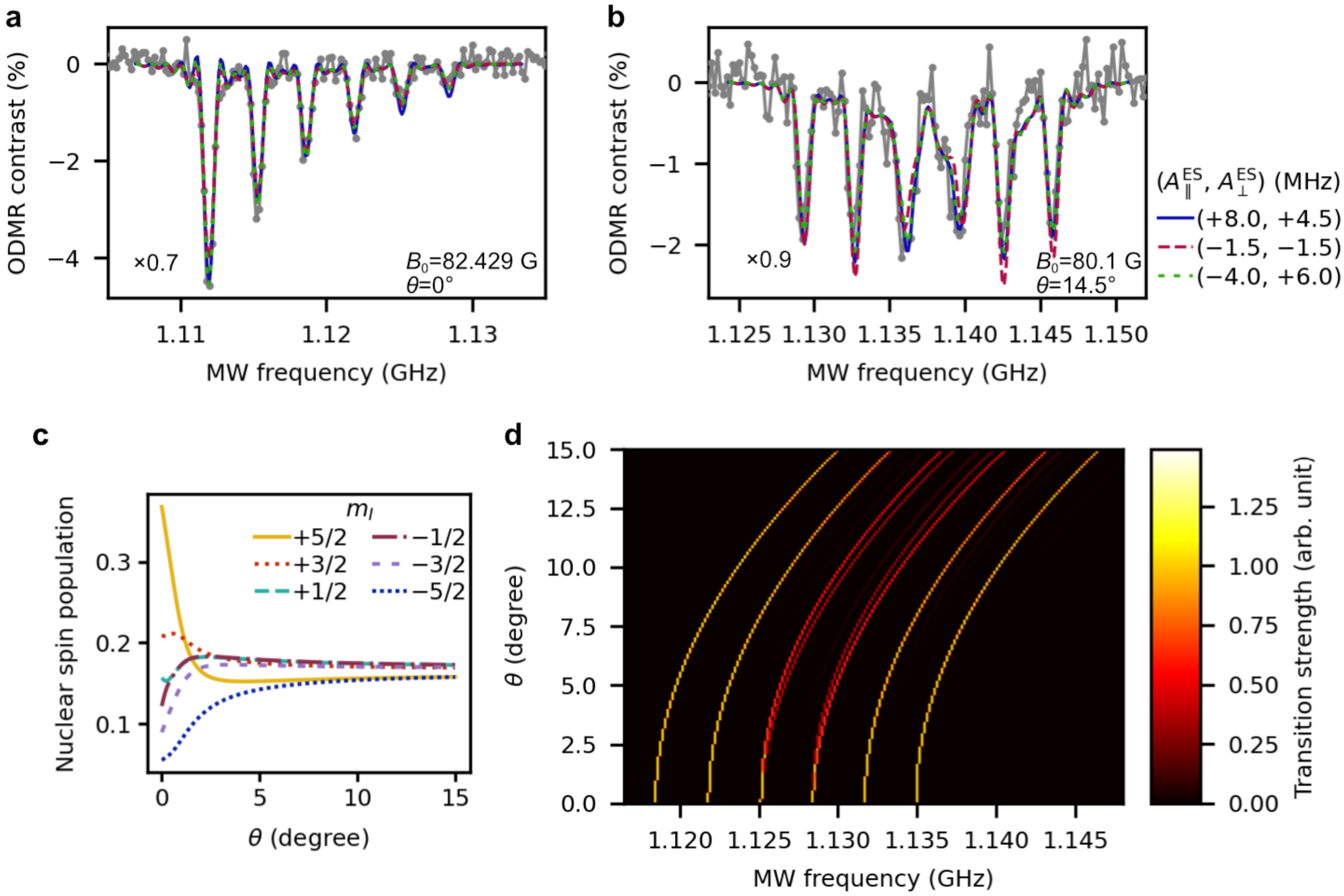

**Supplementary Fig. 19 | ODMR simulations with DNP process.** Simulated ODMR spectra with different $A_\parallel^{\rm ES}$ and $A_\perp^{\rm ES}$ and experimental results of PL6 shown in Figure 4e (corresponding to defect index 11 in Figure 5, grey dots) for **a**, aligned**,** and **b**, misaligned magnetic field conditions**.** For the simulations, contrast magnitudes are scaled as labeled to match the experiments for comparison. The magnetic-field angle dependence of **c**, the nuclear spin polarizations and **d**, the transition strength in the GS at $|B_0| = 80.1$ G. For **c**, we set $A_\parallel^{\rm ES} = 8$ MHz and $A_\perp^{\rm ES} = 4.5$ MHz, respectively.

## Supplementary Note 15: Analysis of PL6 ODMR spectra for hyperfine-splitting statistics

### 15.1 Fitting procedure for the ODMR spectrum in Fig. 5a (defect index 16 in Fig. 5b)

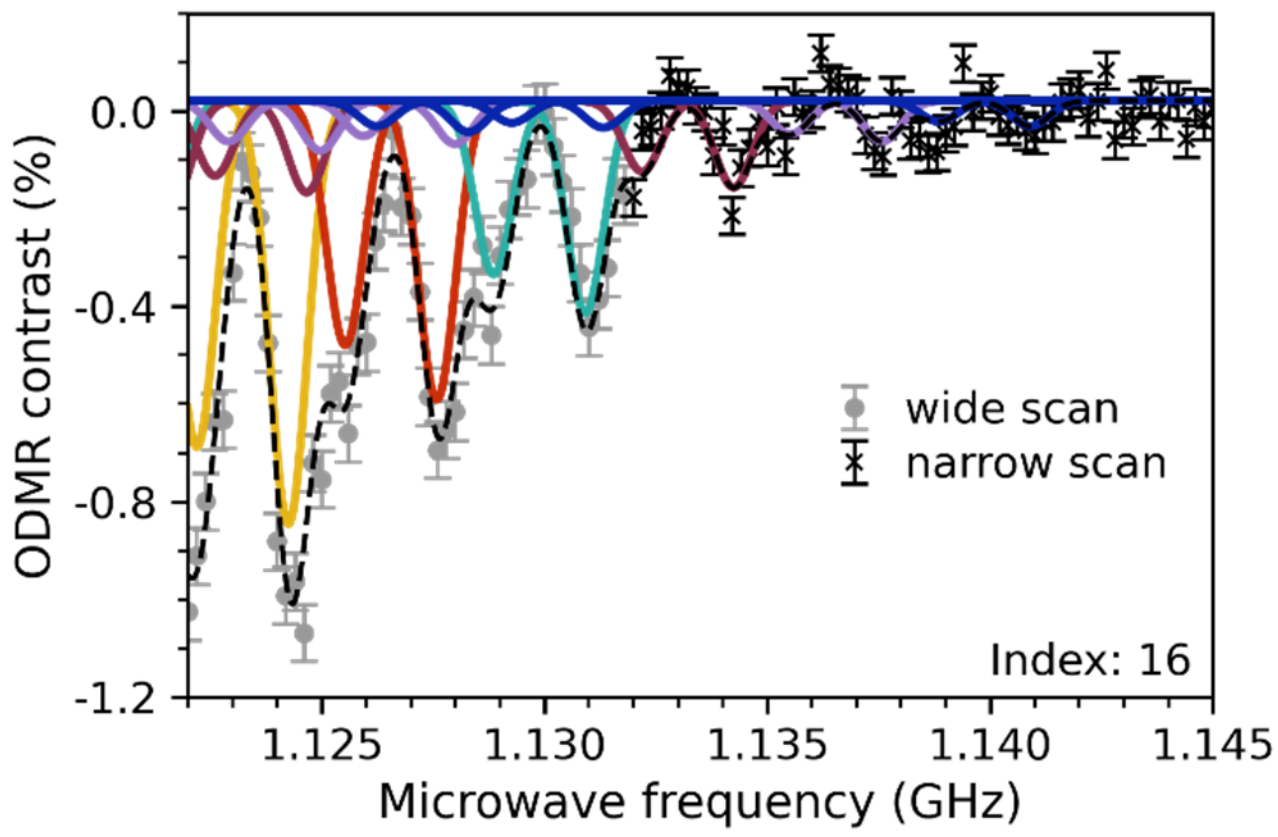

**Supplementary Fig. 20 | Enlarged view of the junction between the wide- and narrow-scan ODMR datasets used for fitting the spectrum in Fig. 5a.**

We performed a separate narrow-scan ODMR measurement in the low-contrast region ($f$ =1.132–1.145 GHz) with a longer integration time to improve the signal-to-noise ratio (inset of Fig. 5a). For fitting, we used the wide-scan data at $f < 1.132$ GHz with the narrow-scan data. The wide- and narrow-scan data were normalized by linear and constant backgrounds, respectively, each estimated from the corresponding baseline regions. The two normalized datasets were concatenated without additional scaling or offset correction. An enlarged view of the combined spectrum around the junction is shown in Supplementary Fig. 20. The combined spectrum was fitted using the procedure described in Supplementary Note 15.2, and the resulting fit is shown in Fig. 5a.

### 15.2 Fitting procedures for hyperfine statistics for Fig. 5

To analyze the hyperfine splittings shown in Fig. 5b, we approximated the six components arising from the $I = 5/2$ nuclear spin as equally spaced for simplicity. The amplitudes of these six components were fitted independently. Additional splittings were modeled using up to three $I = 1/2$ nuclear spins, each producing symmetric splittings of the existing components with a nucleus-specific frequency separation. The amplitude of each resulting component was expressed as the product of the corresponding $I = 5/2$ amplitude and nucleus-specific factors. For each $I = 1/2$ nucleus $j$, the factors for the $m_I = +1/2$ and $-1/2$ branches had a ratio of $r_j$: 1, independent of the states of the other nuclei. A common linewidth was assumed for all Gaussian components, except for the spectrum of defect index 19 measured under the misaligned field condition (Supplementary Fig. 22), for which no $I = 1/2$ splitting was resolved and the six $I = 5/2$ components were fitted with independent linewidths.

The experimental pulsed ODMR spectra of PL6 coupled with $^{17}O$ nuclear spins and corresponding fits are shown in Supplementary Figs. 21–25 for all defects included in Fig. 5b, except for defect index 16 (shown in Fig. 5a). The spectra are grouped by similarity. The defect indices correspond to those on the horizontal axis of Fig. 5b. $t_\pi$ denotes the $\pi$-pulse length.

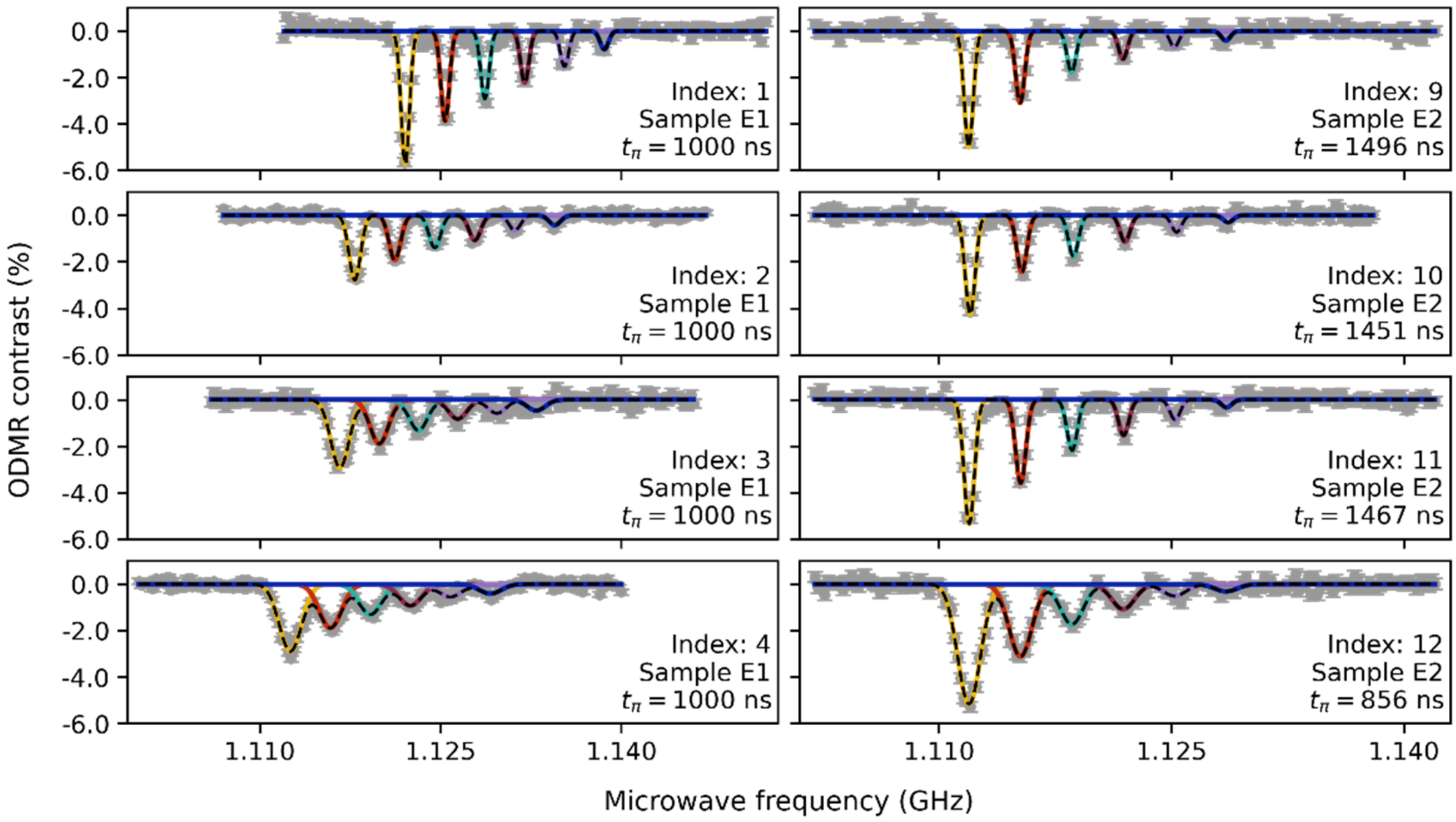

**Supplementary Fig. 21 | Aligned-field ODMR spectra of PL6 without resolved $I = 1/2$ hyperfine splitting.** The hyperfine splittings are consistent across defects, whereas the sextet center frequencies vary in Sample E1 (left; ion implantation) but remain nearly uniform in Sample E2 (right; decoupled engineering). This difference in center-frequency variation between the two engineering methods is consistent with the corresponding difference in ZFS variation observed in $^{16}O$ samples (Figs. 2d and 2f).

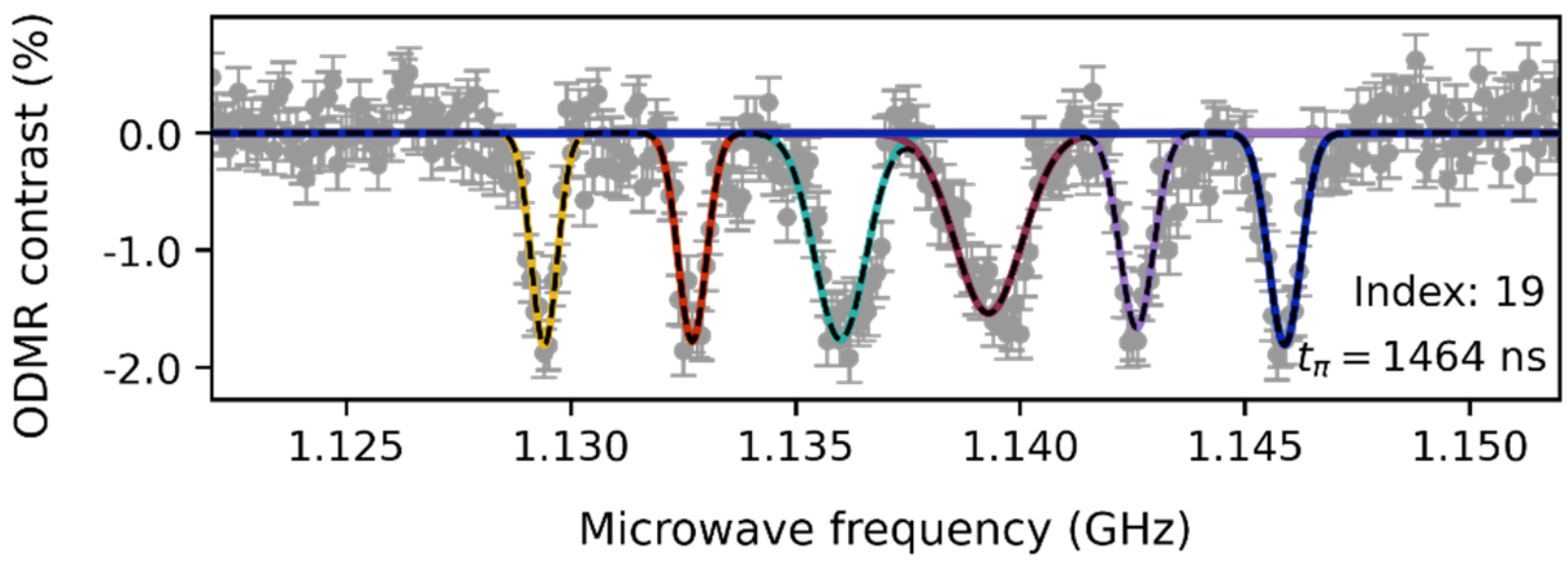

**Supplementary Fig. 22 | Misaligned-field ODMR spectrum of PL6 without resolved $I = 1/2$ splitting.**

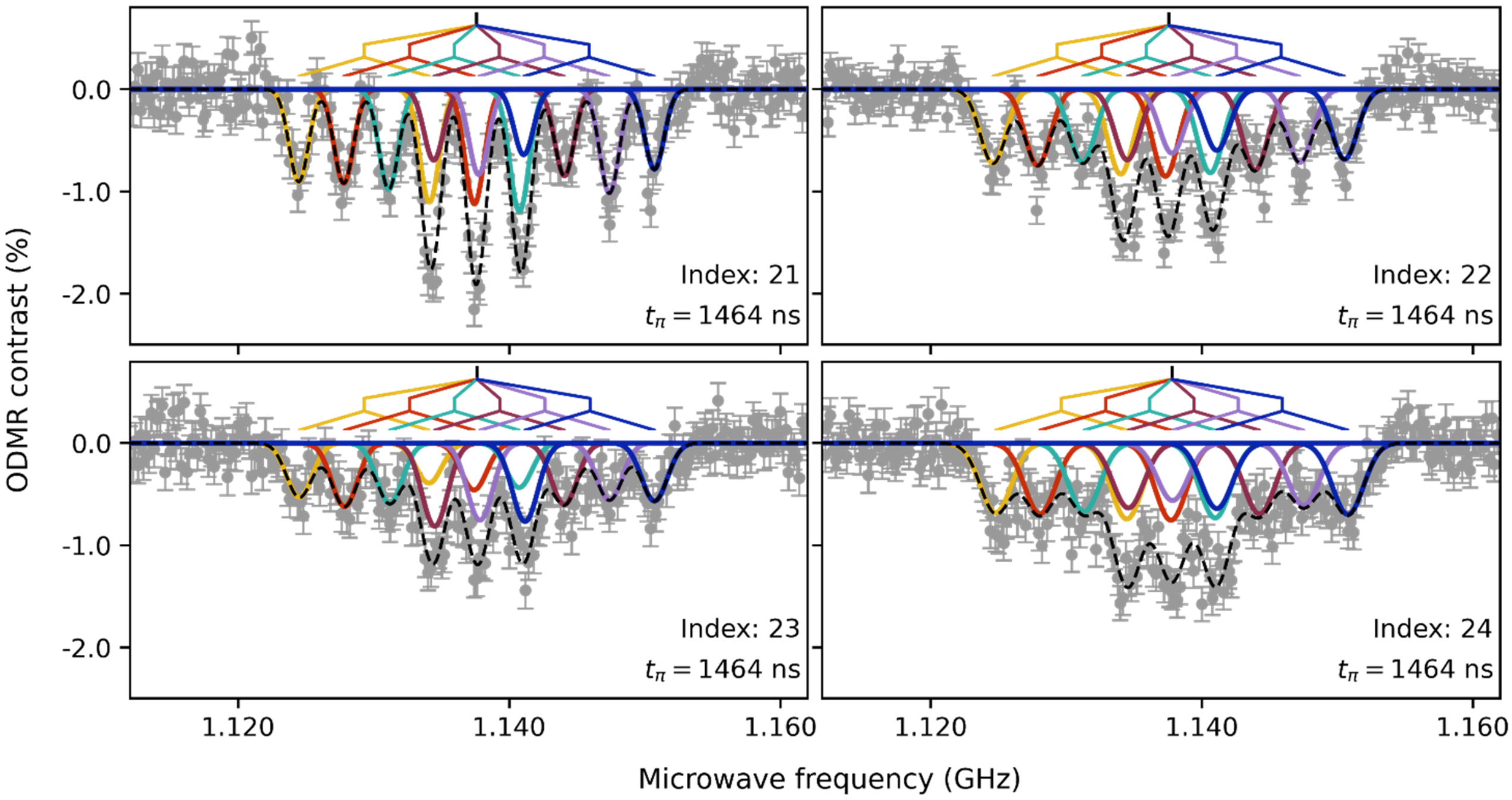

**Supplementary Fig. 23 | Misaligned-field ODMR spectra of PL6 resolving an additional ~9.6 MHz splitting.**

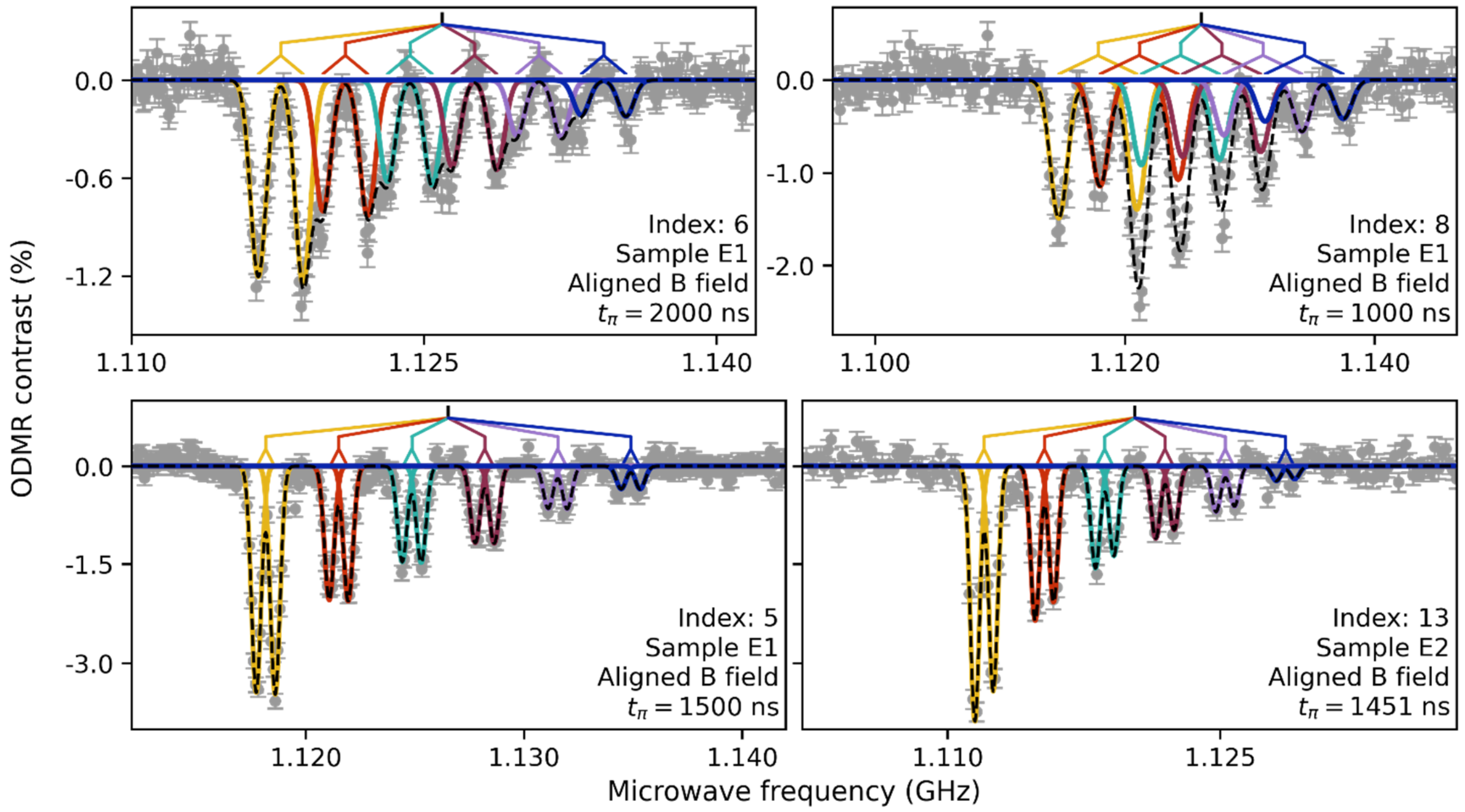

**Supplementary Fig. 24 | Aligned-field ODMR spectra of PL6 resolving an additional splitting.**

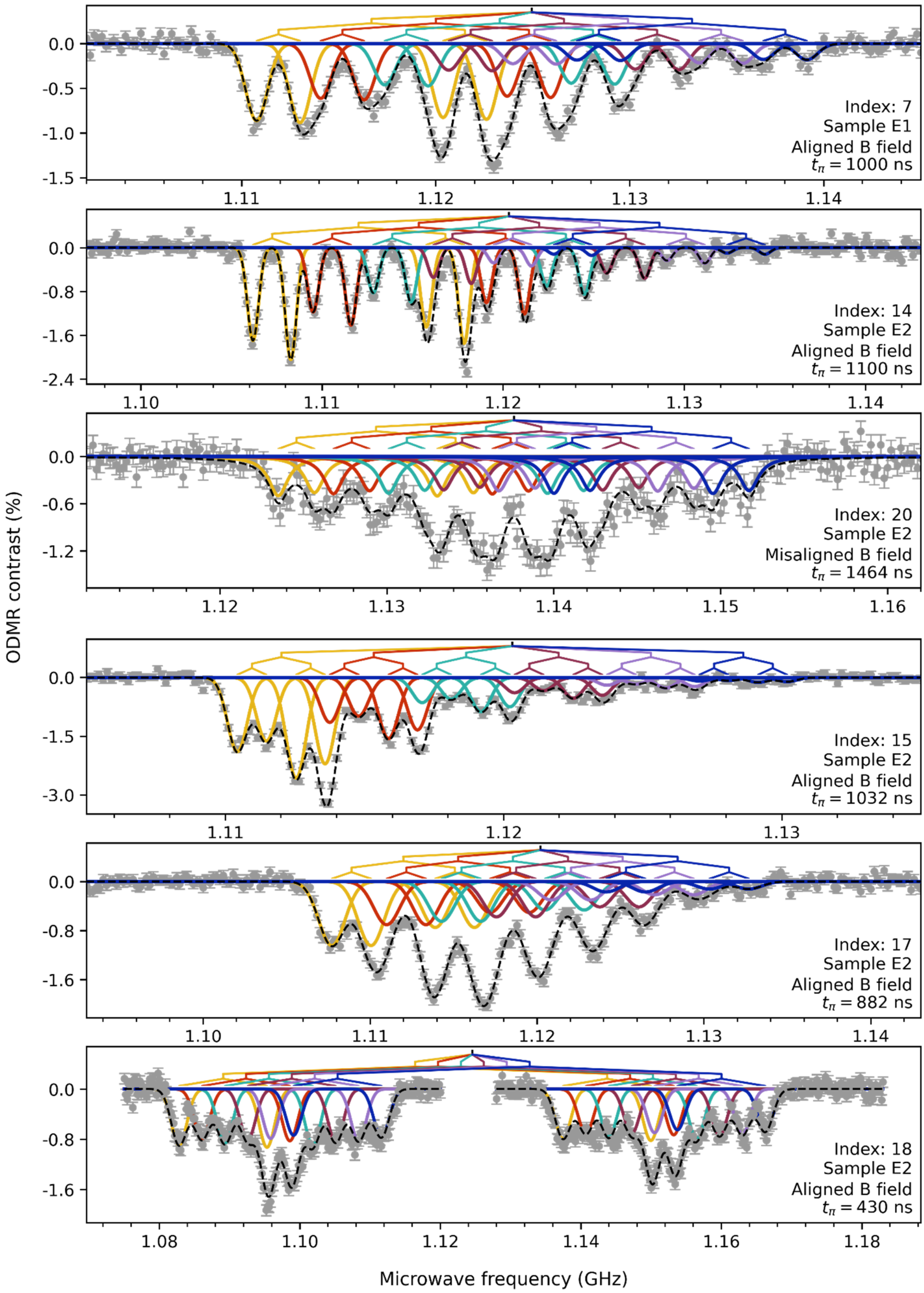

**Supplementary Fig. 25 | ODMR spectra of the PL6s resolved with two $I = 1/2$ hyperfine splittings.**

### 15.3 PL6 defects without a resolved $I = 5/2$ hyperfine signature

As mentioned in the main text, we measured 27 defects and found three PL6 defects that did not show a resolved $I = 5/2$ hyperfine signature. We show their ODMR spectra in Supplementary Fig. 26. The left two spectra exhibited low contrasts and broad linewidths, most probably due to coupling to too many host nuclear spins. The right spectrum exhibited a doublet consistent with PL6 with $I = 0$ isotope of oxygen ($^{16}O$ or $^{18}O$) and 2.2 MHz $I = 1/2$ host nuclear spin, potentially originating from oxygen incorporated to the surface through sacrificial oxidation.[22]

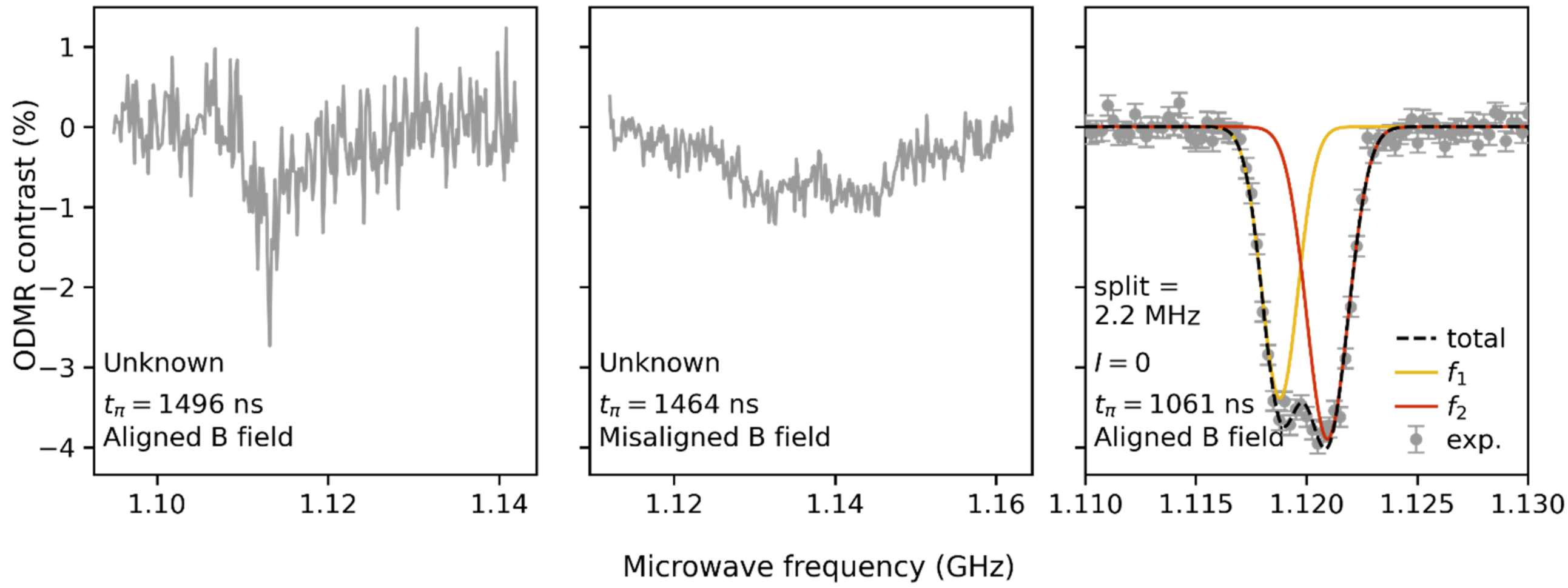

**Supplementary Fig. 26 | ODMR spectra of three PL6s in Sample E2 with unresolved $I = 5/2$ hyperfine structure.**

## Supplementary References

1. Kobayashi, T., Shimura, T. & Watanabe, H. Oxygen-vacancy defect in 4H-SiC as a near-infrared emitter: An ab initio study. *J. Appl. Phys.* **134**, 145701 (2023).

2. Son, N. T. *et al.* Divacancy in 4H-SiC. *Phys. Rev. Lett.* **96**, 055501 (2006).

3. Zhou, J.-Y. *et al.* Experimental Determination of the Dipole Orientation of Single Color Centers in Silicon Carbide. *ACS Photonics* **8**, 2384–2391 (2021).

4. Eserin, S. *et al.* Efficient formation and identification of single emitters in 4H-SiC following maskless heavy ion implantation. Preprint at https://doi.org/10.48550/arXiv.2608.27281 (2026).

5. Hu, Q.-C. *et al.* High-Yield Engineering and Identification of Oxygen-Related Modified Divacancies in 4H-SiC. *Adv. Mater.* **38**, e73419 (2026).

6. Filzmoser, P., Garrett, R. G. & Reimann, C. Multivariate outlier detection in exploration geochemistry. *Comput. Geosci.* **31**, 579–587 (2005).

7. Brouri, R., Beveratos, A., Poizat, J.-P. & Grangier, P. Photon antibunching in the fluorescence of individual color centers in diamond. *Opt. Lett.* **25**, 1294–1296 (2000).

8. Widmann, M. *et al.* Coherent control of single spins in silicon carbide at room temperature. *Nat. Mater.* **14**, 164–168 (2015).

9. Seo, H. *et al.* Quantum decoherence dynamics of divacancy spins in silicon carbide. *Nat. Commun.* **7**, 12935 (2016).

10. Van Oort, E. & Glasbeek, M. Optically detected low field electron spin echo envelope modulations of fluorescent N-V centers in diamond. *Chem. Phys.* **143**, 131–140 (1990).

11. Harris, R. K., Becker, E. D., Cabral De Menezes, S. M., Goodfellow, R. & Granger, P. NMR nomenclature: Nuclear spin properties and conventions for chemical shifts (IUPAC recommendations 2001). *Concepts Magn. Reson.* **14**, 326–346 (2002).

12. Morioka, N., Nishikawa, T., Abe, H., Ohshima, T. & Mizuochi, N. Photoelectrical detection and characterization of divacancy and PL5–PL7 spins in silicon carbide. *Phys. Rev. B* **113**, 104426 (2026).

13. Soltamov, V. A. *et al.* Electron nuclear interactions in spin-3/2 color centers in silicon carbide: A high-field pulse EPR and ENDOR study. *Phys. Rev. B* **104**, 125205 (2021).

14. Kothandaraman, G., Pratt, D. W. & Tinti, D. S. Optically detected magnetic resonance spectra of the lowest triplet state of 1-iodonaphthalene in durene at zero and high magnetic fields. *J. Chem. Phys.* **63**, 3337–3348 (1975).

15. Yamauchi, S. & W. Pratt, D. Oxygen-17 hyperfine structure in the zero-field odmr spectrum of triplet state benzophenone. *Chem. Phys. Lett.* **60**, 145–149 (1978).

16. Jacques, V. *et al.* Dynamic Polarization of Single Nuclear Spins by Optical Pumping of Nitrogen-Vacancy Color Centers in Diamond at Room Temperature. *Phys. Rev. Lett.* **102**, 057403 (2009).

17. Lambert, N. *et al.* QuTiP 5: The Quantum Toolbox in Python. *Phys. Rep.* **1153**, 1–62 (2026).

18. Falk, A. L. *et al.* Optical Polarization of Nuclear Spins in Silicon Carbide. *Phys. Rev. Lett.* **114**, 247603 (2015).

19. Li, Q. *et al.* Room-temperature coherent manipulation of single-spin qubits in silicon carbide with a high readout contrast. *Natl. Sci. Rev.* **9**, nwab122 (2022).

20. Ivády, V. *et al.* Theoretical model of dynamic spin polarization of nuclei coupled to paramagnetic point defects in diamond and silicon carbide. *Phys. Rev. B* **92**, 115206 (2015).

21. Zeledon, C. *et al.* Minute-long quantum coherence enabled by electrical depletion of magnetic noise. Preprint at https://doi.org/10.48550/arXiv.2504.13164 (2026).

22. Hatmanto, A. D. & Kita, K. The kinetics of lattice distortion introduction and lattice relaxation at the surface of thermally-oxidized 4H-SiC (0001). *Appl. Phys. Express* **12**, 055505 (2019).